\documentclass[11pt]{article}

\usepackage{verbatim}
\usepackage{amsmath}
\usepackage{amsfonts}
\usepackage{amssymb}
\usepackage{bbm}
\usepackage{latexsym}
\usepackage{lscape} 
\usepackage{epsfig}
\usepackage{color}
\usepackage{graphicx}
\usepackage{float}
\usepackage{subfig}
\usepackage{slashbox}
\usepackage{multirow}
\usepackage{mathrsfs}

\usepackage{amscd}
\usepackage{cite}

\renewcommand{\d}{\mathrm{d}}

\DeclareMathSymbol{\mg}{\mathrel}{symbols}{"1D}

\newcommand{\ga}{\alpha}
\newcommand{\gb}{\beta}
\renewcommand{\gg}{\gamma}
\newcommand{\gd}{\delta}

\newcommand{\gve}{\varepsilon}

\newcommand{\gk}{\kappa}
\newcommand{\gl}{\lambda}
\newcommand{\gr}{\rho}
\newcommand{\gth}{\theta}

\newcommand{\gt}{\tau}

\newcommand{\gz}{\zeta}
\newcommand{\gp}{\pi}
\newcommand{\gps}{\psi}

\newcommand{\gL}{\Lambda}
\newcommand{\gS}{\Sigma}

\newcommand{\cC}{{\cal C}}

\newcommand{\cP}{{\cal P}}

\newcommand{\cV}{{\cal V}}

\newcommand{\cX}{{\cal X}}
\newcommand{\cY}{{\cal Y}}
\newcommand{\cZ}{{\cal Z}}
\newcommand{\Id}{\text{\small 1}\hspace{-3.5pt}\text{1}}

\newcommand{\ra}{\rightarrow}

\newcommand{\Kh}{K\"{a}hler}
\newcommand{\beq}{\begin{equation}}
\newcommand{\eeq}{\end{equation}}
\newcommand{\barr}{\begin{array}}
\newcommand{\earr}{\end{array}}
\newcommand{\equ}[1]{\begin{gather} #1 \end{gather}}
\newcommand{\equa}[1]{\begin{align} #1 \end{align}}
\newcommand{\items}[1]{\begin{itemize} #1 \end{itemize}}
\newcommand{\enums}[1]{\begin{enumerate} #1 \end{enumerate}}

\newcommand{\arry}[2]{\begin{array}{#1} #2 \end{array}}

\newcommand{\pmtrx}[1]{\begin{pmatrix} #1 \end{pmatrix}}

\newcommand{\non}{\nonumber}
\newcommand{\sfrac}[2]{\mbox{$\frac{#1}{#2}$}}
\newcounter{oldcounter}

\newcommand{\bD}{{\overline D}}

\newcommand{\Intr}{\mathbb{Z}}
\newcommand{\Cplx}{\mathbbm{C}}
\newcommand{\Real}{\mathbbm{R}}

\newcommand{\ba}[2]{\[\begin{array}{#2}\label{#1}}
\newcommand{\ea}{\end{array}\]}
\newcommand{\be}{\begin{equation}}
\newcommand{\ee}{\end{equation}}
\newcommand{\bea}{\begin{eqnarray}}
\newcommand{\eea}{\end{eqnarray}}

\newcommand{\arrystrch}{1.3}

\begin{document}
\thispagestyle{empty}

\begin{flushright}
LMU-ASC 71/11 \\ 
\end{flushright}
\vskip 2 cm
\begin{center}
{\Large {\bf Gauged Linear Sigma Models for toroidal orbifold resolutions} 
}
\\[1cm]

\bigskip
\bigskip {\large
{\bf Michael Blaszczyk$^{a,}$}\footnote{
E-mail: michael@th.physik.uni-bonn.de},
{\bf Stefan Groot Nibbelink$^{b,}$}\footnote{
E-mail: Groot.Nibbelink@physik.uni-muenchen.de},
{\bf Fabian Ruehle$^{a,}$}\footnote{
E-mail: ruehle@th.physik.uni-bonn.de}
\bigskip }\\[0pt]
\vspace{0.23cm}
${}^a$ {\it Bethe Center for Theoretical Physics,\\
~~Physikalisches Institut der Universit\"at Bonn, 
Nussallee 12, 53115 Bonn, Germany
}\\[1ex] 
${^b}$ {\it 
Arnold Sommerfeld Center for Theoretical Physics,\\
~~Ludwig-Maximilians-Universit\"at M\"unchen, 80333 M\"unchen, Germany
 \\} 
\end{center}

\vskip 2cm


\subsection*{\centering Abstract}

Toroidal orbifolds and their resolutions are described within the framework of (2,2) Gauged Linear Sigma Models (GLSMs). 
Our procedure describes two--tori as hypersurfaces in (weighted) projective spaces. The description is chosen such that the orbifold singularities correspond to the zeros of their homogeneous coordinates. The individual orbifold singularities are resolved using a GLSM guise of non--compact toric resolutions, i.e.\ replacing discrete orbifold actions by Abelian worldsheet gaugings. 
Given that we employ the same global coordinates for both the toroidal orbifold and its resolutions, our GLSM formalism confirms the gluing procedure on the level of divisors discussed by L\"ust et al.
Using our global GLSM description we can study the moduli space of such toroidal orbifolds as a whole. In particular, changes in topology can be described as phase transitions of the underlying GLSM. 
Finally, we argue that certain partially resolvable GLSMs, in which a certain number of fixed points can never be resolved, might be useful for the study of mini--landscape orbifold MSSMs. 

\newpage 
\setcounter{page}{1}
\tableofcontents


\section{Introduction}
\label{sc:intro}

\subsection*{Motivation}

Compactifications of the superstring from ten down to four dimensions is conventionally done on so--called Calabi--Yau (CY) manifolds in order to reduce the number of surviving supersymmetries in the low energy theory \cite{Candelas:1985en,gsw_2}. In order to obtain phenomenologically viable models from the heterotic string \cite{Gross:1985fr,Gross:1985rr} one needs to consider a non--trivial vector bundle to break the gauge group down to the Standard Model (SM) or Grand Unified Theory (GUT) gauge group with a semi--realistic particle spectrum. An important consistency requirement is the so--called Bianchi identity which links topological properties of the vector bundle to those of the tangent bundle of the compactification manifold, and ensures that the effective theory is free of dangerous anomalies. Only quite recently it has become possible to construct Minimal Supersymmetric SM (MSSM)--like models along these lines \cite{Donagi:1999ez,Bouchard:2005,Braun:2005ux,Braun:2005bw,Bak:2008ey,Anderson:2008uw,Anderson:2011ns}. Notwithstanding such recent successes, the construction of both vector bundles and their supporting CY spaces remains a highly involved task. Moreover, one typically only has algebraic geometrical and not differential geometrical means to study them, hence their description is necessarily rather abstract and indirect. Therefore, a real string theoretical treatment beyond the supergravity approximation, which requires large volume of all curves and cycles of the CY, is very hard to obtain. 

In the light of this it is quite surprising that there also exist exact string backgrounds which nevertheless are able to break sufficient amounts of supersymmetry in the low energy four dimensional theories. Examples of such exact CFTs are orbifold models \cite{Dixon:1985jw,Dixon:1986jc,Katsuki:1989bf,Kobayashi:1991}, free--fermionic models \cite{Kawai:1986ah,Antoniadis:1989zy}, asymmetric orbifold constructions \cite{Ibanez:1987pj} and Gepner models \cite{Gepner:1987hi}. Given that all these constructions define exact string backgrounds, one can compute the full one--loop partition function and check its modular invariance. This powerful stringy principle is fundamental for the consistency of the string, and therefore guarantees for example the absence of gauge anomalies in the effective theory. Since all these types of constructions allow for systematic searches, they have resulted in various classes of MSSM--like candidates. (See e.g.\ MSSM--like constructions on orbifolds \cite{Buchmuller:2006ik,Buchmuller:2006,Buchmuller:2007zd,Lebedev:2007hv,Lebedev:2008,Kim:2007mt,Blaszczyk:2009in}, 
as free--fermionic models \cite{Faraggi:1989ka,Faraggi:1991jr,Cleaver:2002ps,Faraggi:2006qa}, and as generalizations of Gepner models \cite{Dijkstra:2004cc,Dijkstra:2004ym}, respectively.) Standard orbifolds and free--fermionic models, which represent an alternative description of $\Intr_2\times\Intr_2$ orbifolds at the self--dual radius point, still admit the picture of string compactification \cite{Donagi:2008xy}. On the other hand asymmetric orbifolds or Gepner models do not allow for a straightforward understanding from a target space point of view; they immediately give an effective theory in four dimensions. In this paper we would like to take the possibility of having both an exact string description and the compactification picture seriously, and therefore take orbifolds as the starting point for our investigation. 

Orbifolds can be viewed as CY spaces but with singularities at the orbifold fixed points. These singularities can in general be removed by two different methods: deformations or resolutions. In this work we focus on the resolution procedure, in which one identifies so--called exceptional cycles inside the singularities and subsequently blows up these cycles by giving them finite volumes. In the effective four dimensional theory this corresponds to switching on some non--zero Vacuum Expectation Values (VEVs) for some twisted states located at the orbifold fixed points. Only in rather special non--compact cases it is possible to determine the explicit geometries, in particular the metric, of such blow--ups \cite{Ganor:2002ae,Nibbelink:2007rd,Nibbelink:2008qf}. Fortunately, toric geometry \cite{Oda,Fulton93,Hori:2003ic} provides a general procedure to assemble non--compact CY resolutions \cite{Lust:2006zh,Nibbelink:2007pn}. The resolution of compact toroidal orbifolds can then be described on the level of divisors: One formulates some gluing relations for the inherited torus divisors and the exceptional ones and calculates their intersection ring \cite{Denef:2005mm,Lust:2006zh,Reffert:2006du,Reffert:2007im,Nibbelink:2008tv,Nibbelink:2009sp,Blaszczyk:2010db,Blaszczyk:2011ig}.\footnote{As the intersection ring for K3 is known, heterotic line bundle models can be directly constructed without going through the resolution procedure \cite{Honecker:2006qz}.} 

With such resolution tools in hand one can study what happens to the mini--landscape MSSM models \cite{Buchmuller:2006ik,Lebedev:2006tr,Lebedev:2006kn} when one resolves the orbifold $T^6/\Intr_\text{6--II}$ on which they are based: In full resolution, i.e.\ with all singularities blown up, the SM group (in most cases the hypercharge) gets broken \cite{Nibbelink:2009sp,Blaszczyk:2010db}. The reason for this effect is that all these models have some fixed points where all twisted states are charged under the SM group. These models always require some twisted states to take VEVs to cancel the one--loop induced target space Fayet--Iliopoulos(FI)--term and to decouple exotic states. Hence a partial resolution is necessary. This means that both the original orbifold CFT description as well as standard supergravity techniques cannot be applied reliably in this regime: We need to develop some alternative framework to deal with this kind of partially resolved orbifolds. 

Such a worldsheet framework might be provided by two dimensional Gauged Linear Sigma Models (GLSMs) \cite{Witten:1993yc,Distler:1992gi,Distler:1993mk,Distler:1995mi}. These models are able to capture some of the essential features of CY compactifications, yet avoid most the complications of their non--linear sigma model descriptions. In this work we use GLSMs that possess (2,2) worldsheet supersymmetry, in which the target space coordinates become part of  chiral superfields. Using Abelian gaugings, weighted projective or toric spaces can be described as symplectic quotients, where their radii are set by worldsheet FI--parameters. CY spaces are then defined as complete intersections of hypersurfaces inside these toric ambient spaces. In the GLSM description these hypersurfaces appear as superpotential terms. A result of Beasley and Witten \cite{Beasley:2003fx} shows that the complex FI--parameters and the complex parameters in the superpotential are protected against worldsheet instanton effects \cite{Basu:2003bq,Distler:1994hs,Silverstein:1994ih,Silverstein:1995re} even when (2,0) deformations are considered. The GLSM formulation also allows to study the moduli space of such compactifications. In particular, one can study topological changes as GLSM phase transitions. Such topology changes can range from relatively mild flop--transitions, where one curve is replace by another, to jumps in the target space dimension. The power of the GLSM formulation is that it describes all these processes via continuous variations of the aforementioned variables \cite{Aspinwall:1993xz,Distler:1996tj,Chiang:1997kt}. (For recent developments in this direction see e.g.\ \cite{Melnikov:2005hq,Aspinwall:2010ve,Kreuzer:2010ph}.) 

This brings us to the main subject of study in the present paper: We would like to construct the resolution of compact toroidal orbifolds using GLSM techniques. Our starting point is the well--known observation that any two--torus can be mapped onto an elliptic curve in a (weighted) projective space by the Weierstrass function. Therefore, we consider toroidal orbifolds based on a factorized six--torus $T^6 = T^2 \times T^2 \times T^2$. Describing each $T^2$--factor in this algebraic way does not determine its appropriate weighted projective space uniquely. We use this to our advantage and choose the weighted projective space that is manifestly compatible with the orbifold action and additional discrete translational symmetries of the torus. The fixed points and tori then correspond to simultaneous zeros of some of the homogeneous coordinates. The resolution of these fixed points can be obtained via the GLSM guise of  local $\Cplx^3/\Intr_m$ blow--ups, i.e.\ by the introduction of exceptional coordinates and additional worldsheet gaugings. This procedure results in a GLSM which can describe the fully resolved orbifold. 

Following this procedure one can obtain a whole variety of resolution GLSMs. The model in which essentially all fixed points and tori are resolved independently we call the maximal fully resolvable model. 
Since a given toroidal orbifold consists of a complicated collection of fixed points and tori, the resulting maximal fully resolvable model becomes rather involved. Therefore it is useful to identify the GLSM, which is still able to resolve all singularities with the minimal number of gaugings. In such a so--called minimal fully resolvable model many fixed points or fixed tori are blown up/down simultaneously. In addition to these two extreme cases, our procedure allows to construct a whole variety of intermediate GLSMs. Some of them are only partially resolvable: Their description does not allow to blow--up all fixed points or tori. These types of GLSMs might be very interesting in the light of the mini--landscape models in which not all fixed points should be blown up. Another effect may happen when the resolution gaugings act on more than one two--torus simultaneously: Even though our starting point is always a factorized six--torus we are able to obtain resolutions of orbifolds on non--factorized or even non--factorizable lattices. Finally, as mentioned above the GLSM framework allows to move through the moduli space at large. For the minimal fully resolvable GLSM associated to $T^6/\Intr_3$ we perform this analysis in detail and identify a whole collection of distinct phases corresponding to very different target space geometries.

\subsection*{Outline}

We organized this paper as follows: 

In Section \ref{sc:Background} we lay the necessary foundations: Its three Subsections review some basics of (2,2) GLSMs, toroidal orbifolds and non--compact orbifold resolutions. 

In Section \ref{sc:GenOrbiRes} we outline our construction of toroidal orbifold resolutions as GLSMs. It begins with a non--technical description of two--tori in an algebraic way as an elliptic curve, i.e.\ as a hypersurface in a specific weighted projective space. Next we explain how the compact toroidal orbifold resolutions can naturally be obtained from this description. We distinguish between various types of resolution models: fully and partially resolvable GLSMs, and between minimal and maximal fully resolvable models and many in between. In addition we point out that our resolutions can be applied to certain non--factorizable orbifolds as well. Next, we go beyond the basic construction of toroidal orbifold GLSM resolutions and focus on some particular properties. We explain that such GLSMs may possess many phases which correspond to numerous topologically different target spaces. We describe how one can systematically study the full moduli space of these types of GLSMs. In the final Subsection we identify the divisors of the resolution and compare our results with those obtained by L\"ust et al.~\cite{Lust:2006zh}. Given that this general description is often rather involved we encourage the reader to inspect how these concepts are applied in the concrete examples provided in Sections \ref{sc:ResT6Z3} to \ref{sc:ResT6Z6II}. 

The necessary technical details of the description of elliptic curves and their symmetries, which were eluded to but skipped in Subsection \ref{sc:AlgebraicTori}, are exposed in Section \ref{sc:twotori} in conjunction with Appendix \ref{app:DiscreteActions}. We use the Weierstrass function to establish a mapping between a  torus and an elliptic curve. We provide different descriptions for two--tori that possess different discrete rotational (orbifold) symmetries and identify translational discrete $\Intr_n$ symmetries, n--volutions, that commute with them. 

The next Sections \ref{sc:ResT6Z3} to \ref{sc:ResT6Z6II} illustrate our general toroidal orbifold resolutions in the GLSM language for the specific cases of the $T^6/\Intr_3$, $T^6/\Intr_4$ and $T^6/\Intr_\text{6-II}$ orbifolds, respectively. Because of its relative simplicity the resolution of $T^6/\Intr_3$ is used to demonstrate the main features of GLSM resolutions described in Section \ref{sc:GenOrbiRes} in general. The other two orbifolds exemplify additional specific features which are not present in the $T^6/\Intr_3$ case. In detail: 

Section \ref{sc:ResT6Z3} begins with a description of the maximal and minimal fully resolvable models of the $T^6/\Intr_3$ orbifold. Then it explains how various (partially) resolvable models can be obtained by switching on a limited set of gaugings. These models can correspond to an orbifold on both factorized and non--factorized lattices. In Subsection \ref{sc:MinimalZ3Phases} the various phases of the minimal fully resolvable model is analyzed at length. The final Subsection \ref{sc:Z3dualGLSM} provides a specific fully resolvable model that possesses a symmetry which interchanges its three two--tori \Kh\ parameters by three blow--up parameters. 

Section \ref{sc:ResT6Z4} focuses on GLSMs associated with the $T^6/\Intr_4$ resolution. We show that our formalism has the identification of fixed tori induced by a residual orbifold action built in, both on the orbifold and the resolution level. However, in certain cases our formalism cannot treat all fixed tori independently: In this case one pair of fixed four--tori are blown up/down simultaneously. The Section ends with a discussion of resolution GLSMs corresponding to genuine non--factorizable $\Intr_4$ orbifolds; details of the associated non--factorizable $\Intr_4$ lattices can be found in Appendix \ref{appendixZ4NonfactLattices}. 

Section \ref{sc:ResT6Z6II} discusses some details of GLSM resolutions for the $T^6/\Intr_\text{6-II}$ orbifold. These models provide a second example showing that the identification of fixed tori is built into our GLSM formalism. However, the main significance of this Section is its potential relevance for the so--called mini--landscape MSSMs. In particular, we show that it is possible to construct partially resolvable GLSMs that have certain fixed tori unresolved. Hence, these models may provide the appropriate setting to discuss these mini--landscape models with some twisted VEVs switched on. 

Finally, in Section \ref{sc:concl} we give a summary of our results and speculate on possible future applications.


\section{Background material}
\label{sc:Background} 

\subsection{Gauged linear sigma models}
\label{sc:GLSMs} 

In this work we describe compact orbifold resolutions from the worldsheet perspective using the GLSM language. In this subsection we briefly summarize our notation and conventions. For details please consult e.g.\ \cite{Witten:1993yc}. 

In this paper we are only concerned with worldsheet theories that possess (2,2) supersymmetry. To describe gauged worldsheet theories we need to introduce three types of (2,2) superfields: 
\enums{ 
\item Gauge superfields: $V = (A_0, A_3, A, \gl_+, \gl_-, D)$: 
\\[1ex] 
A gauge superfield in the Wess--Zumino gauge contains gauge fields $A_0, A_3$, a complex scalar $A = A_1 +i A_2$, left-- and right--moving fermions $\gl_\pm$ and a real auxiliary field $D$. 
\item Twisted--chiral superfields: $\gS = \bD_+ D_- V$: 
\\[1ex] 
Here $D_\pm, \bD_\pm$ are the super covariant derivatives. In this work we do not consider twisted--chiral superfields in their own right, but only those which are obtained from the vector multiplets. 
\item Chiral superfields: $\cZ = (z, \gps_{z+}, \gps_{z-}, F_z)$: 
\\[1ex] 
A chiral superfield $\cZ$ contains a complex scalar $z$, left-- and right--moving fermions $\gps_{z\pm}$ and a complex auxiliary field $F_z$. The scalar $z$ has the interpretation of a target space coordinate. Throughout the paper we will employ the convention that letters denoting the superfield and its scalar component coincide, e.g.\ $x,y,z$ are the scalar components of the chiral superfield $\cX,\cY,\cZ$. 
} 
The simplest form of such a theory consists of a set of chiral superfields $\cZ_i$, $i=1,\ldots n$, with positive charges $q_i$ under an Abelian vector multiplet $V$ and another chiral superfield $\cC$ with negative charge $q_c = - \sum_i q_i$. We will assume that all charges are the smallest integers possible. 

Here we do not give the complete action for these superfields, but only give the two ingredients that are essential for the target space interpretation of the GLSM:   
\items{
\item Twisted--superpotential: 
\equ{
W_\text{twisted} = \gr\, \gS\,.
}
\item Superpotential:
\equ{
W_\text{super} = \cC\, P(\cZ)\,.
\label{GeomSP}
}
} 
The complex Fayet--Iliopoulos (FI)--parameter $\gr = a + i \ga$ corresponds to an axionic degree of freedom $\ga$ and a real \Kh\ parameter $a$. The latter interpretation follows from the D--term constraint:
\begin{align}
D= \sum_i q_i\, |z_i|^2 - q_c\,|c|^2 - a=0\,.  
\label{GenDterm} 
\end{align}
The GLSM therefore possesses two phases depending on the value of $a$. In the geometrical regime, where $a> 0$, the size of the cycle is controlled by $a$, which can therefore be interpreted as its K\"ahler parameter. Hence the GLSM describes a weighted projective space $\mathbbm{P}^{n-1}_{q_1,\ldots,q_n}$ with the radial part of the $\Cplx^*$ action fixed by the D--term constraint. 

The set of F--term conditions that result from the superpotential \eqref{GeomSP} reads
\begin{align}
F_c = P(z) = 0\,,
\qquad 
F_{z_i} = c\, P_{,i}(z) = 0\,.
\label{GenFterms} 
\end{align}
In the geometrical regime ($a > 0$) the equation $P(z)=0$ defines a smooth complex codimension one space, provided the transversality condition is fulfilled, i.e.\ not all $P_{,i}(z)$ vanish simultaneously. Hence this describes a degree $q_c$ hypersurface $\mathbbm{P}^{n-1}_{q_1,\ldots,q_n}[q_c]$ within this weighted projective space. In the case $a<0$ we see that the D--term forces $c$ to be non--zero. Consequently the F--terms imply that generically all $z_i=0$, hence the target space is a single point. For this reason we call this the non--geometrical regime. As can been from this rather typical example the target space geometry can be determined in various GLSM phases.

The complete scalar potential of such GLSMs is given by the following expression:
\equ{
V_\text{scalar} = \frac 12\, D^2 + | F_c|^2 + \sum_i | F_{z_i}|^2 + 
|A|^2 \Big( q_c^2 \,|c|^2 + \sum_i q_i^2\, |z_i|^2 \Big)\,. 
}
As in essentially any phase at least one scalar $c$ or $z_i$ is non--vanishing, this potential immediately forces $A=0$. (Only on the boundary between phases this is not necessarily the case, see e.g.\ \cite{Witten:1993yc}.)

\subsection{Toroidal orbifolds}
\label{sc:T6/ZN} 

In this subsection we define the class of toroidal orbifolds we can resolve using GLSM techniques that we expose in this paper. We start from a so--called factorizable torus, i.e.\ a six dimensional torus that can be written as 
\equ{
T^6 = T^2_1 \times T^2_2 \times T^2_3\,.
}
Each of the two--tori $T^2_a$, $a=1,2,3$, is defined via a two dimensional lattice spanned by $1$ and a complex structure $\tau_a$. Hence the corresponding complex coordinate $u_a \in \Cplx$ fulfills the periodic identifications $u_a \sim u_a + 1 \sim u_a + \tau_a\,$. 

Next we demand this $T^6$ admit a $\Intr_N$ orbifold action $\gth$ which acts in the three complex coordinates as discrete $\Intr_N$ rotations 
\equ{
\gth :~ \Big(u_1, u_2, u_3 \Big) \mapsto  
 \Big( e^{2\pi i \frac {N_1}N}\,u_1, e^{2\pi i \frac {N_2}N}\,u_2, e^{2\pi i \frac {N_3}N}\,u_3 \Big)
}
with all $N_i$ positive integers that are relatively prime and $N=\sum_i N_i$. This defines a $T^6/\Intr_N$ orbifold. (When one of the $N_i$ is zero we obtain $T^2 \times T^4/\Intr_N$.) The orbifold action $\gth$ is only well--defined provided that the two dimensional two--tori lattices, i.e.\ their complex structures, are compatible with the action. There are only a finite number of possible two--tori that admit such discrete rotations, based on the Lie lattices $A_2$, $D_2$, $G_2$ and $A_1\times A_1$, which we denote by $T^2(\Intr_3)$, $T^2(\Intr_4)$, $T^2(\Intr_6)$, and $T^2(\Intr_2)$, respectively. These two--tori have a definite complex structure, except for the $T^2(\Intr_2$) which has an arbitrary $\gt$.  In Section \ref{sc:twotori} we describe these tori and their complex structures in detail.

Given a $T^6/\Intr_N$ orbifold, we may ask which finite discrete symmetries it may possess. We call a discrete symmetry of the geometry an n--volution when it is generated by some $\ga$ such that $\ga^n=\mathbbm{1}$. (In particular, a 2--volution is the usual involution.) Such n--volutions $\ga$ should of course be compatible with the orbifold action $\gth$. Given that we started from a factorizable torus $T^6$, we can systematically classify the n--volutions by considering the two--tori $T_a^2$ separately. We denote the generator of the n--volution of $T_a^2$ by $\ga_a$. (For the $\Intr_6$ two--torus there are no n--volutions, while the $\Intr_3$ and $\Intr_4$ two--torus possess one and the $\Intr_2$ two--torus possesses two independent n--volution generators.) 

Next we can identify points in a $T^6/\Intr_N$ orbifold that are related by one or more n--volutions. The resulting space is again a six dimensional orbifold, but depending on the type of n--volutions it may be non--factorized: It certainly remains factorizable when the n--volution only acts on one of the two--tori. Even though the volume of this two--torus is divided by n, it remains a two--torus of the same type. When the n--volution acts on two or more two--tori simultaneously, the result might be a non--factorized orbifold. However, this only happens provided the set of n--volutions cannot be generated by n--volutions that act on only one of the two--tori each. In some cases an orbifold might appear to be non--factorized, but can be turned into a factorizable one by a \Kh\ deformation. We refer to these cases as non--factorized to distinguish them from truly non--factorizable ones. In Section \ref{sc:ResT6Z3} we construct the $T^6/\Intr_3$ orbifold on the Lie lattice of $E_6$ and show that it can be \Kh\ deformed to the factorizable case. While in Section \ref{sc:ResT6Z4} we give examples of truly non--factorizable orbifold resolutions and confirm that their topologies are distinct.

\subsection{Local orbifold GLSM resolutions}
\label{sc:LocalRes} 

In this Subsection we review how the resolution of $\Cplx^3/\Intr_m$ singularities can be described in the GLSM language. Let $(z_1,z_2,z_3)$ denote the complex coordinates of $\Cplx^3$. Consider the $\Intr_m$ orbifold action 
\equ{
\gth :~ \Big(z_1, z_1, z_3\Big) \mapsto 
\Big(e^{2\gp i \frac{m_1}m}\, z_1, e^{2\gp i \frac{m_2}m}\, z_2, e^{2\gp i \frac{m_3}m}\, z_3\Big)
}
with all $m_i$ non--negative integers that are relatively prime and $m=\sum_i m_i$. 

Next we introduce a GLSM with chiral superfields $\cZ_a$ whose scalar components are the coordinates $z_a$. To realize the orbifold symmetry in the GLSM we promote the integers $m_i$ to charges. However, since a gauge symmetry removes degrees of freedom, we have to add new chiral superfields $\cX$ in order to avoid modifying the target space dimension. In detail, for each $1 \leq r \leq m-1$ we define charge vectors $q_r$ whose components are
\equ{
(\tilde q_r)_a = r\,m_a ~\text{mod}~m\,,\quad 
\text{such that}~ 0 \leq (\tilde q_r)_a < m\,, 
\quad 
\sum_a (\tilde q_r)_a = m\,. 
}
The latter condition cannot be satisfied for all $r$. Only if it is, it defines an independent twisted sector and a gauging is introduced on the worldsheet. When all the integral charges $(q_r)_a$ and $m$ can be divided by a common integer $k_r$, we denote the resulting charges by $(q_r)_a = (\tilde q_r)_a/k_r$. We introduce a gauge multiplet $V_r$ and a chiral superfield $\cX_r$ with charge $q_r(\cX_r) = -m_r = m/k_r$. Calling the real parts of the corresponding FI--parameters $b_r$, the D--term constraints of the theory read 
\equ{
(q_r)_{1}\, |z_1|^2 + (q_r)_{2}\, |z_2|^2 + (q_r)_3\, |z_3|^2  - m_r\, |x_r|^2 - b_r = 0\,. 
}
Even though both the superfields $\cX_r$ and the $\cC$ introduced above have negative charges, there is a crucial difference in the reason why they were introduced: The $\cX_r$ realize the orbifold symmetries in the GLSM, while the $\cC$ were introduced to force a hypersurface constraint via its superpotential. For this purpose $\cC$ always appears linearly in the superpotential. (This can be enforced by insisting on an appropriate R--symmetry.) 

By analyzing these D--term conditions, one can demonstrate that such a local resolution GLSM possesses three types of  phases: 
\items{
\item Orbifold phase: 
\\[1ex] 
In the phase where  $b_r < 0$ all scalars $x_r$ are necessarily non--zero and leave the $\Intr_n$ phases of the scalars $z_a$ undetermined. Hence this phase the GLSM describes the original orbifold geometry. 
\item Full resolution phase(s): 
\\[1ex] 
When all the \Kh\ parameters $b_r>0$, the $\Cplx^3/\Intr_n$ singularity has been fully resolved: All the four--cycles have positive volume. Some GLSMs have a unique complete resolution while others may posses various topologically inequivalent resolutions. Their distinction is made by considering which curves exist in a given resolution. 
\item Hybrid phases: 
\\[1ex] 
When there is more than one gauging, there are various mixed phases in which some \Kh\ parameters are positive and some negative. In this case the orbifold geometry has only been partially resolved and so the resulting geometry is still singular. 
}

\section{Construction of toroidal orbifold GLSM resolutions}
\label{sc:GenOrbiRes}

\subsection{Algebraic description of two--tori: elliptic curves}
\label{sc:AlgebraicTori}

We now would like to obtain a GLSM description of toroidal orbifolds. As a toroidal orbifold has various fixed points and possibly fixed tori, we would like to apply the local resolution procedure discussed in Subsection \ref{sc:LocalRes} for each of them. This is unfortunately rather difficult using the periodic torus coordinates $(u_1,u_2,u_3)$ given in Subsection \ref{sc:T6/ZN}: As only the singularity at the fixed point at $(0,0,0)$ has the same description as the non--compact $\Cplx^3/\Intr_N$, it can be treated directly using the local resolution procedure. For the other fixed points that are situated away from the origin, this is not directly possible. Moreover, given that the coordinates $u_a$ are double periodic, one should in principle perform the local resolution procedure at all the images of these fixed points as well. To overcome these complications we use an alternative description of the two--tori as an elliptic curve, i.e.\ hypersurfaces in certain (weighted) projective spaces. To distinguish this algebraic description of a two--torus from the one with periodic coordinates, we refer to the former as the ``elliptic curve'' and to the latter as the ``torus'' formulation. 

The elliptic curves can be described as follows: We denote the homogeneous coordinates of the projective space associated with the $a^\text{th}$ two--torus by $z_{ai}$; i.e.\ $i$ labels the homogeneous coordinates. The hypersurface (or surfaces) are then specified as the vanishing locus of one (or more) homogeneous polynomial(s) $P_a(z_{ai})$ of a certain degree. The connection between the  two--torus and the elliptic curve is encoded in the Weierstrass function $\wp_\tau(u)$ and its derivative $\wp_\tau^\prime(u)$. The properties of this double periodic function depend decisively on the complex structure $\tau$. In Section \ref{sc:twotori} we summarize the features of the relevant projective hypersurfaces and display the Weierstrass mapping between the two--tori and the elliptic curves at length. 

In the GLSM setting the homogeneous coordinates are promoted to (2,2) chiral superfields $\cZ_{ai}$ with certain charges under $U(1)_{R_a}$ gaugings which reflect the weights of the (weighted) projective space. The polynomial constraints that define the hypersurfaces are encoded in a gauge invariant superpotential 
\equ{
W_\text{torus} = \sum_a \cC_a\, P_a(\cZ_{ai})\,.
\label{SPtorus} 
}
(For example, for the $T^6$ with an $\Intr_3$ orbifold symmetry it reads 
$W_\text{torus} = \sum_{a,i} \cC_a \, \cZ_{ai}^3$.) 
The F--terms of the additional chiral superfields $\cC_a$ precisely give the required hypersurface constraints. Their charges are taken to be  
\equ{ 
q(\cC_a) = - \sum_i q(\cZ_{ai})\,, 
}
as this ensures that the resulting complex dimension one geometries have vanishing first Chern class, i.e.\ they indeed define an elliptic curve which is isomorphic to a  two--torus when it is not singular. 

In principle a given two--torus admits various elliptic curve descriptions. We choose that description which reflects the orbifold symmetry $\gth$ manifestly, i.e.\ which acts on each of the homogeneous coordinates as a discrete rotation. (The specific choices we make for projective hypersurface descriptions of the two--tori possessing different $\Intr_n$ rotational symmetries are listed in Table \ref{tab:SummaryTori}.) For the purpose of orbifold resolutions this has the added bonus that for each label $i$ the equation $z_{ai}=0$ identifies a fixed point of the orbifold action $\gth$ on the $a^\text{th}$ torus. This motivates to refer to the orbifold fixed point at $z_{1i} = z_{2j} = z_{3k} = 0$ as $(i,j,k)$. Similarly we refer to the fixed torus with $z_{1i} = z_{2j} = 0$ as $(i,j)$, etc.  As we explain in the next subsection, the fact that the fixed points and tori are given by zeros of some homogeneous coordinates in the description which uses elliptic curves is of crucial importance to us. It allows us to utilize the local resolution procedure not only for the singularity at the origin, but for many (and in most case even all) orbifold fixed points and tori. 

As each two--torus is described by a $U(1)_{R_a}$ gauging, each of their GLSMs contains an FI--parameter $\gr_a$ whose real part $a_a$ defines a \Kh\ modulus. A six dimensional torus $T^6$ compatible with a $\Intr_N$ orbifold may posses additional off--diagonal \Kh\ deformations. The possible \Kh\ deformations are in one--to--one correspondence with the invariant (1,1)--forms $\d \bar{u}_a \wedge \d u_b$: For any $\Intr_N$ orbifold the diagonal two--forms $\d \bar{u}_a \wedge \d u_a$ are invariant under the orbifold twist $\gth$, and in certain cases some off--diagonal forms,  $\d \bar{u}_a \wedge \d u_b$ with $b \neq a$, as well. In our factorized GLSM description of a six--torus these additional off--diagonal \Kh\ deformations do not appear as FI--parameters in the GLSM. However, we find that these moduli may appear in a CFT fashion as gauge invariant kinetic terms, which would correspond to allowed marginal deformations of the theory.

\subsection{Toroidal orbifold resolution GLSMs}
\label{sc:ResGLSMs} 

\begin{table}[t!] 
\centering 
\renewcommand{\arraystretch}{1.3}
 \begin{tabular}{|c||c|c|c|c|c|c|}
\hline
  Point  & Orbifold twist  & $T^6$ torus  & Exceptional & \multicolumn{2}{|c|}{Invisible moduli} & Indistinguishable  \\
   group &  vector(s) &  lattice & gaugings &  $h^{1,1}_\text{off--diag}$ &  $h^{1,2}_\text{twisted}$ & fixed points/tori \\
\hline \hline
$\Intr_3$ & $\frac13 (1,1,-2)$ & $A_2^3$ & $27$ & $6$ & $0$ & $0$ 
\\
\hline
\multirow{3}{*}{$\Intr_4$} & \multirow{3}{*}{$\frac14 (1,1,-2)$} & $D_2^2\times A_1^2$ & $23$ & $2$ & $6$ & $1\times2\,$FT 
\\
& & $D_2\times A_1 \times A_3$ & $6$ & $2$ & $2$ & $2\times 8\,$FP, $2\times 2\,$FT
\\
& & $A_3^2$ & $8$ & $2$ & $0$ & $4\times 4\,$FP
\\
\hline 
$\Intr_{6-\rm I}$ & $\frac16 (1,1,-2)$ & $G_2^2 \times A_2$ & $17$ & $2$ & $5$ & $1\times 3\,$FT , $1\times 2\,$FP 
\\
\hline
$\Intr_{6-\rm II}$ & $\frac16 (1,2,-3)$ & $G_2 \times A_2 \times A_1^2$ & $32$ & $0$ & $10$ & $0$ 
\\
\hline 
\multirow{2}{*}{$\Intr_{2} \times \Intr_{2}$} & $\frac12 (1,-1,0),$  & \multirow{2}{*}{$A_1^6$} & \multirow{2}{*}{$48$} & \multirow{2}{*}{$0$} & \multirow{2}{*}{$0$} & \multirow{2}{*}{$0$} \\
& $\frac12(0,1,-1)$ & & & & & \\
\hline
\multirow{2}{*}{$\Intr_{2} \times \Intr_{4}$} & $\frac12(0,1,-1),$ & \multirow{2}{*}{$D_2^2 \times A_1^2$} & \multirow{2}{*}{$57$} & \multirow{2}{*}{$0$} & \multirow{2}{*}{$0$} & \multirow{2}{*}{$1\times 2\,$FT} \\
& $\frac14 (1,-1,0)$ & & & & & \\ 
\hline
\multirow{2}{*}{$\Intr_{2} \times \Intr_{6-\rm I}$} & $\frac12(0,1,-1),$ & \multirow{2}{*}{$G_2^2 \times A_2$} & \multirow{2}{*}{$26$} & \multirow{2}{*}{$0$} & \multirow{2}{*}{$0$} & \multirow{2}{*}{$3 \times 3\,$FT , $1 \times 2$FP} \\
& $ \frac16 (1,1,-2)$ & & & &  & \\ 
\hline
\multirow{2}{*}{$\Intr_{2} \times \Intr_{6-\rm II}$} & $\frac16 (1,-1,0),$ & \multirow{2}{*}{$G_2^2 \times A_1^2$} & \multirow{2}{*}{$46$} & \multirow{2}{*}{$0$} & \multirow{2}{*}{$2$} & \multirow{2}{*}{$1 \times 3\,$FT} \\
 & $\frac12(0,1,-1)$ & & & & & \\ 
\hline
\multirow{2}{*}{$\Intr_{3} \times \Intr_{3}$} & $\frac13 (1,-1,0),$ & \multirow{2}{*}{$A_2^3$} & \multirow{2}{*}{$81$} & \multirow{2}{*}{$0$} & \multirow{2}{*}{$0$} & \multirow{2}{*}{$0$} \\
& $\frac13(0,1,-1)$ & & &  & & \\ 
\hline
\multirow{2}{*}{$\Intr_{3} \times \Intr_{6}$} & $\frac13(0,1,-1),$ & \multirow{2}{*}{$G_2^2 \times A_2$} & \multirow{2}{*}{$65$} & \multirow{2}{*}{$0$} & \multirow{2}{*}{$1$} & \multirow{2}{*}{$2 \times 2\,$FT , $3 \times 2\,$FP} \\
& $ \frac16 (1,-1,0)$ & &  && &  \\ 
\hline
\multirow{2}{*}{$\Intr_{4} \times \Intr_{4}$} & $\frac14 (1,-1,0),$ & \multirow{2}{*}{$D_2^3$} & \multirow{2}{*}{$87$} &  \multirow{2}{*}{$0$} & \multirow{2}{*}{$0$} & \multirow{2}{*}{$0$} \\
& $ \frac14(0,1,-1)$ & & & & & \\
\hline
\multirow{2}{*}{$\Intr_{6} \times \Intr_{6}$} & $\frac16 (1,-1,0),$ & \multirow{2}{*}{$G_2^3$} & \multirow{2}{*}{$80$} & \multirow{2}{*}{$0$} & \multirow{2}{*}{$0$} & \multirow{2}{*}{$1 \times 2\,$FP} \\
 & $\frac16(0,1,-1)$ & & & & & \\ 
\hline 
 \end{tabular}
 \renewcommand{\arraystretch}{1}
\caption{This Table shows toroidal orbifolds for which we can construct fully resolvable GLSMs. It shows all maximal, factorized models plus two non--factorizable $\Intr_4$ models. The column labeled ``exceptional gaugings'' indicates the number exceptional coordinates (which come with the same number of gaugings and FI-parameters). The two columns labeled ``invisible moduli'' indicate the amount of geometric moduli which are not explicitly visible in the GLSM formulation: off--diagonal untwisted K\"ahler moduli and twisted complex structure moduli. The final column gives the number of indistinguishable fixed points (FP) and fixed tori (FT) in this construction. 
\label{tab:ClassificationModels}} 
\end{table}

After the preparation work described above, obtaining GLSMs that describe resolutions of toroidal orbifolds is in principle rather straightforward. However, in this process one finds that one can distinguish various types of resolution GLSM models. As many of these different classes of models exhibit interesting features, we define them by specifying the differences in the GLSM resolution process and comment on some of their generic properties. In Table \ref{tab:ClassificationModels} we have collected the complete list of toroidal orbifolds for which we can construct GLSM realizations using the methods explained below. A comparison with all the possible toroidal orbifolds classified in \cite{Erler:1992ki} reveals that we are lacking a description for the orbifolds with the point groups $\Intr_7, \Intr_\text{8--I}, \Intr_\text{8--II}, \Intr_\text{12--I}$ and $\Intr_\text{12--II}$ all together and the $\Intr_\text{6--II}$ on the non--factorizable lattices. 

The initial data for the GLSM resolution process of a toroidal $T^6/\Intr_N$ orbifold are the elliptic curves compatible with the orbifold twist $\gth$. Using the information collected in Section \ref{sc:twotori} we can see the how orbifold actions $\gth, \gth^2, \ldots$ act on the homogeneous coordinates $z_{ai}$. The orbifold action only acts on one of the homogeneous coordinates in each torus $T^2_a$. (In the notation conventions chosen in that Section these are typically denoted by $z_{a1}$.) However, using the n--volutions $\ga$ possibly combined with discrete subgroups of the $U(1)_{R_a}$ groups, we can transport the orbifold action $\gth$ to act on one of the other homogeneous coordinates $z_{ai}$. 

At fixed points of the (transported) orbifold action, where some of the coordinates $z_{ai}=0$ vanish, we can employ the local GLSM resolution procedure discussed in Subsection \ref{sc:LocalRes}: Each of these actions is replaced by an ``exceptional'' $U(1)_E$ gauging and an additional chiral superfield $\cX$ is introduced. Given that some of the chiral superfields $\cZ_{ai}$ become charged under the exceptional $U(1)_E$ gauge symmetries in the resolution process, the torus superpotential \eqref{SPtorus} is no longer gauge invariant. This is easily repaired by inserting the superfields $\cX$ at the appropriate places. For example, considering the orbifold $T^6/\Intr_3$, the introduction of the gauging $U(1)_{E_{111}}$ modifies the torus superpotential to 
\equ{
W = \cC_1 \Big( \cZ_{11}^3\, \cX_{111} + \cZ_{12}^3 + \cZ_{13}^3 \Big) 
+ \cC_2 \Big( \cZ_{21}^3\, \cX_{111} + \cZ_{22}^3 + \cZ_{23}^3 \Big) 
+ \cC_3 \Big( \cZ_{31}^3\, \cX_{111} + \cZ_{32}^3 + \cZ_{33}^3 \Big)\,. 
}
The resolution superpotential can quickly become rather involved because a resolution GLSM often contains many exceptional gaugings. 

Since there are many fixed points and fixed tori which will be resolved in this process, we need to introduce some systematic notation: We denote by $U(1)_{E_{r,ijk}}$ and $\cX_{r,ijk}$ the $U(1)$ gauging and the extra chiral superfield, respectively, that are introduced to resolve the $r^\text{th}$ sector at fixed point $(i,j,k)$. (When the (transported) orbifold action only acts on two homogeneous coordinates, i.e.\ corresponds to a fixed torus, we use the corresponding two labels as described above.) As this procedure guarantees that for each twisted sector at all fixed points there is a blow--up \Kh\ parameter $b_{r,ijk}$, the resulting GLSM describes a completely smooth geometry for appropriate choices of the \Kh\ parameters. For this reason we refer to this model as a fully resolvable GLSM.

\subsubsection*{Built in fixed points/tori identification}

Our description has the very useful feature that the identification of fixed points or tori is already built in for most cases. As we will show explicitly in Sections \ref{sc:ResT6Z4} and \ref{sc:ResT6Z6II} for the $T^6/\Intr_4$ and $T^6/\Intr_\text{6--II}$ orbifolds, the technical reason for this is that the roots of algebraic equations, like $z_{ai} = 0$, get mapped to each other by the residual orbifold action. Such identifications of fixed points had to be built in by hand in the blow--up description of \cite{Lust:2006zh}. First, in the local resolution, one has to introduce ordinary and exceptional divisors for all fixed points on the orbifold, including those which are mapped onto each other. Later, when gluing the local patches together, one has to build suitable linear combinations of the divisors which are mapped onto each other in order to account for this action. After this, the linear equivalences of these divisors have to be added up accordingly for each of the equivalence classes, while in addition taking into account the order of the subgroup under which the divisors are identified. Such complications do not occur in the GLSM approach, because such redundancies in the fixed point identifications are automatically avoided.

\subsubsection*{Indistinguishable fixed points and tori}

However, it may happen that the labels $(i,j)$ do not uniquely identify some fixed tori. This happens when $z_{1i} = 0$ and $z_{2j} =0$ both have multiple roots which are not all mapped to each other via the residual orbifold action. In the following we refer to such fixed tori as indistinguishable. This always happens in non-maximal models. But also in maximal models this can happen when the factorization of $T^6$ includes two $\Intr_4$-- or $\Intr_6$--tori. These cases are $\Intr_4$, $\Intr_{6-\rm I}$, $\Intr_2\times\Intr_4$, $\Intr_2 \times \Intr_{6-\rm I}$, $\Intr_2 \times \Intr_{6-\rm II}$, $\Intr_3 \times \Intr_6$ and $\Intr_6 \times \Intr_6$, see Table \ref{tab:ClassificationModels}. In this Table we have indicated the number of indistinguishable fixed points or tori for each of the orbifolds we are able to describe using our GLSM methods. In heterotic orbifold theories it turns out that the matter content of these fixed tori is always the same. Under a symmetry which exchanges these fixed points, the states combine to pairs of one invariant and one or more non-trivially transforming states. In the GLSM description we only identify the invariant blow--up mode with an associated  FI--parameter. The $T^6/\Intr_4$ orbifold provides an example of this situation. We use its resolution to illustrate this issue in Section \ref{sc:ResT6Z4}.

\subsection{Maximal fully resolvable GLSMs}
\label{sc:MaxFullRes}

The GLSM obtained by the procedure described in the previous Subsection contains the maximal number of gaugings and all singularities are resolved, hence we call it the maximal fully resolvable model. Such a GLSM might nevertheless contain less FI--parameters than the total number of \Kh\ parameters predicted by the Hodge number $h_{1,1}$. The reason for this mismatch can be two--fold: Firstly, as mentioned above, contrary to the diagonal \Kh\ deformations, the off--diagonal ones are not associated to FI--parameters in the GLSM. This we illustrate in Section \ref{sc:ResT6Z3} using the maximal fully resolvable GLSM of the $T^6/\Intr_3$. Secondly, as mentioned at the end of the previous Subsection, a consequence of our description of the full GLSM resolution is that we are not always able to distinguish all fixed points and tori separately, even though all singularities got smoothed out: The indistinguishable fixed tori get resolved with the same size simultaneously. We illustrate this case in Section \ref{sc:ResT6Z4} by GLSM resolutions of the $T^6/\Intr_4$ orbifold. 

The construction of a maximal fully resolvable GLSM associated with a given $T^6/\Intr_N$ is important in the light of the Beasley--Witten result \cite{Beasley:2003fx}: All \Kh\ and complex structure deformation which appear as FI-- and superpotential parameters in a GLSM are protected from worldsheet instanton effects even when (2,0) deformations are considered, and hence they constitute genuine moduli. A maximal fully resolvable model therefore gives the maximal number of protected \Kh\ parameters within this description. The two types of \Kh\ deformations mentioned above that are not explicitly contained in our GLSM description, i.e.\ the off--diagonal torus deformations and those associated with indistinguishable fixed points or tori, are not protected and might therefore be lifted by non--perturbative worldsheet effects \cite{Silverstein:1994ih,Silverstein:1995re}. 

In the light of this it is worthwhile to summarize the unprotected moduli in the various resolutions. In Table  \ref{tab:ClassificationModels} we have indicated the number of untwisted $(1,1)$--forms and twisted $(1,2)$--forms that are not explicitly visible in the maximal GLSMs, and the number of identified fixed points and tori. For our three prime orbifold resolutions we have in particular: 
\items{
\item 
The $T^6/\Intr_3$ orbifold resolution:
\\[1ex] 
The resolution of $T^6/\Intr_3$ is rigid (i.e.\  $h_{2,1} = 0$) but admits $h_{1,1} = 9+27 =36$ \Kh\ deformations\footnote{The first number in the sum gives the so--called untwisted moduli, the second the twisted ones.}. As we show in Section \ref{sc:ResT6Z3} in its maximal fully resolvable GLSM we recover $3+27=30$ of them as FI--parameters. The maximal full--resolution GLSM has far less ``elusive'' \Kh\ deformations than the Aspinwall--Plesser resolution \cite{Aspinwall:2011us} (to which we  return in Subsection \ref{sc:SingleGauging}.) where only $3+1=4$ of them appear in the GLSM description. 
\item 
The $T^6/\Intr_4$ orbifold resolution: 
\\[1ex] 
The $T^6/\Intr_4$ resolution has $h_{1,1} = 5 + 26$ and $h_{2,1} = 1+6$, see Section \ref{sc:ResT6Z4}. In the maximal fully resolvable factorized orbifold on the lattice $D_2^2 \times A_1^2$ we encounter three untwisted \Kh\ parameters associated with its two--tori and $25$ exceptional ones: Because one pair of fixed four--tori is indistinguishable, we miss one \Kh\ deformation. We were unable to identify hints of the twisted complex structure deformations.  
\item 
The $T^6/\Intr_\text{6-II}$ orbifold resolution: 
\\[1ex] 
The $T^6/\Intr_\text{6-II}$ resolution has $h_{1,1} = 3 + 32$ and $h_{2,1} = 1+10$, see Section \ref{sc:ResT6Z6II}. Its maximal fully resolvable GLSM realizes all \Kh\ deformations as FI--parameters. The untwisted complex structure appears in the superpotential explicitly, but the twisted complex structure deformations remained elusive. 
}

\subsection{Minimal fully resolvable GLSMs}
\label{sc:MinFullRes} 

It can often be rather demanding to have to work with a large multitude of \Kh\ parameters. Therefore it is interesting to define another extreme of toroidal orbifold resolution GLSMs. Namely those with a minimal number of gaugings that still can describe a fully resolved orbifold. 

Such minimal fully resolvable GLSMs can be obtained by the following considerations. In a model that can describe a fully resolved orbifold geometry, we do not need to introduce individual gaugings for all the fixed points under the (transported) orbifold twists separately. If a set of fixed points that can be obtained from each other by transporting the twist using n--volutions, it is sufficient to introduce the resolving gaugings for only one of them. This can be confirmed by investigating the consequences of the F--term constraints. Technically, this arises as setting a certain coordinate to zero gives rise to a collection of different roots. 

The GLSM constructed recently by Aspinwall--Plesser \cite{Aspinwall:2011us} defines an example of a minimal fully resolvable GLSM for the $T^6/\Intr_3$ orbifold: Any fixed point can be obtained from the fixed point $(1,1,1)$ by 3--volutions $\ga_a$, hence only a single exceptional gauging is required. Consequently, all 27 fixed points of the $T^6/\Intr_3$ orbifold are resolved simultaneously. We return to this model in more detail in Section \ref{sc:ResT6Z3}.

\subsection{Partially resolvable GLSMs}
\label{sc:PartialGLSMs} 

From the presentation so far one might get the impression that the geometry becomes less singular the more local resolution gaugings one has introduced. However, this is definitely not always the case: As an example we start with a minimal fully resolvable GLSM as defined above. By introducing an additional gauging one obtains two independent \Kh\ parameters to control certain blow--up cycles as expected, but surprisingly some other singularities are not resolved anymore for any value of these parameters! We illustrate this effect explicitly for the resolution GLSM of $T^6/\Intr_3$ with two $U(1)_E$ groups in Section \ref{sc:ResT6Z3}. We refer to these type of GLSMs as partially resolvable GLSMs since they are incapable of describing a fully resolved orbifold for any value of the \Kh\ parameters. 

A possible intuitive way of understanding this effect is the following: To define the minimal fully resolvable models we used that collections of fixed points or tori were equivalent since they could be obtained from each other via n--volutions. When we now turn on one additional gauging, certain fixed points or tori become distinguishable and hence certain n--volutions are not available anymore to make these identifications. Now if this induces discrete rotational symmetries which do not arise from a single $U(1)_E$ gauging, there will be some fixed points or tori that cannot be resolved since there is no cooresponding \Kh\ parameter. If one wants to resolve all fixed points, one has to introduce one (or more) suitably chosen additional \Kh\ parameters and gaugings, corresponding to the fixed points that could be reached by n--volutions before.

We will illustrate this in detail in Section \ref{sc:ResT6Z3} by the following example:  Consider the Aspinwall--Plesser model, i.e.\ the minimal fully resolvable GLSM of $T^6/\Intr_3$ with a single $U(1)_{E_{111}}$ gauging. When we introduce a second gauging, say $U(1)_{E_{211}}$, then there are two fields $x_{111}$ and $x_{211}$ which can give rise to two $\Intr_3$ gauge symmetries. One of them we may interpret as the orbifold action $\Intr_{3,\rm Orbi}$, and the other corresponds to a 3--volution $\Intr_{3-\rm vol}$, spanned by $\ga_1$, in the first torus. This divides the 27 fixed points in three sets of 9 fixed points. As we will demonstrate in that Section two of these sets can be resolved but the third set remains singular. To obtain a fully resolvable GLSM in this case we need to introduce all the gaugings that correspond to the fixed points that can be reached from the point $(1,1,1)$ via the 3--volution $\ga_1$, i.e.\ the points $(2,1,1)$ and $(3,1,1)$. Hence, we obtain a complete resolution when we also introduce the gauging $U(1)_{E_{311}}$. 

Maybe precisely because partially resolvable GLSMs never describe completely smooth geometries, such GLSMs might be very interesting: They potentially define a framework in a regime where both supergravity as well as orbifold CFT techniques loose their validity. This is not only of theoretical interest, but might eventually be relevant for string phenomenology: It has been shown that any of the ``mini--landscape MSSMs'' \cite{Buchmuller:2006ik,Lebedev:2006tr} would break the hypercharge in full blow--up \cite{Nibbelink:2009sp}. The partial resolvable GLSMs might provide a possible valid description for such situations. In Section \ref{sc:ResT6Z6II} we consider this point in more detail for some specific mini--landscape benchmark MSSMs.

\subsection{Non--factorized and non--factorizable orbifold resolutions}
\label{sc:NonFacRes} 

In Subsection \ref{sc:T6/ZN} we saw that we can produce non--factorized orbifolds from factorizable ones by modding out n--volutions that act on two or three two--tori simultaneously. We have just shown that switching on gaugings in a resolution GLSM that corresponds to fixed points which are related by an n--volution possesses an orbifold limit where this n--volution has been modded out. For this argument it is immaterial whether this n--volution acts only in a single two--torus or in various tori at the same time. Putting these findings together, we see that we can also construct GLSMs which correspond to resolutions of non--factorized orbifolds. 

However, as mentioned in Subsection \ref{sc:T6/ZN}, some orbifolds admit off--diagonal \Kh\ deformations which bring the lattice back to a factorized form. Whether this is also possible in blow--up cannot be said with certainty, as precisely these off--diagonal \Kh\ deformations do not appear explicitly in our GLSM formulation. When the underlying orbifold theory does not admit off--diagonal \Kh\ deformations, a non--factorized orbifold cannot be turned into a factorized one, hence in this case the orbifold resolution is really non--factorizable. 

In this paper we provide examples of both non--factorized and non--factorizable orbifolds. In Section \ref{sc:ResT6Z3} we illustrate this feature using resolutions of the simple $T^6/\Intr_3$ orbifold. We construct resolution GLSMs on non--factorized lattices like $F_4 \times A_2$, $E_6$, and even identify a lattice which is not even a Lie algebra lattice. In Section \ref{sc:ResT6Z4} we show that truly non--factorizable $\Intr_4$ orbifolds can be obtained in this fashion. Nevertheless, as can be seen from Table \ref{tab:ClassificationModels} most of the orbifolds for which we have a GLSM description, are factorizable.

\subsection{Phase structure}
\label{sc:Phases} 

Any of the GLSMs of toroidal orbifold resolutions possesses a large variety of phases. They depend on the different possible regimes of the (relative) values of the \Kh\ parameters. From the target space perspective these phase transitions are rather drastic: They correspond to changes of the topology of the target space. There are various levels of topology changes measured by the number of codimensions involved: 
\enlargethispage{\baselineskip}
\items{
\item[i)] Modification of the intersection properties of divisors:
\\[1ex] 
Examples are the so--called flop transitions, where in one phase two divisors intersect, while in the next they do not anymore. Such transitions are relatively mild, because even though the intersection numbers in target space changes, the dimension of the moduli space remains the same. 
\item[ii)] Appearance or disappearance of divisors: 
\\[1ex] 
Examples are the blow--ups of cycles of orbifold singularities. Here the dimension of the moduli space may change: The divisors are in correspondence with the number of harmonic (1,1)--forms the geometry admits. 
\item[iii)] Alteration of the target space dimension:  
\\[1ex] 
Probably the easiest and best known example of this is the shrinking of the quintic to a point as discussed in \cite{Witten:1993yc}: The target space dimension jumps from three to zero. 
}
It is the power of the GLSM formalism that it is able to describe such topology changes in a smooth fashion and thus to connect topological spaces which do not seem to have much in common \cite{Aspinwall:1993xz,Witten:1993yc,Aspinwall:1994ev,Blumenhagen:2011sq}. 

Given that our toroidal orbifold GLSM resolutions in general possess a large number of FI--parameters their phase structures inevitably become rather involved. To illustrate the different regimes we will present only a schematic picture here: We distinguish only two types of \Kh\ parameters $a$ and $b$ corresponding to the torus cycles and exceptional cycles, respectively. Moreover, in the classification below we only make a distinction between phases when their transition is rather severe, i.e.\ for those of types ii) or iii) defined above. (In other words we ignore flop--like transitions here.) Using these simplifying assumptions we may distinguish the following types of phases: 
\enums{ 
\item {\bf Non--geometrical regime} ($a,b < 0$): 
\\[1ex] 
In this regime all the coordinates $z_{ai}$ have to vanish. The resulting space is therefore just a point. 
\item 
{\bf Orbifold regime} ($b <  0 < a$): 
\\[1ex] 
In this regime the GLSM describes our starting point: the toroidal orbifold. As all the exceptional coordinates are non--vanishing, the vacuum is determined only up to some discrete group actions. 
\item 
{\bf Blow--up regime} ($0 < b < \gg_0\, a$): 
\\[1ex] 
In this regime all exceptional cycles have finite size, but they are parametrically smaller than the torus cycles. This means that they do not intersect with each other unless they stem from the same fixed set. 
\item 
{\bf Critical blow--up regime} ($0 < \gg_0\, a < b < \gg_\infty\, a$): 
\\[1ex]
In this regime the exceptional and torus cycles have comparable sizes. In particularly this entails that different exceptional cycles which in the blow--up regime are disjunct, now start to intersect. In a given resolution GLSM this is typically not just a single phase, but involves a cascade of phase transitions. 
\item 
{\bf Over--blow--up regime} ($0 < \gg_\infty\, a < b$): 
\\[1ex] 
In this regime the roles of $a$ and $b$ have essentially become interchanged as compared to the blow--up regime: Here the idea that the geometry is built around a six torus has to be abandoned; the geometry is rather constructed starting from the base space that is defined through the exceptional gaugings. 
\item 
{\bf Singular over--blow--up regime} ($a < 0 < b$): 
\\[1ex] 
In this regime the geometry is defined through the exceptional gaugings. 
Moreover, the original torus cycles have disappeared and the non--zero VEVs of the $c$ fields induce singularities in the target space geometry. 
} 
Here the parameters $0 < \gg_0 < \gg_\infty$ define the boundaries surrounding the critical blow--up regime; their values depend on the GLSM in question. 

This is far from being a complete classification. For example we could have some of the torus and exceptional cycles large while others are small. It may happen that some of the phases that exist on the level of the ambient space do not lie on the zero loci of the various F--term conditions, and hence do not correspond to phases of the resulting Calabi--Yau. In Subsection \ref{sc:MinimalZ3Phases} we describe the phase structure of the minimal fully resolvable $T^6/\Intr_3$ GLSM in detail. We characterize each of these phases and investigate some of their basic properties. In particular we show that the dimension of the target space jumps from three to one between the blow--up and over--blow--up regimes: The critical phase in the middle turns out to contain both one and three dimensional components. In addition we identify two blow--up phases that are distinguished by a flop--like transition.

\subsubsection*{Phase structure analysis}

As outlined above the phase structure of a given GLSM, in particular one that corresponds to a toroidal orbifold resolution, quickly becomes rather involved. Therefore it is necessary to have some systematic way of analyzing its structure. The phase structure of the GLSMs is determined by the D--term equations. They decide which sets of coordinates need to have at least one non--vanishing element. This thus determines the phase structure on the ambient space. The restriction to the physical geometry is made by the F--terms. These F--terms could in principle exclude certain phases, but typically the different phases simply descent from the ambient space to the target space geometry. 

Below we give some details of this analysis. However, given that it becomes rather lengthy and involved in concrete cases, we preform it for one particular GLSM only: The minimal fully resolvable GLSM of $T^6/\Intr_3$ discussed in Subsection \ref{sc:MinimalZ3Phases}.

\subsubsection*{Ambient space phases} 

To analyze the phases of the ambient space we have to investigate the complete set of equations that can be obtained by building all possible linear combinations of the D--terms that result from the gaugings defining a given GLSM. In practice one only needs to form those linear combinations in which some of the coordinates drop out. This gives a finite set of equations. 

We can order the terms in each of the equations in this set such that all coordinates appear on the left--hand--side and all \Kh\ parameters on the right. Moreover, on each side we first write down all the terms with positive coefficients and then all terms with negative coefficients. When the set of D--term equations is represented in this way, we can easily read off the following important information: 
\items{
\item Phases of the ambient space:
\\[1ex]
Each time the right--hand--side of one of these equations changes sign, the GLSM goes through a phase transition, which in target space typically results in a change of topology. Hence the vanishing of the right--hand--sides of these equations determine the boundaries between the various possible \Kh\ cones. 
\item Sets of coordinates that cannot all vanish simultaneously: 
\\[1ex] 
In a given phase of the GLSM (e.g.\ a given \Kh\ cone) it is straightforward to read off which sets of coordinates need to have at least one non--zero member, depending on whether the combinations of the \Kh\ parameters on the right--hand--side is positive or negative. This  determines which coordinate patches exist.  
}

\subsubsection*{Restriction to the target space geometry}

To go from the ambient space to the target space geometry the F--term constraints have to be implemented. Taken at face value one finds that this gives an overcomplete set of equations, which would force all coordinates to zero. This is not the case, because some of the F--terms are redundant in the geometric phases. However, the subset of trivially fulfilled F--terms strongly depends on the non--vanishing coordinates in a given GLSM phase. 

Here, our specific way to describe two--tori  simplifies the analysis considerably: As will be described in detail in the next Section \ref{sc:twotori}, we use a formulation in which each superfield $\cZ_{ai}$ appears in a single monomial of the superpotential. Consequently its F--term is a monomial of coordinates. Therefore, vanishing of this F--term implies that at least one of the coordinates appearing in this monomial has to vanish. The analysis of a given phase of the ambient space provides a set of coordinates that cannot all vanish at the same time. This information combined often uniquely implies which coordinates necessarily vanish in the phase under investigation. 

The knowledge of the coordinates that need to vanish in a given phase often leads to considerable simplifications of the other F--terms which typically contain polynomials: It might happen that such an F--term reduces to a single monomial itself, in which case its consequences can be analyzed in the same way as the F--terms of the $\cZ_{ai}$ fields above. A second possibility is that only two monomials remain. This often implies that a certain pair of coordinates is either simultaneously vanishing or non--vanishing (because the equation implies that they are equal up to some phase factor). Combining this with the knowledge of sets of coordinates that cannot vanish all simultaneously leads to further restrictions. 

The effective target space dimension $d$ is determined from the dimension of the ambient space minus the number of coordinates that are forced to zero and minus the number of non--trivial left--over F--term constraints. Since the number of coordinates that are forced to zero depends on the set of coordinates that cannot vanish simultaneously, the target space dimension $d$ can in principle vary over the various phases of the GLSM.

\subsection{Identifying divisors}
\label{sc:Divisors} 

In the works of L\"ust et al.~\cite{Lust:2006zh,Reffert:2006du} the geometrical properties of toroidal orbifold resolutions were discussed. It is interesting to compare their description with the GLSM formalism outlined in the present work. 

To this end we briefly recall the essential ingredients of their gluing description of the topology of orbifold resolutions. They introduced three types of divisors: The so--called inherited divisors $R_a$ correspond to the divisors of the underlying six--torus. To describe the geometry near the orbifold singularities they introduce the so--called ordinary divisors $D_{ai}$. Finally, the local resolutions introduce the exceptional divisors $E_{r,ijk}$ and $E_{r,ij}$. Special care has to be taken for ordinary and exceptional divisors associated to fixed points or tori that get identified by a residual orbifold action. Moreover, since a global description was not available to these authors, they had to ingeniously shift between the global torus perspective and the local description of non--compact $\Cplx^3/\Intr_n$ resolutions to establish linear equivalence relations between these divisors and their intersection properties. 

One of the strong points of the GLSM formalism is that it provides a global description of orbifold resolutions, which is reflected in the identification of divisors. First of all, each superfield introduced in the GLSM can be associated with a divisor: 
\equ{ 
S_a := \{ c_a = 0 \}\,,
\qquad
D_{ai} := \{z_{ai} = 0\}\,, 
\quad\text{and}\quad  
E_{r,ijk} := \{x_{r,ijk} = 0\}\,,
}
etc., in the maximal fully resolvable model. (The divisors of non--maximal models are sums of the divisors of the maximal resolution GLSMs.) The description of the inherited divisors $R_a$ is a bit more involved. Following the description in  \cite{Lust:2006zh}  they are introduced on the  torus as so--called sliding divisors 
\equ{
R_a :=  \Big\{ u_a^{N_a} - \tilde{u}_{a}^{N_a} = 0 \Big\}\,, 
}
which depend on constants $\tilde{u}_{a}$. This realizes the hyperplane equation $u_a = \tilde{u}_{a}$ and its images under the space group; $N_a$ is the order of the orbifold action $\gth$ on the $a$--th two--torus. Using the Weierstrass functions $\wp_{\gt_a}(u_a)$ associated with the two--tori these equations can be written in terms of the homogeneous coordinates $z_{ai}$ of the elliptic curves. This results in zero loci of polynomials which have the same degrees as the $F$--terms of the superfields $\cC_a$; their coefficients encode the positions $\tilde{u}_{a}$ on the torus. In Subsection \ref{sc:InheritedDivisors} we show this explicitly for the GLSM description of the maximal resolution of $T^6/\Intr_3$. 

Not all of these divisors are present in all phases of the GLSM: The geometry of a given phase might entail that some homogeneous coordinates are identically zero; the corresponding divisors are part of the definition of the geometry. For example, in the standard geometrical phases, e.g.\ orbifold and blow--up regimes, the $S_a$ are part of the definition of the geometry since in these phases $c_a = 0$. Another option is that some coordinates cannot vanish in a given phase; the associated divisors do not exist in this phase. For example, the exceptional divisors $E_{r,ijk}$ are not present in the orbifold phase because $x_{r,ijk}$ is necessarily non--zero there. Moreover, as we observed at the end of Subsection \ref{sc:ResGLSMs},  the identification of the orbifold images of fixed tori is automatically build in. Hence essentially all divisors introduced in \cite{Lust:2006zh} can be realized as hyperplane equations.\footnote{As mentioned above in some cases, like the $T^6/\Intr_4$, our formalism is not able to distinguish all fixed tori separately. Hence in such cases our identification of divisors is slightly coarser than that presented in  \cite{Lust:2006zh}.} 

Two divisors $D$ and $D'$ are said to be linear equivalent, denoted as $D\sim D'$, when they are characterized by the same transition functions. These linear equivalence relations turn out to be insensitive to possible choices of triangulation. (The triangulation corresponds to a choice of GLSM phase, but does not affect the GLSM charge assignment.) In weighted projective spaces the transition functions result from using the $\Cplx^*$--scalings to interpolate from one coordinate patch to the next. In the GLSM description $U(1)$ subgroups of the $\Cplx^*$--scalings are realized as gauge symmetries. Hence two polynomials $P$ and $P'$ define linear equivalent divisors $D := \{P=0\}$ and $D' := \{ P'=0\}$ when all their gauge charges are the same. With this observation we can determine many linear equivalence relations: First of all, because of the gauge invariance of the superpotential, each of its monomials results in a linear equivalence relation involving a divisor $S_a$ and some divisors $D_{ai}$ and $E_{r,ijk}$. Moreover, as observed above the degrees of the polynomial defining $R_a$ is the same as $F_{c_a}$, and it follows that $S_a $ and $R_a$ are inverse to each other and can thus never appear simultaneously as effective divisors. The definition of the linear equivalences is formal in the sense that as observed above in a given phase certain divisors may not be present. However, since they only depend on the charges of the defining polynomials, they are independent of the GLSM phase, i.e.\ of the triangulation. In this way toroidal orbifold GLSM resolutions reproduce the linear equivalence relations from \cite{Lust:2006zh}. 

Finally, the intersections of divisors and the existence of curves can be worked out in the various GLSM phases by analyzing the resulting worldsheet scalar potential.


\section{GLSM description of two--tori}
\label{sc:twotori}

\begin{table}[t]
\renewcommand{\arraystretch}{\arrystrch}
\centering
\begin{tabular}{|l||l|l|l|l|l|l|l|l|}
\hline
Torus & Projective 	& \multicolumn{6}{c|}{$U(1)$ charges of} & Superpotential \\
\cline{3-9}
	 & hypersurface & $\mathcal{Z}_1$ & $\mathcal{Z}_2$ & $\mathcal{Z}_3$ & $\mathcal{Z}_4$ & $\mathcal{C}$ & $\mathcal{C}'$ & $W =$
\\ \hline\hline 
$T^2(\mathbbm{Z}_3)$&$\mathbbm{P}_{1,1,1}^2[3]$&1&1&1&--&-3&--& 
$\cC(\cZ_1^3+\cZ_2^3+\cZ_3^3)$ 
\\\hline 
$T^2(\mathbbm{Z}_4)$&$\mathbbm{P}_{1,1,2}^2[4]$&1&1&2&--&-4&--&
$\cC(\cZ_1^4+\cZ_2^4+\cZ_3^2)$ 
\\\hline 
$T^2(\mathbbm{Z}_6)$&$\mathbbm{P}_{1,2,3}^2[6]$&1&2&3&--&-6&--&
$\cC(\cZ_1^6+\cZ_2^3+\cZ_3^2)$ 
\\\hline 
$T^2(\mathbbm{Z}_2)$&$\mathbbm{P}_{1,1,1,1}^3[2,2]/\Intr_2^2$&1&1&1&1&-2&-2& 
$\cC  (\gk\, \cZ_1^2 + \cZ_2^2 +  \cZ_3^2 ) 
+
\cC' (\cZ_1^2 + \cZ_2^2 + \cZ_4^2 )$
\\\hline 
\end{tabular}
\renewcommand{\arraystretch}{1.0}
\caption{Summary of the weighted projective spaces and the order of the equations which are used to cut out all possible two--tori and their underlying $\Intr_N$ symmetry.}
\label{tab:SummaryTori}
\end{table}

In this Section we describe two dimensional tori as (2,2) GLSMs. To streamline the presentation we first give the summary of their elliptic curve description in Table \ref{tab:SummaryTori}, and then explain in subsequent subsections in detail how these descriptions can be obtained using the Weierstrass function. In addition we give there the possible orbifold and other discrete symmetries these tori may possess. 

Two dimensional tori may possess four different discrete orbifold symmetries: $\mathbbm{Z}_2$, $\mathbbm{Z}_3$, $\mathbbm{Z}_4$, and $\mathbbm{Z}_6$. Each of these two--tori can be conveniently described by a specific GLSM corresponding to a hypersurface defined by a polynomial $P(\cZ)$ inside a weighted projective space. The degree of these polynomials (specified in the square brackets of the respective weighted projective spaces) are given for the various two--tori in Table~\ref{tab:SummaryTori}. The $U(1)$ charges of the superfields are specified in this table for the different tori. From the charge assignments it is immediately obvious that the resulting objects are indeed  two--tori, since: 
\begin{align}
\chi(X)=c_1(X)=\sum\limits_i q_i + q_c =0\,.
\label{eq:ChernClassTori}
\end{align}
The first equality holds as the first Chern class is the top Chern class in two dimensions, and the second follows from the fact that the fields $\mathcal{Z}, \mathcal{C}$ couple to the tangent bundle of the variety.

We always take the polynomial $P(\cZ)$ as a sum of pure monomials. In this way we make the discrete symmetries that can act on each of the coordinates $z_i$ manifest. In particular, the fixed points of the orbifold action correspond to zeros of these coordinates. For example the $T^2/\Intr_3$ has three $\Intr_3$ singularities corresponding to the zeros of $z_1, z_2$ and $z_3$. For the $T^2/\Intr_4$ we have two $\Intr_4$ singularities and one $\Intr_2$ singularity. On the fundamental domain of the two--torus it looks as if there are two such fixed points, but they get identified by the residual orbifold action. This already shows that our formalism has the identification of fixed points under the orbifold action built in. 

The two--tori GLSMs possess two phases depending on the value of $a$. In the geometrical regime, where $a> 0$, the size of the torus is controlled by $a$, which can therefore be interpreted as the K\"ahler parameter of the torus. Since in our conventions the polynomial $P(z)$ will always be given as sum of monomials of a single coordinate, i.e.\ $z_i^{n_i}$, it is guaranteed that $c$ vanishes in the geometrical phase. In the non--geometrical phase, where $a< 0$, the D--term \eqref{GenDterm} implies that $c$ is non--vanishing. Consequently, again since $P(z)$ is a sum of pure monomials, the $\cZ_i$ F--terms imply that all $z_i$ vanish. This means that the target space geometry in the non--geometrical phase is just a single point. 

Finally, notice that the GLSM that describes the two--torus that possesses a $\Intr_2$ symmetry is special for the following reasons: It is described by four rather than three chiral superfields $\cZ_i$ and by two homogeneous polynomials associated to the two chiral superfields $\cC$ and $\cC'$. This description ensures that each of the $\Intr_2$ orbifold fixed points corresponds to a homogeneous coordinate $z_i = 0$. Secondly the complete intersection of two hypersurfaces $\mathbbm{P}^3_{1,1,1,1}[2,2]$ is modded out by an additional $\Intr_2\times\Intr_2$ symmetry. The reason for this is that we insist on having pure monomials of $\cZ_{i}$ in the superpotential with powers that corresponding to the order of the orbifold singularities in that torus. For the $\Intr_2$--torus we therefore need four coordinates that all appear quadratic in the superpotential. As will be explained in detail in Subsection \ref{sc:Z2torus} the mapping between this description and the torus description is not one--to--one unless one mods out these $\Intr_2$ symmetries. After these generalities we discuss below how to obtain these descriptions of the various two--tori in detail.

\subsection{The Weierstrass mapping for two dimensional tori}

Let $u$ be a double--periodic complex coordinate on a two--torus $\mathbbm{C}/\Lambda$, i.e.\ $u \sim u + 1 \sim u+\tau$. The lattice $\Lambda$ is spanned by $1$ and a fixed complex structure $\tau$, i.e.\ $\Lambda=\{m+n\tau|~m,n\in\mathbbm{Z}\}$. A generic lattice always possesses a $\Intr_2\times\Intr_2$ symmetry generated by $m\ra -m$ and $n \ra -n$. By requiring that the complex structure $\tau = \tau_1 + i \tau_2$ lies in the upper half--plane, i.e.\ $\tau_2 > 0$, the latter symmetry is removed. 

There are two instances in which the lattice $\Lambda$ possesses an enhanced symmetry:  i) The lemniscatic case has $\tau = i$, so that this lattice of so--called Gaussian integers exhibits a rotational $\mathbbm{Z}_4$ symmetry. This lattice underlies the $T^2/\mathbbm{Z}_4$ orbifold. ii) The equianharmonic case has $\tau=\gz= e^{2\pi i/3}$. The corresponding lattice, the so--called Eisenstein integers, has a rotational symmetry group $\mathbbm{Z}_6$ (with a subgroup $\mathbbm{Z}_3$). Therefore, this lattice underlies the $T^2/\mathbbm{Z}_6$ and $T^2/\mathbbm{Z}_3$ orbifolds. Some characteristics of these lattices are summarized in Table~\ref{tb:SpezialLattices} in terms of properties of the Weierstrass function discussed next.

\begin{table}[t]
\renewcommand{\arraystretch}{\arrystrch}
\begin{center}
\begin{tabular}{|c||c|c|c|c|c|}
\hline 
Lattice $\gL$ & Symmetry & Complex structure & Roots of $\wp_\tau^\prime(u)$ & Modular invariants 
\\ \hline\hline 
General & $\Intr_2$ & $\tau = \tau_1 + i \,\tau_2$ & $\gve_1+\gve_2+\gve_3 = 0$ & $f$, $g$ 
\\ \hline 
Gaussian Integers & $\Intr_4$ & $i = e^{i \gp/2}$ & $\gve_3=0$\,, $\gve_2=-\gve_1$ & $g=0$\,, $f=4\gve_1^2$ 
\\ \hline 
Eisenstein Integers & $\Intr_6\supset \Intr_3$ & $\gz = e^{2\gp i/3}$ & $\gve_i = \gz^{i-1} \gve_1$ & $f=0$\,, $g = 4\gve_1^3$
\\ \hline 
\end{tabular}
\end{center}
\caption{Special torus lattices with additional rotational symmetries.
\label{tb:SpezialLattices}}
\renewcommand{\arraystretch}{1.0}
\end{table}

We briefly review basic properties of the Weierstrass elliptic function and explain how it can be used to define elliptic curves in weighted projective spaces. A detailed discussion of the properties of Weierstrass functions can be found in textbooks like \cite{Ahlfors,Koblitz}. The Weierstrass function  $\wp_\tau(u)$ is an even double--periodic meromorphic function 
\begin{align}
 \wp_\tau(u)=\wp_\tau(u+1)=\wp_\tau(u+\tau)
 \label{periodicities} 
\end{align}
on the two--torus. Explicitly, the Weierstrass function $\wp_\tau(u)$ and its derivative $\wp_\tau^\prime(u)$ can be defined through its pole expansion 
\begin{align}
 \wp_\tau(u)=\frac{1}{u^2} + \sum\limits_{(m,n)\neq (0,0)} \Big\{ \frac{1}{(u+m+n\tau)^2}-\frac{1}{(m+n\tau)^2}\Big\}\,,
 \qquad 
 \wp_\tau^\prime(u)=-2\sum\limits_{m,n}\frac{1}{(u+m+n\tau)^3}\,.
\label{Weierstrass_poles}
\end{align}
The Weierstrass function satisfies the following differential equation
\begin{align}
 \left[\wp_\tau^{\prime}(u)\right]^2=4(\wp_\tau(u)-\varepsilon_1)(\wp_\tau(u)-\varepsilon_2)(\wp_\tau(u)-\varepsilon_3)\,.
 \label{eq:WeierstrassGeneral2}
\end{align}
The three zeros of $\wp_\tau^\prime(u)$ at $\varepsilon_1=\wp_\tau(\frac12)$, $\varepsilon_2=\wp_\tau(\frac\tau2)$, and $\varepsilon_3=\wp_\tau(\frac{1+\tau}{2})$ satisfy $\varepsilon_1+\varepsilon_2+\varepsilon_3=0$. Equivalently, this differential equation can be written as
\begin{align}
 \left[\wp_\tau^{\prime}(u)\right]^2 = 4\left[\wp_\tau(u)\right]^3- f(\tau) \wp_\tau(u) - g(\tau)\,.
 \label{eq:WeierstrassGeneral}
\end{align}
In terms of the roots $\varepsilon_i$ of $\wp_\tau^\prime(u)$ the functions $f(\tau)$ and $g(\tau)$ are expressed as
\begin{align}
 f(\tau)=-4 (\varepsilon_1\varepsilon_2+\varepsilon_1\varepsilon_3+\varepsilon_2\varepsilon_3)\,,
 \qquad 
 g(\tau)=4 \varepsilon_1\varepsilon_2\varepsilon_3\,.
  \label{eq:WeierstrassPrimeZeros}
\end{align}

The Weierstrass function defines an isomorphism between a two--torus with complex structure $\tau$ and an elliptic curve in the weighted projective space $\mathbbm{P}^2_{p,q,1}$ for given non--negative integers $p$ and $q$. The homogeneous coordinates $(x,y,v)$ of $\mathbbm{P}^2_{p,q,1}$ are subject to the $\Cplx^*$--scalings 
\begin{align}
\big(x,y,v\big) \quad\sim\quad 
\big(\lambda^p x, \lambda^q y, \lambda v\big)\,, 
\qquad 
\lambda\in\Cplx^*\,. 
\label{C*scalings}
\end{align} 
Equation \eqref{eq:WeierstrassGeneral} then defines the elliptic curve inside a weighted projective space. In the next Subsections we discuss the various possible two--tori in detail.

\subsection{Generic two--torus}
\label{sc:Gentorus}

The description of a generic two--torus is similar to that in \cite{Blaszczyk:2010db}, hence we will be brief here. The starting point for this discussion of a two--torus with generic complex structure $\tau$ is 
\begin{align}
 y^2= 4v (x-\varepsilon_1\, v) (x-\varepsilon_2 \,v) (x-\varepsilon_3 \,v)\; 
 \label{eq:Z2TorusWeierstrass}
\end{align}
in $\mathbbm{P}_{1,2,1}^{2}$. The mapping of the  torus coordinate $u$ to $(x,y,v)$ is given by 
\begin{align}
\arry{rcl}{
T^2 = \Cplx/\Lambda &\ra& \mathbbm{CP}_{1,2,1}^2 
 \\ 
 u &\mapsto& (x,y,v) =  
 \begin{cases} 
  \Big(\wp_\gt(u),\wp_\gt^\prime(u)/2,1\Big)\,, & \text{away from lattice points~,}  
    \\
   \big(1,0,0\big)\,, & u \in \Lambda\,, 
  \\ 
   \Big(1, \frac{\wp_\gt^\prime(u)}{2(\wp_\gt(u))^2},\frac{1}{\wp_\gt(u)}\Big)\,, & \text{near lattice points~.}
\end{cases} 
 }
 \label{eq:Z2Mapping}
\end{align}

There are various discrete symmetries that can act on the generic two--torus. First of all there is a $\Intr_2$ orbifold action $\gth$. On the torus parameterized by the variable $u$ it is realized as $\gth :~ u \mapsto -u$. On the elliptic curve parameterized by $(x,y,v)$ the orbifold action reads
\equ{
\gth :~ (x,y,v) \mapsto (x,-y,v)\,. 
\label{GenTorusOrbiAct}
}
This follows directly from the mapping \eqref{eq:Z2Mapping} and the fact that the Weierstrass function is even and so its derivative odd. The four fixed points of this action have $y=0$ and 
\equ{
(x,v) =  N_0 = (1,0)\,,~
N_{i} = (\gve_i, 1)\,,
\qquad 
i =1,2,3\,. 
}

In addition, there are three $\Intr_2$ involutions $\ga_i$, with $i=1,2,3$ that are compatible with this orbifold action $\gth$. On the  torus they act as translations over half lattice vectors: 
\equ{
\ga_1 :~ u \mapsto u + \sfrac 12\,, 
\quad 
\ga_2 :~ u \mapsto u + \sfrac \tau2\,, 
\quad 
\ga_3 :~ u \mapsto u + \sfrac {1+\tau}2\,, 
\label{Z2translation} 
}
hence clearly $\ga_3 = \ga_1 \ga_2 = \ga_2 \ga_1$, and $\ga_i^2 =\Id$ since their squares are torus lattice translations. Their actions on the elliptic curve can be compactly described as 
\equ{
\ga_i :~ \pmtrx{x \\ v} \mapsto 
\frac1{\sqrt{-\gve_{i+1}\gve_{i+2} - 2\,\gve_i^2}}
\pmtrx{\gve_i & \gve_{i+1}\gve_{i+2} + \gve_i^2 \\
1 & - \gve_i } \pmtrx{x \\ v}\,, 
\qquad 
\ga_i :~ y \mapsto \begin{cases} - y\,,  \quad &i=1,2 \\ y\,,  &i=3\end{cases}\,, 
\label{WL_Z2} 
}
using the roots $\gve_i$ of the Weierstrass equation defined \eqref{eq:WeierstrassGeneral2}. This is established in Appendix \ref{app:Z2torus}. In the subscripts here we perform addition modulo 3 such that the entries are either 1,2 or 3. It is not difficult to explicitly check that they constitute $\Intr_2$ involutions, as their square acts trivially on $x,y,v$. Finally, their actions on the fixed points $N_0, N_i$ of the orbifold action $\gth$ can be summarized as the following pairwise permutations
\equ{
\ga_i :~ N_0 \leftrightarrow N_i\,, 
\qquad 
\ga_i :~ N_{i+1} \leftrightarrow N_{i+2} \,, 
\label{Z2fixedpointmapping}
}
see Appendix \ref{app:Z2torus}. These are precisely the actions one expects from translations over half lattice vectors on the  torus, which permutes the fixed points pairwise in this way.

\subsection[Two--torus ${T^2(\Intr_6)}$ possessing a ${\mathbbm{Z}_6}$ symmetry]{Two--torus $\mathbf{T^2(\Intr_6)}$ possessing a $\mathbf{\mathbbm{Z}_6}$ symmetry}

The two--torus with a $\Intr_6$ symmetry has $f=0$ and $\gt = \gz$  according to Table \ref{tb:SpezialLattices}. It is mapped to an elliptic curve in $\mathbbm{P}^2_{2,3,1}$ via 
\begin{align}
\arry{rcl}{
T^2 = \Cplx/\Lambda &\ra& \mathbbm{P}_{2,3,1}^2 
 \\ 
 u &\mapsto& (x,y,v) =  
 \begin{cases} 
  \Big(\wp_\tau(u),\wp_\tau^\prime(u)/2,\gve_1^{1/2}\Big)\,, & \text{away from lattice points~,}  
    \\
   \big(1,1,0\big)\,, & u \in \Lambda\,, 
  \\ 
   \Big(\frac{4(\wp_\tau(u))^3}{(\wp_\tau^\prime(u))^2}, \frac{4(\wp_\tau(u))^3}{(\wp_\tau^\prime(u))^2},\frac{2\wp_\tau(u)}{\wp_\tau^\prime(u)}  \gve_1^{1/2} \Big)\,, & \text{near lattice points~.}
\end{cases} 
 }
\end{align}
This mapping brings the Weierstrass equation \eqref{eq:WeierstrassGeneral} to the form 
\begin{align}
 y^2=x^3 - v^6\,,
 \label{eq:Z6TorusWeierstrass}
\end{align}
where the parameter $\gve_1$ no longer appears explicitly. To make the symmetries that the description of the two--torus as an elliptic curve in $\mathbbm{P}^2_{2,3,1}$ possesses manifest, we perform a linear change of coordinates 
\begin{align} 
(x,y,v)\mapsto (-z_2,z_3, z_1)\,. 
\end{align} 
to transform the hypersurface equation to 
\begin{align}
 z_1^6+z_2^3+z_3^2=0\;
 \label{eq:Z6Torus}
\end{align}
in $\mathbbm{P}^2_{1,2,3}$. This makes the $\Intr_6\times \Intr_3\times \Intr_2$ symmetries, $z_1 \ra -\gz^2\, z_1$, $z_2 \ra \gz\, z_2$ and $z_3 \ra - z_3$, that act on each of these homogeneous coordinates separately, manifest. 

Let us give an interpretation of this set of discrete symmetries. First of all, the $\Intr_6$ symmetry that acts on all three coordinates simultaneously, i.e.\ $(z_1,z_2,z_3) \mapsto (-\gz^2\, z_1, \gz\, z_2, - z_3)$ is part of the $\Cplx^*$ group that defines the weighted projective space $\mathbbm{P}^2_{1,2,3}$, and is hence not physical. On the  torus the orbifold action reads $\gth :~ u \mapsto -\gz^2\, u$. The pole expansions \eqref{Weierstrass_poles} of the Weierstrass function and its derivative show that $\wp_\gz(-\gz^2\, u) = \gz^2\, \wp_\gz(u)$ and $\wp_\gz^\prime(-\gz^2\, u) = - \wp_\gz^\prime(u)$. Hence we conclude that on the homogeneous coordinates it acts as: 
\equ{
\gth :~ (z_1,z_2,z_3) \mapsto (z_1, \gz^2\, z_2, - z_3) = (-\gz^2\, z_1, z_2, z_3)\,. 
}
As this generates the remainder of the $\Intr_2\times \Intr_3\times \Intr_6$ that can act on the homogeneous coordinates associated with the $\Intr_6$ torus, there are no additional discrete group actions in this case.

\subsection[Two--torus ${T^2(\Intr_3)}$ possessing a ${\mathbbm{Z}_3}$ symmetry]{Two--torus $\mathbf{T^2(\Intr_3)}$ possessing a $\mathbf{\mathbbm{Z}_3}$ symmetry}
\label{sc:Z3torus}

The two--torus with a $\Intr_3$ symmetry has $f=0$ and $\gt =\gz$ according to Table \ref{tb:SpezialLattices}. It is mapped to an elliptic curve in $\mathbbm{P}^2_{1,1,1}$ via 

\begin{align}
\arry{rcl}{
T^2 = \Cplx/\Lambda &\ra& \mathbbm{P}_{1,1,1}^2 
 \\ 
 u &\mapsto& (x,y,v) =  
 \begin{cases} 
  \Big(\wp_\gz(u),\gve_1^{-1/2}\wp_\gz^\prime(u)/2,\gve_1\Big)\,, & \text{away from lattice points~,}  
    \\
   \big(0,1,0\big)\,, & u \in \Lambda\,, 
  \\ 
   \Big(\frac{2 \gve_1^{1/2}\wp_\gz(u)}{\wp_\gz^\prime(u)},1,\frac{2 \gve_1^{3/2}}{\wp_\gz^\prime(u)}\Big)\,, & \text{near lattice points~.}
\end{cases} 
 }
 \label{WeierstrassMappingZ3}
\end{align}
This mapping has been chosen so as to bring the Weierstrass equation \eqref{eq:WeierstrassGeneral} to a form
\begin{align}
 y^2 v = x^3  -  v^3\,, 
\label{eq:Z3TorusWeierstrass}
\end{align}
where the parameter $\gve_1$ no longer appears. To make the symmetries that the description of the two--torus as an elliptic curve in $\mathbbm{P}^2_{1,1,1}$ possesses manifest, we perform a  linear change of coordinates 
\begin{align}
\left(
\begin{array}{c}
z_1\\
z_2\\
z_3
\end{array}
\right)
=
\left(
\begin{array}{ccc}
  -1 & 0 & 0 \\
  \phantom{-}0 & \phantom{-}\frac{1}{\sqrt{3}} & 1\\
  \phantom{-}0 & -\frac{1}{\sqrt{3}} & 1
\end{array}
\right)
\left(
\begin{array}{c}
x\\
y/2^{1/3}\\
v / 2^{1/3} 
\end{array}
\right)\,,
\label{eq:Z3TorusFieldRedefinition}
\end{align}
to transform the hypersurface equation to 
\begin{align}
z_1^3+z_2^3+z_3^3=0\,, 
 \label{eq:Z3Torus}
\end{align}
in $\mathbbm{P}^2_{1,1,1}$. In this form three separate $\mathbbm{Z}_3$ symmetries $z_j \mapsto \gz\, z_j$ for each of the homogeneous coordinates $z_j$ of the torus are manifest. 

Let us give an interpretation of these three $\Intr_3$ symmetries. First of all the $\Intr_3$ symmetry that acts on all three coordinates simultaneously, i.e.\ $(z_1,z_2,z_3) \mapsto (\gz\, z_1, \gz\, z_2,\gz\, z_3)$, is just part of the $\Cplx^*$ action defining the projective space $\mathbbm{P}^2_{1,1,1}$ and hence is unphysical. Next, on the torus the orbifold map acts as $\theta: u \mapsto \zeta u$. The pole expansions \eqref{Weierstrass_poles} shows that 
$ \wp_\gz(\zeta u) = \zeta \wp_\gz(u)$ and $\wp_\gz'(\zeta u) = \wp_\gz'(u)$, from which we infer that
\begin{align}
\theta:~ (z_1,z_2,z_3) \mapsto (\zeta z_1,z_2,z_3).
\label{OrbiActionZ3}
\end{align}
The third independent $\Intr_3$ action can be taken to be: 
\begin{align}
\alpha:~ (z_1,z_2,z_3) \mapsto (z_1,\zeta^2 z_2,\zeta z_3) \,.
\label{WL_Z3}
\end{align}
This corresponds to a discrete translation on the torus 
\equ{ 
\alpha:~ u \mapsto u + \sfrac {\gz-1}3\,, 
\label{Z3translation}
}
which maps the three $\Intr_3$ fixed points, $0$, $\frac{\gz-1}{3}$ and $2 \frac{\gz-1}{3}$ onto each other. In Appendix \ref{app:Z3torus} we show this equivalence using summation properties of the Weierstrass function.

\subsection[Two--torus ${T^2(\Intr_4)}$ possessing a ${\mathbbm{Z}_4}$ symmetry]{Two--torus $\mathbf{T^2(\Intr_4)}$ possessing a $\mathbf{\mathbbm{Z}_4}$ symmetry}
\label{sc:Z4torus} 

The torus with a $\Intr_4$ symmetry has $g=0$ and $\gt = i$ according to Table \ref{tb:SpezialLattices}. It is mapped to an elliptic curve in $\mathbbm{P}^2_{1,2,1}$ via 
\begin{align}
\arry{rcl}{
T^2 = \Cplx/\Lambda &\ra& \mathbbm{P}_{1,2,1}^2 
 \\ 
 u &\mapsto& (x,y,v) =  
 \begin{cases} 
  \Big(\gve_1^{-1/4}\wp_i(u-u_0),\wp_i^\prime(u-u_0)/2,\gve_1^{3/4}\Big)\,, & \text{away from lattice points~,}  
    \\
   \big(1,0,0\big)\,, & u \in \Lambda\,, 
  \\ 
   \Big(1,\gve_1^{1/2} \frac{\wp_i^\prime(u-u_0)}{2(\wp_i(u-u_0))^2},\frac{\gve_1}{\wp_i(u-u_0)}\Big)\,, & \text{near lattice points~.}
\end{cases} 
 }
 \label{eq:Z4Mapping}
\end{align}
As explained in detail in Appendix \ref{app:Z4torus} the reason for including $u_0 = \frac {1+i}4$ here is to ensure that the orbifold actions on the  torus and elliptic curve are the standard ones. This mapping brings the Weierstrass equation \eqref{eq:WeierstrassGeneral} to the form 
\begin{align}
 y^2=x^3v-xv^3\,. 
 \label{eq:Z4TorusWeierstrass}
\end{align}
To make the symmetries that the description of the two--torus as an elliptic curve in $\mathbbm{P}^2_{1,2,1}$ possesses manifest, we perform a  linear change of coordinates 
\begin{align}
\left(
\begin{array}{c}
z_1\\
z_2\\
z_3
\end{array}
\right)
=
\left(
\begin{array}{ccc}
  \frac {(-1)^{3/4}}2	& 0 & -\frac{(-1)^{1/4}}2 \\
 \frac i2 	& 0 & \frac 12 \\
  0 & \frac{(-1)^{1/4}}{\sqrt 2} & 0
\end{array}
\right)
\left(
\begin{array}{c}
x\\
y\\
v
\end{array}
\right)\,.
\label{eq:Z4TorusFieldRedefinition}
\end{align}
to bring it to the form: 
\begin{align}
z_1^4+z_2^4+z_3^2=0\,, 
 \label{eq:Z4Torus}
\end{align}
in $\mathbbm{P}^2_{1,1,2}$. This equation is manifestly invariant under two $\Intr_4$ symmetries $z_i \ra i\, z_i$, for $i=1,2$ and a $\Intr_2$ symmetry $z_3 \ra -z_3$. 

Let us also in this case investigate these discrete symmetries in a bit more detail. First of all they contain a non--physical $\Intr_4$ symmetry 
$(z_1,z_2, z_3) \mapsto (i\, z_1, i\, z_2, - z_3)$ as it is part of the $\Cplx^*$ action which defines $\mathbbm{P}^2_{1,1,2}$. The $\Intr_4$ orbifold action on the  torus $\gth :~ u \mapsto i\, u$ translates into the action 
\equ{
\gth :~ (z_1,z_2, z_3) \mapsto (i\, z_1, z_2, z_3)\,, 
}
as is shown in Appendix \ref{app:Z4torus}. This leaves a single $\Intr_2$ symmetry. This involution $\alpha$ acts on the elliptic curve as  
\equ{
\ga :~ (z_1,z_2,z_3) \mapsto ( i z_1, - i z_2, z_3)\,, 
\label{WL_Z4}
}
while on the  torus it is simply a translation 
\equ{ 
\ga :~ u \mapsto u + \sfrac{1+i}2\,,
\label{Z4translation} 
}
as is also derived in Appendix \ref{app:Z4torus}.

\subsection[GLSM for the two--torus ${T^2(\Intr_2)}$ possessing a ${\mathbbm{Z}_2}$ symmetry]{GLSM for the two--torus $\mathbf{T^2(\Intr_2)}$ possessing a $\mathbf{\mathbbm{Z}_2}$ symmetry}
\label{sc:Z2torus} 

In this Section we have presented descriptions of two--tori that possess $\Intr_3$, $\Intr_4$ and $\Intr_6$ symmetries. For two--tori that possess a $\Intr_2$ symmetry, we have not yet given such a description. We do so in this final Subsection. It turns out that the GLSM language is convenient for this purpose. From the discussion in Subsection \ref{sc:Gentorus} a two--torus with a generic complex structure $\gt$ can be described by the GLSM with the superfields
\equ{
\arry{c|c|c|c}{
& \cX, \cV & \cY & \cC_0
\\ \hline 
q & 2 & 4  & -8 
}
}
and the superpotential 
\equ{ 
W_\text{gen.\ torus} = \cC_0 \big[\cY^2  - \cV(\cX - \gve_1\, \cV)(\cX - \gve_2\,\cV)(\cX -\gve_3\cV)\big]\,. 
\label{SPgen} 
}
We claim that this theory is equivalent to the following GLSM with superfields  
\equ{ 
\arry{c|c|c|c|c|c}{
& \cX, \cV & \cY & \cZ_1,\ldots, \cZ_4 & \cC_0' & \cC_1,\ldots,\cC_4 
\\ \hline 
q  & 2 & 4 & 1 & -4 & -2 
} 
\label{RedundantZ2fields}
} 
and superpotential 
\equ{ 
W_{\Intr_2\ \text{torus}} = \cC_0' \big( \cY - a_0\,\cZ_1 \cZ_2 \cZ_3 \cZ_4 \big)  
+ \cC_1\big( a_1\,\cZ_1^2 - \cV\big) 
+ \cC_2\big( a_2\, \cZ_2^2 - \cX+\gve_1 \cV \big) 
\non \\[2ex] 
+ \cC_3\big( a_3\, \cZ_3^2 - \cX+\gve_2 \cV \big) 
+ \cC_4\big( a_4\, \cZ_4^2 -\cX+\gve_3 \cV \big)\,, 
\label{SPZ2}
}
(with $a_0, \ldots, a_4$ some for now arbitrary non--zero complex coefficients) subject to the $\Intr_2\times\Intr_2$ identifications generated by 
\equ{ 
\beta_1: (\cZ_1,\cZ_2,\cZ_3,\cZ_4) \mapsto (\cZ_1,-\cZ_2,-\cZ_3,\cZ_4)\,, 
\qquad 
 \beta_2: (\cZ_1,\cZ_2,\cZ_3,\cZ_4) \mapsto (\cZ_1,\cZ_2,-\cZ_3,-\cZ_4)\,.
 \label{Z22sym}
}

To show this, we first integrate out the Lagrange multiplier superfields $\cC_1, \ldots, \cC_4$ to remove the superfields $\cZ_1,\ldots,\cZ_4$ via their equations of motion
\equ{ 
a_1\, \cZ_1^2 = \cV\,, 
 \quad 
a_2\, \cZ_2^2 = \cX- \gve_1 \cV\,,
\quad 
a_3\, \cZ_3^2 = \cX-\gve_2 \cV\,, 
\quad 
a_4\, \cZ_4^2 =\cX-\gve_3 \cV\,. 
\label{EoMC}
}
But since their charges are not of the same opposite size, this would lead to an anomaly, unless we perform a field redefinition 
\equ{ 
\cC_0' = (\cY + a_0\, \cZ_1 \cZ_2\cZ_3\cZ_4 ) \cC_0
}
at the same time. Inserting this field redefinition and using the equations of motions \eqref{EoMC}, we obtain \eqref{SPgen} from \eqref{SPZ2}, provided that we set $a_0^2 = a_1a_2a_3a_4$. The description in terms of the $\cZ_i$ has a $\Intr_2^4$ redundancy since each of the equations in \eqref{EoMC} determine $\cZ_i$ up to $\cZ_i \mapsto -\cZ_i$. The $\Intr_2$ transformation 
\(
(\cZ_1,\cZ_2,\cZ_3,\cZ_4) \mapsto - (\cZ_1,\cZ_2,\cZ_3,\cZ_4) 
\) 
is part of the $U(1)$ that defines the GLSM. Secondly the $\Intr_2$ orbifold symmetry \eqref{GenTorusOrbiAct} can be realized as 
\equ{ 
\gth :~ (\cZ_1,\cZ_2,\cZ_3,\cZ_4) \mapsto (-\cZ_1,\cZ_2,\cZ_3,\cZ_4)\,. 
}
The remaining sign ambiguities precisely correspond to the $\Intr_2\times\Intr_2$ symmetries generated by \eqref{Z22sym}. Using these symmetries and the $\Intr_2\subset U(1)$ we can realize the $\Intr_2$ orbifold action as a sign--flip of any of the four superfields $\cZ_i$. 

The description which uses the set of superfields given in \eqref{RedundantZ2fields} is somewhat redundant: As can be seen from the superpotential the superfield pair $(\cY, \cC_0')$ form a massive multiplet that decouples in the IR. Similarly, the superfields $\cX$ and $\cV$ form massive multiplets with two linear combinations of $\cC_1,\ldots,\cC_4$: 
\equ{
W_{\Intr_2\ \text{torus}} \supset 
\cX \big( \cC_2 + \cC_3 + \cC_3 \big) 
+ \cY \big( -\cC_1 + \gve_1 \cC_2 + \gve_2 \cC_3 + \gve_3 \cC_4\big)\,. 
}
Hence, the decoupling of these massive multiplets leaves the perpendicular combinations $\cC_2 = - \cC_3 - \cC_4$ and 
$\cC_1 = (\gve_2 -\gve_1) \cC_3 + (\gve_3 - \gve_1) \cC_4$. Inserting them back into the superpotential gives 
\equ{
W_{\Intr_2\ \text{torus}} = 
\cC_3 \Big[ (\gve_2-\gve_1) a_1\, \cZ_1^2 - a_2\, \cZ_2^2 + a_3\, \cZ_3^2 \Big]
+
\cC_4 \Big[ (\gve_3-\gve_1) a_1\, \cZ_1^2 - a_2\, \cZ_2^2 + a_4\, \cZ_4^2 \Big]\,. 
}
By making the choices $a_1 = 1/(\gve_3-\gve_1)$, $-a_2=a_3=a_4=1$ this reduces to 
\equ{
W_{\Intr_2\ \text{torus}} = 
\cC  (\gk\, \cZ_1^2 + \cZ_2^2 +  \cZ_3^2 ) 
+
\cC' (\cZ_1^2 + \cZ_2^2 + \cZ_4^2 )\,,
}
where we renamed $\cC = \cC_3$ and $\cC' = \cC_4$. The parameter $\gk = (\gve_2-\gve_1)/(\gve_3-\gve_1)$ encodes the complex structure 
$\gt$ of the two--torus in this description. The remaining field content is
\equ{ 
\arry{c|c|c|c|c|c}{
&  \cZ_1,\ldots, \cZ_4 & \cC & \cC'
\\ \hline 
q  & 1 & -2 & -2 
} 
\label{NonRedundantZ2fields}
} 
This is precisely the description indicated in Table~\ref{tab:SummaryTori}. 

Finally, we want to study discrete involutions on this torus. For this, we consider the translations $\ga_1$ and $\ga_2$ by half lattice vectors on the  torus, defined in \eqref{Z2translation}, 
and translate them via the Weierstrass map \eqref{eq:Z2Mapping} to the $\mathbbm{P}_{1,2,1}[4]$ torus and then via \eqref{EoMC} to the $T^2(\Intr_2)$ torus. Here we just show the result and work out the details of the calculation in Appendix \ref{app:Z2torus}, 
\begin{align}
 \alpha_1 :~ \big(z_1,z_2,z_3,z_4\big) & \mapsto \big( \kappa^{-1/4}\, z_2 ~,~  \kappa^{1/4} \, z_1 ~,~  \kappa^{1/4}\, z_4 ~,~ -\kappa^{-1/4} \, z_3 \big) \,, 
 \label{involutionsZ2Torus} \\ \nonumber
 \alpha_2 :~ \big(z_1,z_2,z_3,z_4\big) &\mapsto \left(i(\kappa(\kappa-1))^{-1/4}\, z_3 ~,~  \left(\frac{\kappa}{\kappa-1}\right)^{1/4}\, z_4 ~,~ i(\kappa(\kappa-1))^{1/4}\, z_1 ~,~  \left(\frac{\kappa}{\kappa-1}\right)^{-1/4}\, z_2 \right) \,.
\end{align}
We observe that the action of $\alpha_1$ and $\alpha_2$ is basically to permute the coordinates, and thus the fixed loci, pairwise, as is expected, since the fixed loci on the  torus sit at half lattice vectors. However, also the prefactors have a certain meaning. When we apply the shifts twice, we recover the $\Intr_2 \times \Intr_2$ action \eqref{Z22sym} we already modded out, i.e.\ $\alpha_1^2 = \beta_1$ and $\alpha_2^2 = \beta_2$. This shows that $\beta_1$ and $\beta_2$ can be interpreted as shifts by $1$ and $\tau$, respectively, of a  torus with periods $2$, $2\tau$, i.e.\ a torus with same complex structure but doubled radii.


\section{GLSM resolutions of $\mathbf{T^6/\Intr_3}$ orbifolds}
\label{sc:ResT6Z3}

\begin{table}[tb]
\centering
\renewcommand{\arraystretch}{1.1}
\begin{tabular}{|c||c|c|c|c|c|c|}
\hline
Exceptional &\multirow{2}{*}{Discrete Group} & \multicolumn{2}{|c|}{FP Sets} &  \multirow{2}{*}{Lattice}     & Fully resolvable\\
\cline{3-4}
coordinates                      &  & Groups        & Singular      &                               &                        by adding\\
\hline 
\hline 
$x_{111}$                     & --                              &$1 \times 27$  & $0$           &  $A_2 \times A_2 \times A_2$  & --       \\
\hline
\hline 
$x_{111},x_{211}$ &\multirow{3}{*}{$\mathbb{Z}_3$} & \multirow{3}{*}{$3 \times 9$}& \multirow{3}{*}{$9$} &  $A_2 \times A_2 \times A_2$ & $x_{311}$ \\
\cline{1-1} \cline{5-6}
$x_{111},x_{221}$ &                                &                              &                      &  $F_4 \times A_2$            & $x_{331}$ \\
\cline{1-1} \cline{5-6}
$x_{111},x_{222}$ &                                &                              &                      &  $E_6$                       & $x_{333}$ \\
\hline 
\hline 
\multirow{2}{*}{$x_{111},x_{211},x_{121}$}  &\multirow{6}{*}{$\mathbb{Z}_3\times\mathbb{Z}_3$} & \multirow{6}{*}{$9 \times 3$} & \multirow{6}{*}{$18$} &  \multirow{2}{*}{$A_2 \times A_2 \times A_2$} & $x_{131},x_{221},x_{231}$,\\ &&&&&$x_{311},x_{321},x_{331}$\\
\cline{1-1} \cline{5-6}
\multirow{2}{*}{$x_{111},x_{221},x_{112}$}  &						  &                               &                       &  \multirow{2}{*}{$F_4 \times A_2$} &$x_{113},x_{222},x_{223}$,\\&&&&& $x_{331},x_{332},x_{333}$\\
\cline{1-1} \cline{5-6}
 \multirow{2}{*}{$x_{111},x_{221},x_{212}$}  &                                                 &                               &                       &  \multirow{2}{*}{no Lie lattice} &$x_{123},x_{133},x_{232}$,\\&&&&& $x_{313},x_{322},x_{331}$\\
\hline 
\hline 
$x_{111},x_{211},x_{121},x_{112}$ & $\mathbb{Z}_3 \times \mathbb{Z}_3 \times \mathbb{Z}_3$ & $27 \times 1$            & 23           &   $A_2 \times A_2 \times A_2$ & rest \\
\hline
\end{tabular}
\renewcommand{\arraystretch}{1}
\caption{This table summarizes the structures that can be obtained upon inclusion of exceptional gaugings. The first column specifies which exceptional fields $x_{\ga\gb\gg}$ have to be introduced for each class. The second column gives the number of discrete $\Intr_3$ actions which are induced in addition to the orbifold action. The third column specifies how the 27 fixed points are grouped and how many fixed points cannot be blown up and thus stay singular. The fourth column specifies the Lie lattice underlying the torus in the various cases. In the fifth column, we specify which additional exceptional coordinates have to be included in order to be able to completely blow up the geometry.}
\label{tab:T6Z3Summary}
\end{table}

In this section we discuss resolutions of the $T^6/\Intr_3$ orbifold. As we shall see there are various ways to describe $T^6/\Intr_3$ and its resolutions as GLSMs. The minimal model, which was recently used by Aspinwall and Plesser in \cite{Aspinwall:2011us}, treats all fixed points in a symmetric way and therefore gives the simplest description. The maximal model is in some sense the most complicated one, because it can resolve each of the 27 fixed points independently. In between there are many models which differ in the amount of gaugings and $x$--fields and therefore in the way the $27$ fixed points, or their resolutions, are organized. In the following we present the classification of these models. We will see that discrete symmetries play an essential role and can even lead to non--factorized lattices in some cases.

In Table \ref{tab:T6Z3Summary} we give a systematic overview of all the possible $T^6/\Intr_3$ GLSMs we can construct using the methods outlined in Section \ref{sc:GenOrbiRes}. Various details of most of these models can be found in the remainder of this section. We classify the models according to the minimal amount and kind of $U(1)_{E_{\ga\gb\gg}}$ gaugings that have been switched on, to mod out a $\Intr_3^n$ ($n=0,\ldots 3$) discrete symmetry group of 3--volutions. We give the exceptional coordinates $x_{\ga\gb\gg}$, which are accompanied by the corresponding exceptional gaugings and FI--parameters. (Here we use the freedom to perform a relabeling of the two--tori and their three homogeneous coordinates to present one representative for each of class of models.) As we will see below in detail, these discrete groups split the 27 fixed points into smaller groups of equivalent ones. It turns out that quite a few of them cannot be resolved unless we include additional gaugings. In Table \ref{tab:T6Z3Summary} we indicate both the number of fixed points that are singular in each case, and which exceptional coordinates we have to add to make the model fully resolvable. 

The orbifold resolution GLSMs discussed in this section possess a vast amount of different phases. Studying these different regimes is more convenient in a setting where less \Kh\ parameters are around. In most parts of this section we confine ourselves to studying the orbifold and blow--up regimes only. However, in Subsection \ref{sc:MinimalZ3Phases} we attempt to gain an overview of the moduli space of the minimal fully resolvable model.

\subsection{The $T^6/\Intr_3$ orbifold}
\label{sc:Z3orbi} 

The $T^6/\Intr_3$ orbifold is probably the most--studied orbifold in the literature, being discussed from the very beginning of orbifold theories \cite{Dixon:1985jw}. The orbifold action on the three  two--tori with complex coordinate $u_a$ is defined as
\begin{align}
 \theta:(u_1,u_2,u_3)\mapsto (\zeta u_1,\zeta u_2,\zeta u_3)\,,\qquad\zeta:=e^{2\pi i/3}\,.
 \label{eq:T6Z3OrbifoldAction}
\end{align}
The orbifold has $27$ fixed points which we label by $(\alpha,\beta,\gamma)$, each running from one to three. These fixed points all have the local structure of a $\Cplx^3/\Intr_3$ singularity. 

In principle any $T^6/\Intr_3$ orbifold is defined on the root lattice of $A_2^3$ because of the $\Intr_3$ orbifold symmetry. However, if one has fixed the factorized structure of the torus coordinates $u_a$ as above, then one can obtain $\Intr_3$ orbifolds on non--factorized lattices by modding out 3--volutions that act on two or three complex torus directions simultaneously. This can produce lattices like $F_4\times A_2$, $E_6$ or even non--Lie algebra lattices. An overview of the possibilities up to relabeling of the coordinates is given in Table \ref{tab:T6Z3Summary}. Since there is really only one $T^6/\Intr_3$ orbifold, one can find a non--diagonal \Kh\ deformation to take any of these lattices to the standard $A_2^3$ factorized form. In Appendix \ref{app:LatticeAnayses} we give a detailed account of the possible lattices, and give the \Kh\ deformations that bring them back to the factorized form. Hence when in the remainder of this section we talk about non--factorized $\Intr_3$ orbifolds and their resolutions, we always mean non--factorized w.r.t.\ our specific choice of two--tori. 

In order to prepare for the resolution procedure, we represent each of the three two--tori with the appropriate $\Intr_3$ symmetry as the complete intersection space $\mathbbm{P}^2[3]$, as described in Section \ref{sc:Z3torus}. In the GLSM description the homogeneous coordinates $z_{ai}$ become part of chiral superfields $\cZ_{ai}$, $a,i =1,2,3$, with charges given in Table \ref{tab:T6Charges}. Following Section \ref{sc:Z3torus} to implement the three hypersurface constraints we introduce the superpotential 
\equ{ 
W_\text{torus} = \sum_{a,i} \cC_a \, \cZ_{ai}^3\,, 
\label{SPtorusZ3} 
} 
where we have absorbed all possible coefficients. 

\begin{table}[tb]
\centering
\renewcommand{\arraystretch}{\arrystrch}
\begin{tabular}{|c||c|c|c||c|c|c|}
\hline
Superfield & $\mathcal{Z}_{1i}$ 	& $\mathcal{Z}_{2j}$ & $\mathcal{Z}_{3k}$ 		& $\mathcal{C}_1$ & $\mathcal{C}_2$ & $\mathcal{C}_3$\\
$U(1)$ charges & $z_{1i}$		& $z_{2j}$	& $z_{3k}$	& $c_1$		  & $c_2$& $c_3$\\
\hline
\hline
${R_1}$ & $1$		& $0$			& $0$						& $-3$		  & $0$& $0$\\
\hline
${R_2}$  & $0$		& $1$			& $0$						& $0$	  	  & $-3$& $0$\\
\hline
${R_3}$  & $0$		& $0$			& $1$					& $0$	  	  & $0$& $-3$\\
\hline
\end{tabular}
\renewcommand{\arraystretch}{1}
\caption{Charges of the superfields in the GLSM description of the $T^6$ torus possessing a $\mathbbm{Z}_3$ orbifold symmetry.}
\label{tab:T6Charges}
\end{table}

On the $3\times 3$ homogeneous coordinates $z_{ai}$ the orbifold action can be chosen to be 
\equ{ 
\gth :~ z_{a1} \mapsto \zeta\, z_{a1}\,, 
\quad z_{a2} \mapsto z_{a2}\,, 
\quad z_{a3} \mapsto z_{a3}\,.
}
The three 3-volutions in the three torus directions act as 
\equ{ 
\ga_a :~ \big(z_{a1}, z_{a2}, z_{a3}\big) \mapsto 
\big(z_{a1}, \gz^2\, z_{a2}, \gz\, z_{a3}\big)\,, 
\label{Z3action}
}
and trivially on the other coordinates. In addition for use in the subsequent discussion we introduce the 27 $\Intr_3$ actions 
\begin{align}
 \theta_{\alpha\beta\gamma}: (z_{1\alpha},z_{2\beta},z_{3\gamma}) \rightarrow (\zeta\, z_{1\alpha},\zeta\, z_{2\beta},\zeta\, z_{3\gamma}) \,.
 \label{Z3reduction}
\end{align}
Even though this naively looks like $(\mathbb{Z}_3)^{27}$, it turns out that only a $(\mathbb{Z}_3)^4$ subgroup acts non--trivially on the coordinates. This subgroup is generated by the orbifold action and the three 3-volutions
\equ{ 
\gth_{\ga\gb\gg} = \gth\, 
\ga_1^{\ga-1}\, 
\ga_2^{\gb-1}\, 
\ga_3^{\gg-1}\,.  
}

\subsection{The maximal fully resolvable model}
\label{sc:MaxFullResZ3} 

Next we investigate the resolution models. To present them in a systematic fashion we first describe the so--called maximal fully resolvable model, because all the other $T^6/\Intr_3$ resolution GLSMs are constructed in a similar way, just with less gaugings. 

To construct compact resolutions we recall how to resolve a single $\Cplx^3/\Intr_3$ singularity. The orbifold action on its local coordinates is $\gth(z_1,z_2,z_3) = (\gz\, z_1,\gz\, z_2, \gz\, z_3)$. Hence, according to Subsection \ref{sc:LocalRes}, we need to consider the GLSM with a single $U(1)_E$ gauging with charges 
\begin{center}
 \begin{tabular}{c|c|c|c|c}
$U(1)$ charge & $\mathcal{Z}_1$ &    $\mathcal{Z}_2$ &    $\mathcal{Z}_3$ &    $\mathcal{X}$  \\
\hline
${E}$ & 1 & 1 & 1 & -3   
 \end{tabular}
\end{center} 
When $x\neq 0$ this induces the $\Intr_3$ on the local coordinates. 

Therefore, in the maximal fully resolvable GLSM we introduce Abelian gaugings $U(1)_{E_{\ga\gb\gg}}$ with $\ga,\gb,\gg = 1,2,3$  for each of the 27 fixed points. In order to keep the number of degrees of freedom unchanged we also introduce exceptional coordinates $x_{\alpha\beta\gamma}$. The charges of the resulting GLSM are summarized in Table~\ref{tab:T6Z3MaximalCharges}. These gaugings give rise to $3+27$ D--terms 
\begin{align}
\begin{alignedat}{2}
  |z_{a1}|^2 + |z_{a2}|^2 + |z_{a3}|^2 -3\, |c_a|^2 &= a_a \,, & a&=1,2,3 \,,
  \\[1ex] 
    |z_{1\alpha}|^2 + |z_{2\beta}|^2 + |z_{3\gamma}|^2 - 3\, |x_{\alpha\beta\gamma}|^2 &=  b_{\alpha\beta\gamma} \,, \qquad& \alpha,\beta,\gamma&=1,2,3 \,.
\end{alignedat}
\label{DtermsZ3} 
\end{align}
Here $a_a$ and $b_{\ga\gb\gg}$ are the real parts of the FI--parameters of the $U(1)_{R_a}$ and $U(1)_{E_{\ga\gb\gg}}$ gaugings, respectively. 

Because of the exceptional gaugings $U(1)_{E_{\ga\gb\gg}}$ the six--torus superpotential \eqref{SPtorusZ3} is no longer gauge invariant. But this is easily repaired by introducing the superfields $\cX_{ijk}$ at the appropriate places. This results in the superpotential
\begin{align}
 \mathcal{W} = \mathcal{C}_1 \sum_{i} \mathcal{Z}_{1i}^3 \prod_{j,k} \mathcal{X}_{ijk} +  \mathcal{C}_2 \sum_{j} \mathcal{Z}_{2j}^3 \prod_{i,k} \mathcal{X}_{ijk} +  \mathcal{C}_3 \sum_{k} \mathcal{Z}_{3k}^3 \prod_{i,j} \mathcal{X}_{ijk} \,.
\label{eq:Z3superpot}
\end{align}
Note that the various $U(1)_E$ symmetries forbid mixed monomials which fixes the complex stricture to $\tau = \zeta$. The superpotential leads to a large set of F--term conditions: We find three geometrical F--terms $F_{c_{a}}$: 
\begin{align}
\begin{alignedat}{2}
 &z_{11}^3 \prod_{\beta,\gamma}x_{1\beta\gamma} + z_{12}^3 \prod_{\beta,\gamma}x_{2\beta\gamma} + z_{13}^3 \prod_{\beta,\gamma}x_{3\beta\gamma} = 0\,,\\
 & z_{21}^3 \prod_{\alpha,\gamma}x_{\alpha1\gamma} +  z_{22}^3 \prod_{\alpha,\gamma}x_{\alpha2\gamma} +  z_{23}^3 \prod_{\alpha,\gamma}x_{\alpha3\gamma} = 0\,,\\
 & z_{31}^3 \prod_{\alpha,\beta}x_{\alpha\beta1} + z_{32}^3 \prod_{\alpha,\beta}x_{\alpha\beta2} + z_{33}^3 \prod_{\alpha,\beta}x_{\alpha\beta3} = 0\,,
 \label{Fca} 
\end{alignedat}
\end{align}
$9$ F--terms $F_{z_{ai}}$: 
\begin{align}
 c_1\, z_{1i}^2 \prod_{\beta,\gamma}x_{i \beta \gamma} = 0\,,\qquad 
 c_2\,  z_{2i}^2 \prod_{\alpha,\gamma}x_{\alpha i \gamma} = 0\,,\qquad
 c_3\, z_{3i}^2 \prod_{\alpha,\beta}x_{\alpha \beta i} = 0\,,
\label{Fzai} 
\end{align}
and finally, $27$ F--terms $F_{x_{ijk}}$: 
\begin{align}
c_1\, z_{1i}^3 \prod_{(\beta,\gamma)\neq (j,k)}x_{i\beta\gamma} + c_2 \, z_{2j}^3 \prod_{(\alpha,\gamma)\neq(i,k)}x_{\alpha j \gamma} + c_3\, z_{3k}^3 \prod_{(\alpha,\beta)\neq(i,j)} x_{\alpha \beta k} = 0 \,.
\label{Fxijk}  
\end{align}
The combination of these F--term and D--term conditions determine the target geometry described by the GLSM in a given phase.

\begin{table}[tb]
\centering
\renewcommand{\arraystretch}{\arrystrch}
\begin{tabular}{|c||c|c|c||c|c|c||c|}
\hline
Superfield & $\mathcal{Z}_{1i}$ & $\mathcal{Z}_{2j}$ & $\mathcal{Z}_{3k}$ 	
& $\mathcal{C}_1$ & $\mathcal{C}_2$ & $\mathcal{C}_3$
& $\mathcal{X}_{ijk}$
\\
$U(1)$ charges & $z_{1i}$ & $z_{2j}$	& $z_{3k}$	
& $c_1$ & $c_2$ & $c_3$
& $x_{ijk}$			
\\
\hline
\hline
${R_1}$ & $1$ & $0$ & $0$ 		
& $-3$ & $0$ & $0$
& $0$
\\
\hline
${R_2}$ & $0$ & $1$ & $0$			
& $0$ & $-3$& $0$
& $0$
\\
\hline
${R_3}$ & $0$	& $0$ & $1$	
& $0$ & $0$ & $-3$
& $0$\\
\hline\hline 
${E_{\alpha\beta\gamma}}$ & $\delta_{i \alpha}$& $\delta_{j \beta}$ & $\delta_{k \gamma}$	
& $0$ & $0$ & $0$ 
& $-3\, \delta_{i \alpha}\delta_{j \beta}\delta_{k \gamma}$
\\
\hline
\end{tabular}
\renewcommand{\arraystretch}{1}
\caption{Charges of the fields needed in the maximal description of the resolved $T^6/\mathbbm{Z}_3$ orbifold. The indices $i,j,k,\alpha,\beta,\gamma$ run from 1 to 3.}
\label{tab:T6Z3MaximalCharges}
\end{table}

\subsubsection*{Inherited divisors}
\label{sc:InheritedDivisors}

In \cite{Lust:2006zh} three types of divisors were introduced: ordinary, exceptional and inherited divisors. There are the ordinary and the exceptional divisors, which in the GLSM description can be described nicely as zero loci of homogeneous coordinates, i.e.\ ordinary divisors $D_{ai} := \{z_{ai} = 0\}$ and exceptional divisors $E_{r,ijk} := \{x_{r,ijk} = 0\}$. Since our GLSM description also involves superfields $\cC_a$ we can further define the divisors $S_a := \{c_a = 0\}$. For the inherited divisors $R_a$ the situation is more complicated. 

To deduce the proper form of the inherited divisors $R_a$ in the GLSM and show that it has the desired properties, we start from their definitions on the  torus 
\begin{align}
 R_a :=  \bigcup_{k,l=0}^2 \Big\{ u_a = \gz^k\, \tilde u_{a} + l\, \frac{\gz-1}3 \Big\}\,,  
\label{eq:InheritedDivisorFlat}
\end{align}
where $\tilde u_{a}$ are some constants. The union over the nine distinct pieces is necessary for $\Intr_3 \times \Intr_{3}$ invariance. Via the Weierstrass map \eqref{WeierstrassMappingZ3}, we can translate them into 
\equ{ 
{y_a}= p_a \, {v_a}\,, \quad y_a \left( 1 - i\, p_a \right) = v_a \left( p_a-3i\right)\,, \quad y_a \left( 1 + i p_a \right) = v_a \left( p+3i\right)\,, 
\label{xyWeierstrass} 
}
with $p_a:= \frac{\wp_\gz'(\tilde u_{a})}{2\sqrt{\epsilon_1}}$, using the properties of the Weierstrass function given above \eqref{OrbiActionZ3}. Each of these terms is $\Intr_{3, \rm Orbi}$ invariant and thus represents three pieces inside \eqref{eq:InheritedDivisorFlat}. To make the equations invariant under $\Intr_{3,\rm Orbi} \times \Intr_{3-\rm vol}$ we multiply them to obtain 
\equ{
0 = -( v_a^3 + v_a y_a^2 )\left( p_a^3 + 9 p_a \right) + (9v_a^2 y_a + y_a^3 ) \left( 1 + p_a^2 \right)\,.
}
Using the Weierstrass equation and the substitution \eqref{eq:Z3TorusFieldRedefinition} we can turn these equations into polynomials in $z_{ai}$:  
\begin{align}
0 =  \left( p_a^3 + 9 p_a \right) z_{a1}^3 + 6 \sqrt{3} \left( 1 + p_a^2 \right) \cdot \left( z_{a2}^3 - z_{a3}^3 \right) \,.
\end{align}
In the GLSM, to make it fully gauge invariant, we also have to include the $x_{ijk}$ coordinates. Then this corresponds to a homogeneous equation of the same degree as the F--term which cuts out the torus. The result is%
\begin{align}
 R_1 := \left\{ a_{11}(\tilde u_{1})\, z_{11}^3 \prod_{j,k} x_{1jk} + a_{12}(\tilde u_{1})\, z_{12}^3 \prod_{j,k} x_{2jk} + a_{13}(\tilde u_{1})\, z_{13}^3 \prod_{j,k} x_{3jk} = 0 \right\} \,,
\label{eq:InhDivAlg}
\end{align}
and $R_2$, $R_3$ similarly. Now the coefficients $a_{ai}(\tilde u_{a})$ encode the information about the position of the inherited divisors set by the constants $\tilde u_{a}$. 

From this we can easily read off the linear equivalence relations as each monomial defines one particular sum of divisors: 
\begin{align}
 R_1 \sim 3\, D_{11} + \sum_{j,k} E_{1jk} 
 \sim 3\, D_{12} + \sum_{j,k} E_{2jk} 
  \sim 3\, D_{13} + \sum_{j,k} E_{3jk} 
 \,,
\end{align}
and similarly for the divisors $R_2$ and $R_3$. In this way we recover the linear equivalences from \cite{Lust:2006zh}.

\subsubsection*{The intersection ring}

With the divisors and linear equivalences at hand we can determine the intersection ring in a GLSM manner. This allows us to prove the results on the intersection ring in \cite{Lust:2006zh} which were obtained by the means of an auxiliary polyhedron. For this we first obtain the intersection numbers of three distinct divisors by counting the number of solutions of the D-- and F--term equations. Then we use linear equivalences to determine the self--intersection numbers. In fact it is sufficient to give intersection numbers of a real basis of divisors. Such a basis is given by the divisors $R_a$, $a=1,2,3$ and all $E_{ijk}$. We will systematically derive their intersection numbers here in the blow--up phase only. This phase is roughly determined by $0 < b \ll a$, where $b$ and $a$ represent any FI parameter of the respective type. In this phase the geometry is a smooth Calabi--Yau space for which intersection numbers can be compared to the results of L\"ust et.\ al.\  \cite{Lust:2006zh}.

First of all, we find that in the blow--up regime, $R_a$ does not intersect with $E_{ijk}$. For an explanation we take e.g.\ $E_{111}$ and $R_1$. We insert the condition $x_{111}=0$ into the defining equation  \eqref{eq:InhDivAlg} for $R_a$ and the F-term \eqref{Fca} and find the conditions
\begin{align}
z_{12}^3 \prod_{j,k} x_{2jk} = z_{13}^3 \prod_{j,k} x_{3jk} = 0 \,.
\end{align}
Thus, in each monomial one field must be zero which gives us three different classes of cases, roughly described by $z=z=0$, $z=x=0$ or $x=x=0$. The first one is $z_{12}=z_{13}=0$. Then, the D-term of $U(1)_{E_{111}} - U(1)_{R_1}$ implies that $b_{111}>a_1$, which is beyond the blow--up regime. In the second class we choose e.g.\ $z_{12}=x_{311}=0$ to find that the D-term of $U(1)_{E_{111}} + U(1)_{E_{311}} - U(1)_{R_1}$ implies $b_{111} + b_{311} > a_1$, which again violates the blow--up phase condition. The third class is handled analogously. In the same way we see that $R_a$ and  $D_{ai}$ do not intersect. This implies via the linear equivalences that $R_a$ does not self--intersect. 

Next we focus on the intersection $R_1R_2R_3$. For this we impose three conditions of the form \eqref{eq:InhDivAlg}, one for each torus. As argued above on $R_a$ all $x_{ijk}$ have non--zero VEVs, so we fix their phases against the $U(1)_E$ gaugings, which leaves a $\Intr_{3,\rm Orbi} \times \Intr_{3-\rm vol}^3$ symmetry acting on the $z_{ai}$. Then, the inherited divisor constraints together with the three F--terms can be linearly combined to eliminate all $z_{a1}$ and we get three equations which are cubic in $z_{a2}$ and $z_{a3}$. Thus they factorize into three pieces, giving three solutions each. Now we insert these solutions into the original equations, which are cubic in $z_{a1}$ and thus again give three solution for each $z_{a1}$. Altogether this gives $3^6=729$ solutions. Now the discrete symmetries identify them in groups of $3^4=81$ each, so that the intersection number is $R_1R_2R_3=9$.

Furthermore, an intersection of $E_{ijk}$ and $E_{i'j'k'}$ where $(i,j,k)\neq(i',j',k')$ would also lead to a violation of the blow--up phase condition. The argumentation is similar to the one of $R_a E_{ijk}=0$. Finally, in the blow--up phase, we find the intersection number $D_{1i}D_{2j}D_{3k}=0$, since at a common intersection the D-term of $U(1)_{E_{ijk}}$ would contradict the blow--up condition $b_{ijk}>0$. Inserting linear equivalences shows
\begin{align}
\begin{split}
 27\, D_{1i}D_{2j}D_{3k} & = \Big( R_1 - \sum_{\gb,\gg} E_{i\gb\gg} \Big) \cdot \Big( R_2 - \sum_{\ga,\gg} E_{\ga j\gg} \Big)  \cdot \Big( R_3 - \sum_{\ga,\gb} E_{\ga\gb k} \Big) 
= R_1 R_2 R_3 - E_{ijk}^3~, 
\end{split}
\end{align}
which implies that $E_{ijk}^3=9$. Thus we have determined the complete intersection ring. The non--vanishing intersection numbers of inherited and exceptional divisors in the blow--up phase ($0 < b \ll a$) are
\begin{align}
 E_{ijk}^3 = 9 \,, \qquad R_1 R_2 R_3 = 9\,.
\end{align}
Intersection numbers containing ordinary divisors are obtained by using linear equivalences. Hence we have reproduced the intersection numbers obtained in \cite{Lust:2006zh} using GLSM methods.

\subsection{The minimal fully resolvable model} 
\label{sc:SingleGauging} 

This brings us to the other extreme as compared to the previous Subsection: the minimal fully resolvable model. This GLSM, discussed previously by Aspinwall and Plesser~\cite{Aspinwall:2011us}, contains just a single  exceptional divisor. Up to relabeling we may take the corresponding coordinate to be $x_{111}$. The charge assignment is therefore as shown in Table \ref{tab:T6Z3MaximalCharges} but with just a single gauging $U(1)_{E_{111}}$. 

In the orbifold and blow--up regimes we have $a_a > 0$ and $b_{111} < a_a$ such that the F--terms \eqref{Fzai} force  $c_a=0$ for $a=1,2,3$. (In Subsection \ref{sc:MinimalZ3Phases} we explain this in detail, and study the other phases which this model possesses as well.) The remaining non--trivial D-- and F--terms are
\begin{align}
\begin{alignedat}{2}
& |z_{a1}|^2 + |z_{a2}|^2 + |z_{a3}|^2 = a_a \,, 
& 
 z_{a1}^3 x_{111} + z_{a2}^3 + z_{a3}^3 = 0 \,,
 &  \qquad 
 a=1,2,3 \,,
 \\[2ex] 
& |z_{11}|^2 + |z_{21}|^2 + |z_{31}|^2 - 3\, |x_{111}|^2 = b_{111} \,.
 \end{alignedat}
\label{eq:DFAspinwallPlesser1}
\end{align}
As $b_{111}<0$ in the orbifold regime, we find that $x_{111}$ has to have a VEV breaking the $U(1)_{E_{111}}$ to $\Intr_3$. This discrete symmetry acts on the $z_{a1}$ as
\begin{align}
\theta:(z_{11},z_{21},z_{31})\mapsto(\zeta\, z_{11},\zeta\, z_{21},\zeta\, z_{31})\,.
\label{eq:T6Z3OrbifoldActionAlgebraic}
\end{align}
Using the mapping \eqref{WeierstrassMappingZ3} and properties of the Weierstrass function, it can be seen that \eqref{eq:T6Z3OrbifoldActionAlgebraic} indeed corresponds to the orbifold action in \eqref{eq:T6Z3OrbifoldAction}. Obviously, the action \eqref{eq:T6Z3OrbifoldActionAlgebraic} has fixed points at $z_{11}=z_{21}=z_{31}=0$. To see that there are indeed $27$ of them we look at the F--terms at $z_{a1}= 0$ for $a=1,2,3$: 
\begin{align}
z_{a2}^3  + z_{a3}^3 = 0\,.
\label{Z3fixedPoints}
\end{align}
Each of these equations has three solutions, 
\equ{
z_{a3} = - \zeta^{n_a} z_{a2}\,,
\label{Z3roots} 
}
for $n_a=0,1,2$, which altogether result in 27 distinct fixed loci. 

In the blow--up regime $0 < b_{111}< a_a$, the equation set~\eqref{eq:DFAspinwallPlesser1} allows for $x_{111}=0$. The second equation in~\eqref{eq:DFAspinwallPlesser1} then reduces to a symplectic quotient description of an ordinary $\mathbb{P}^2$, with homogeneous coordinates $z_{a1}$ and a volume growing with $b_{111}$. Since this $\mathbb{P}^2$ appears at each of the 27 positions determined by \eqref{Z3fixedPoints}, it consists of $27$ disconnected pieces replacing the $27$ fixed points. As they are all controlled by a single FI--parameter $b_{111}$, all the resolved $27$ singularities that make up exceptional divisor $E_{111} :=\{x_{111}=0\}$, can only be blown up or blown down simultaneously.

\subsection{Factorized resolutions with multiple exceptional coordinates} 
\label{sc:FactResZ3}

In the previous Subsections we have described the maximal and minimal fully resolvable GLSMs associated with $T^6/\Intr_3$. They provides two specific examples of factorized resolutions. In this Subsection we give a systematic treatment of all possible factorized resolutions. As we explained in the general discussion in Subsection \ref{sc:T6/ZN} we mean here factorized w.r.t.\ to the factorized description of $T^6 = (T^2(\Intr_3))^3$ we have chosen in our discussion. 

When we introduce more than one gauging then their relative orientation, i.e.\ which $U(1)_{E_{\ga\gb\gg}}$ are switched on, matters: In the orbifold regime, each of these gaugings induces a $\Intr_3$ action $\gth_{\ga\gb\gg}$ defined in \eqref{Z3reduction}. This results in further discrete actions, 3--volutions, in addition to the orbifold action. Now, depending on the orientation choice, these 3--volutions either act on different homogeneous coordinates within one two--torus or act on the coordinates of two or all three two--tori simultaneously. The former leads to factorized orbifolds, while the later leads to non--factorized compactification lattices. So concretely, let $\gth = \gth_{111}$ define the $\Intr_3$ orbifold action. Then $\gth_{211}$ gives a factorized orbifold while $\gth_{221}$ or $\gth_{222}$ give a non--factorized one. In the present Subsection we consider only the factorized compactifications; the non--factorized ones are the subject of Subsection \ref{sc:NonFact}. 

These discrete $\Intr_3$ actions map the existing fixed points onto each other; they get identified in groups of three, nine or 27. However, at the same time they create new fixed points in a such way that the total number of fixed points always remains $27$. Thus the single group of $27$ fixed points discussed above gets split up into subgroups of fewer indistinguishable fixed points. The resulting pattern is as follows: Two exceptional gaugings give one additional $\Intr_3$ which splits the $27$ fixed points into $3$ groups of $9$ fixed points each. Three exceptional coordinates can yield two additional discrete actions such that they are split into $9$ groups of $3$. Finally, upon introduction of a fourth exceptional coordinate, we can obtain three $\Intr_3$ actions, and all fixed points can be distinguished from one another. This behavior is expected for orbifolds equipped with Wilson lines associated to discrete actions. Below we study this in some specific examples.

\subsubsection*{Partially resolvable model with exceptional coordinates $\mathbf{x_{111}}$ and $\mathbf{x_{211}}$}
\label{FacResTwoX}

We take the GLSM with two exceptional coordinates $x_{\alpha\beta\gamma}$ where two of the three indices coincide. Without loss of generality we take them to be $x_{111}$ and $x_{211}$. The charge assignment can again be read off from Table \ref{tab:T6Z3MaximalCharges}. Here we do not aim to give a complete picture of the moduli space, but assume that $a_a > 0$ and take all $c_a=0$. To facilitate the subsequent discussion, we quote the D-- and F--term equations under these assumptions:
\begin{subequations} 
\begin{align}
 & |z_{a1}|^2 + |z_{a2}|^2 + |z_{a3}|^2 = a_a \,,\qquad a=1,2,3 \,,
 \\[1ex]
 & |z_{11}|^2 + |z_{21}|^2 + |z_{31}|^2 - 3\, |x_{111}|^2 = b_{111} \,,
 \\[1ex]
 & |z_{12}|^2 + |z_{21}|^2 + |z_{31}|^2 - 3\, |x_{211}|^2 = b_{211} \,.
\\[2ex]
 & z_{11}^3 \,x_{111}\phantom{x_{211}} + z_{12}^3\, x_{211} + z_{13}^3 = 0\,,
 \\[1ex]
 & z_{21}^3 \,x_{111}x_{211} + z_{22}^3\phantom{\,x_{211}} + z_{23}^3 = 0\,,
 \\[1ex]
 & z_{31}^3\, x_{111}x_{211} + z_{32}^3\phantom{\,x_{211}} + z_{33}^3 = 0\,.
\label{eq:DFAspinwallPlesser2}
\end{align}
\end{subequations} 

We first investigate this model in the blow down regime, $b_{111},b_{211} < 0$, where both $x_{111}$ and $x_{211}$ take VEVs. These VEVs induce two $\Intr_3$ actions given by
\begin{align}
\begin{split}
 \theta_{111}:~ &(z_{11},z_{21},z_{31})\mapsto(\zeta\, z_{11},\zeta\, z_{21},\zeta\, z_{31})\,,\\
 \theta_{211}:~ &(z_{12},z_{21},z_{31})\mapsto(\zeta\, z_{12},\zeta\, z_{21},\zeta\, z_{31})\,.
\end{split}
\label{eq:T6Z3TwoIntrActions}
\end{align}
It is instructive to describe the fixed points of these actions. The fixed points of $\theta_{111}$ are defined by $z_{11}=z_{21}=z_{31}=0$. Since $x_{211}$ has a non--zero VEV, we can write $x_{211}:= y_{211}^3$. However, this does not determine $y_{211}$ uniquely as $y_{211} \mapsto \gz\, y_{211}$ leaves this definition invariant. Evaluating the last three equations in \eqref{eq:DFAspinwallPlesser2} yields
\begin{align}
 (y_{211} z_{12})^3   + z_{13}^3 = 0\,, 
 \qquad 
  z_{22}^3 + z_{23}^3 = 0\,, 
 \qquad 
  z_{32}^3  + z_{33}^3 = 0\,. 
\end{align}
Naively this corresponds to 27 roots, as in \eqref{Z3roots}. However, because of the phase ambiguity of $y_{211}$ the roots in the first torus get identified; consequently, there are only nine independent solutions. A similar analysis  reveals nine further fixed points corresponding to the action of $\theta_{211}$.

We can uncover a third set of nine fixed points by inspecting the combined action $\theta_{111}^2\theta_{211}^2$
\begin{align*}
\theta_{111}^2\theta_{211}^2 :~ (z_{11},z_{12},z_{21},z_{31})
\mapsto 
(\zeta^2\, z_{11},\zeta^2\, z_{12},\zeta \,z_{21}, \zeta\, z_{31})\,.
\end{align*}
Using the element $\gth_{R_1}$ of  $\Intr_3 \subset U(1)_{R_1}$, this can be brought to the form
\begin{align}
\theta_{111}^2\theta_{211}^2\theta_{R_1} :~ (z_{13},z_{21},z_{31})
\mapsto 
(\zeta\,z_{13},\zeta\, z_{21},\zeta\, z_{31})\,, 
\label{eq:T6Z3TwoIntrActionsInduced}
\end{align}
which acts precisely as $\theta_{311}$ on the $z$ coordinates. Writing $x_{111} = y_{111}^3$ and $x_{211} = y_{211}^3$, we find that these fixed points are located at
\equ{
(y_{111} z_{11})^3 + (y_{211} z_{21})^3 =0\,, 
\qquad 
z_{22}^3 + z_{23}^3 = 0\,, 
\qquad 
z_{32}^3 + z_{33}^3 = 0\,. 
\label{theta311fp}
}
This corresponds to nine fixed points because the $\Intr_3$ phases of $y_{111}$ and $y_{211}$ are undetermined. 

Finally, to show that the combined 27 fixed points are all the possible ones, we have to investigate the remaining $\theta_{111}^2\theta_{211}$ sector. It acts as
\begin{align}
\theta_{111}^2 \theta_{211} :~ 
(z_{11},z_{12},z_{21},z_{31})\mapsto
 (\zeta^2\, z_{11},\zeta\, z_{12}, z_{21}, z_{31})\,.
\end{align}
Using again the discrete $\gth_{R_1}$ element of $U(1)_{R_1}$ this can be rewritten as 
\begin{align}
\theta_{111}^2\theta_{211}\theta_{R_1} :~ 
(z_{11},z_{12},z_{13})\mapsto 
(z_{11},\zeta^2\, z_{12},\zeta\, z_{13})\,.
\end{align}
This is the 3--volution $\ga_1$ in the direction of the first torus, see \eqref{Z3action}. This action is free in the blow--up regime, because 
fixed points of this action would require two of the three coordinates $z_{11}$, $z_{12}$, $z_{13}$ to be zero. Under the assumption $b_{111}, b_{211} < a_a$, which avoids over--blow--ups, we find that this cannot lead to a solution of the F-- and D--term equations in \eqref{eq:DFAspinwallPlesser2}.

Next we investigate what happens to the fixed points in the blow--up regime where $b_{111},b_{211}>0$ but small compared to $a_a$. For the two groups of nine fixed points induced by $\theta_{111}$ and $\theta_{211}$ the situation is as expected: The D--term in \eqref{eq:DFAspinwallPlesser2} involving $b_{111}$ forces at least one of the coordinates $z_{11}, z_{21}$ or $z_{31}$ to be non--zero. This mean that the fixed points of $\gth_{111}$ no longer exist. Moreover, setting $x_{111}=0$ reproduces the equations of $\mathbb{P}^2$'s glued into the previous nine fixed point positions of $\gth_{111}$ to resolve the singularities. Hence in this case the exceptional divisor $E_{111}$ consists of nine disjoint pieces. Using similar arguments we find that also the fixed points of $\gth_{211}$ are replace by an exceptional divisor, $E_{211}$ in this case, which contains nine $\mathbb{P}^2$'s. 

For the final set of nine fixed points, those of $\gth_{111}^2\gth_{211}^2 \gth_{R_1}$, the situation is totally different: These fixed points are not resolved in the blow--up regime. The easiest way of confirming this statement is to observe that $z_{13}=z_{21}=z_{31}=0$ still gives rise to the same solutions \eqref{theta311fp} of the equations \eqref{eq:DFAspinwallPlesser2} which are zero dimensional, i.e.\ points.

\subsubsection*{Fully resolvable model with three exceptional coordinates 
$\mathbf{x_{111}}$, $\mathbf{x_{211}}$ and $\mathbf{x_{311}}$}   

Let us summarize what we have seen so far. In Subsection \ref{sc:SingleGauging} we found that with a single exceptional gauging we can resolve all fixed points simultaneously. In Subsection \ref{FacResTwoX} we discussed a partially resolvable model with two gaugings in which nine of the 27 fixed points could not be resolved. 

The cause of this curious effect can be traced back to the following: The $\Intr_3$ actions of \eqref{eq:T6Z3TwoIntrActions} are the unbroken discrete subgroups that result from the VEVs of $x_{111}$ and $x_{211}$ breaking the underlying $U(1)_{E_{111}}$ and $U(1)_{E_{211}}$ groups. On the contrary, the action \eqref{eq:T6Z3TwoIntrActionsInduced} does not result from a separate gauging, and therefore, there is no exceptional field $x_{311}$, which can directly generate this action via its VEV. Moreover, because of the lack of this gauging, there is also no FI--parameter that can serve as the resolution modulus in this case.  

Hence one can interpret the non--resolvable fixed points as a consequence of the lack of the coordinate $x_{311}$ and the gauging $U(1)_{E_{311}}$. Indeed, if we include them, the $\gth_{311}$ fixed points are found in the orbifold regime. They get replaced by an exceptional divisor $E_{311} := \{x_{311}=0\}$ when the corresponding FI--parameter $b_{311} > 0$.

What we have seen here turns out to be a general pattern: If we introduce multiple gaugings which allow us to distinguish various collections of fixed points, then there will be a certain set of fixed points that are non--resolvable, unless we include all the possible gaugings in the corresponding torus directions. The number of non--resolvable fixed points is determined by the following consideration: The number of gaugings determine the number of independent \Kh\ parameters that can be used to resolve sets of equivalent fixed points. For example, when we have three exceptional coordinates, say $x_{111}, x_{221}, x_{212}$, then the fixed points fall into nine sets of three equivalent points. Three of these nine sets we can resolve, i.e.\ we have nine resolvable fixed points, and hence 18 which cannot be blown up in this model. In Table \ref{tab:T6Z3Summary} we have specified various gaugings by giving the corresponding exceptional coordinates. In addition, we indicated which additional gaugings have to be included in order to arrive at the corresponding fully resolvable model.

\subsection{Non--factorized resolutions}
\label{sc:NonFact} 

In this Subsection we explain how also non--factorized orbifolds and their resolutions naturally arise from our GLSM resolution procedure. Since a single gauging leads to the minimal fully resolvable model studied in Subsection \ref{sc:SingleGauging}, which describes a factorized orbifold resolution, we need at least two exceptional coordinates. For this reason we return again to studying the case of two exceptional gaugings. However, this time they are introduced such that the resulting free discrete $\Intr_3$ action acts on more than one torus simultaneously. The resulting types of non--factorized toroidal orbifolds, each of which we briefly describe below, are given in Table \ref{tab:T6Z3Summary}.

\subsubsection*{Resolutions on the $\mathbf{F_{4}\times A_2}$ lattice}

If we introduce the exceptional fields $x_{111}$ and $x_{221}$, the tori can only be factorized into $F_{4}\times A_2$ due to the discrete action. As the discussion of the three groups of 9 fixed points carries over from the factorized cases, we only outline here how to identify the free discrete action. In the orbifold regime, the gaugings induce the $\Intr_3$ actions $\gth_{111}$ and $\gth_{221}$. They can be combined to the free action
\begin{align}
\theta_{111}\theta_{221}^2:(z_{11},z_{12},z_{21},z_{22},z_{31})\mapsto (\zeta z_{11},\zeta^2 z_{12}, \zeta z_{21}, \zeta^2 z_{22}, z_{31})\,.
\label{eq:T6Z3NonFactorizedF4}
\end{align}
As fixed points of this action are not allowed by the D-- and F--terms in the orbifold or blow--up regime, this action is indeed free.  Its explicit form shows that it acts identically in both two--tori at the same time. This action is identical to the action of the 3--volution $\ga_1\ga_2$ defined in \eqref{Z3action}. In Appendix \ref{app:Z3LatticeF4} we show that this induces the $F_4 \times A_2$ lattice. 

We obtain the same lattice if we use a triple gauging with exceptional coordinates $x_{111}, x_{221}$ and $x_{112}$. The effect of the third gauging $U(1)_{E_{112}}$ is to distinguish the fixed points in third two--torus.

\subsubsection*{Resolutions on the $\mathbf{E_6}$ lattice} 

Upon introduction of $x_{111}$ and $x_{222}$, we obtain a completely non--factorized free discrete action. This case is equivalent to a description of a torus based on the $E_6$ lattice. The fixed point structure is the same as in the previous cases. The discrete action 
\equ{
\theta_{111}\theta_{222}^2: (z_{a1},z_{a2},z_{a3})\mapsto (z_{a1},\zeta z_{a2},\zeta^2 z_{a3})\,, \qquad \text{ for all } a=1,2,3\,.
}
clearly acts on all three tori identically at the same time. This is the action of the 3--volution $\ga_1\ga_2\ga_3$, see \eqref{Z3action}. In Appendix \ref{app:Z3LatticeE6} we show that this leads to the $E_6$ lattice.

\subsubsection*{Resolutions on a non--Lie lattice}

The final non--factorized lattice occurs when we take three exceptional coordinates, $x_{111}, x_{221}, x_{212}$. Here we run into an interesting situation where the resulting lattice does not correspond to a Lie--algebra lattice. This case is in a sense  twice the $F_4\times A_2$ case; once induced by $x_{111}, x_{221}$ and once by $x_{111}, x_{212}$, but the $F_4$'s get intertwined. In Appendix \ref{app:Z3LatticeNonLie} we determine the corresponding lattice and show that it does not fit within the Lie--algebra classification.

\subsubsection*{Off--diagonal \Kh\ deformations}

On a six--torus the K\"ahler deformation are associated to the $(1,1)$-forms $\d \bar u_{\bar a} \wedge \d {u}_b$, $a,b \in \{1,2,3\}$. As was explained in Subsection \ref{sc:AlgebraicTori}, on the orbifold only those torus \Kh\ deformations survive that are invariant under the orbifold action. Obviously always the three diagonal deformations, i.e.\ those with $b = \bar a$, are left invariant. Since the defining action \eqref{eq:T6Z3OrbifoldAction} of the $T^6/\Intr_3$ orbifold acts on all three coordinates $u_a$ in the same way, the six off--diagonal \Kh\ deformations, i.e.\ $b \neq \bar a$, also survive the orbifold projection. (For most other orbifolds this is not the case, and as a result they have less or often no off--diagonal \Kh\ deformations.) 

Now, the non--factorized lattices found above can be turned into the factorized lattice $A_2^3$ via a specifically chosen off--diagonal \Kh\ deformation. In Appendix \ref{app:LatticeAnayses} we give their explicit form which leads to the factorization for each of the lattices.


\begin{figure}[H]
\[
\includegraphics[width=7cm]{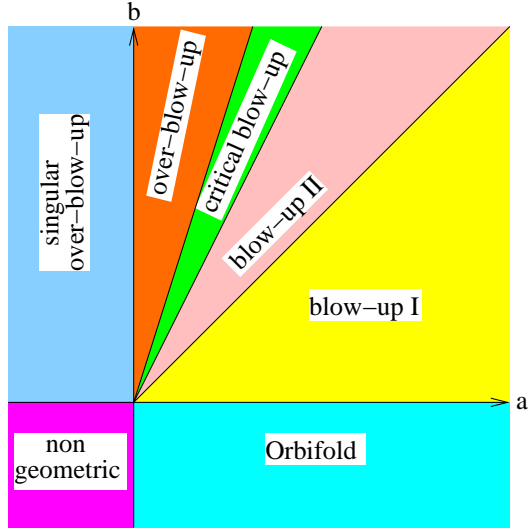} 
\]
\caption{This picture displays the different phases that the minimal fully resolvable GLSM possesses. These phases are characterized in Table \ref{tb:MinimalPhases} and discussed in detail in Subsection \ref{sc:MinimalZ3Phases}.
\label{fg:MinimalPhases}}
\end{figure}
\begin{table}[H] 
\[
\renewcommand{\arraystretch}{1}
\arry{|c||c|cc|c|c|c|c|}{
\hline 
\multirow{2}{*}{\text{Phase}} & \multirow{2}{*}{\text{Dim}} & \multirow{2}{*}{Vanishing} & \multirow{2}{*}{coordinates} & \text{Fixed} & \multirow{2}{*}{$D_{a1}$} & \multirow{2}{*}{$E_{\phantom{b1}}$}   & \multirow{2}{*}{$C_{a}$}  
\\ & & & & \text{Points} & & & 
\\\hline\hline 
 \multirow{2}{*}{\text{Non geom.}} & \multirow{2}{*}{0} & 
 \multirow{2}{*}{$z_{bj},$} & \multirow{2}{*}{$b,j = 1,2,3$} 
 & \multirow{2}{*}{1} & \multirow{2}{*}{--} & \multirow{2}{*}{--} & \multirow{2}{*}{--} 
 \\ & & & & & & & 
  \\ \hline 
 \multirow{2}{*}{\text{Orbifold}} & \multirow{2}{*}{3} & \multirow{2}{*}{$c_b, $} & \multirow{2}{*}{$b=1,2,3$} & \multirow{2}{*}{$3 \times 9$} & \multirow{2}{*}{$\checkmark$} & \multirow{2}{*}{--} & \multirow{2}{*}{$\checkmark$} 
 \\ & & & & & & & 
 \\ \hline 
 \multirow{2}{*}{\text{Blow--up I}} & \multirow{2}{*}{3}  & \multirow{2}{*}{$c_b, $} & \multirow{2}{*}{$b=1,2,3$} & \multirow{2}{*}{\text{none}} & \multirow{2}{*}{$\checkmark$} & \multirow{2}{*}{$\checkmark$} & \multirow{2}{*}{$\checkmark$} 
 \\ & & & & & & & 
 \\ \hline 
  \multirow{2}{*}{\text{Blow--up II}} & \multirow{2}{*}{3}  & \multirow{2}{*}{$c_b, $} & \multirow{2}{*}{$b=1,2,3$} & \multirow{2}{*}{\text{none}} & \multirow{2}{*}{$\checkmark$} & \multirow{2}{*}{$\checkmark$} & \multirow{2}{*}{--} 
  \\ & & & & & & & 
\\ \hline 
\multirow{2}{*}{\text{Critical}} & 3 & c_b, & b=1,2,3 & \text{none} & \checkmark & \checkmark  & \text{--}  
\\ 
& 1 & x, c_a, z_{bj}, & b\neq a, j\neq 1 & \text{none} & \checkmark & \text{--} & \text{--} 
\\ \hline 
\multirow{2}{*}{\text{Over--blow--up}} & 1 & x,z_{bj}, & b=1,2,3, j\neq 1 & \text{none} & \checkmark & \text{--} & \text{--} 
\\ 
& 1 & x, c_a, z_{bj}, & b\neq a, j\neq 1 & \text{none}  & \checkmark & \text{--} & \text{--} 
\\ \hline 
\text{Singular} & \multirow{2}{*}{1} & \multirow{2}{*}{$x,z_{bj}, $} & \multirow{2}{*}{$ b=1,2,3, j\neq 1$} & \multirow{2}{*}{3} & \multirow{2}{*}{$\checkmark$} & \multirow{2}{*}{\text{--}} & \multirow{2}{*}{--} 
\\ \text{over--blow--up}& & & & & & & 
\\ \hline 
}
\renewcommand{\arraystretch}{1}
\]
\caption{In this Table we give a set of data which distinguishes the different phases of the minimal fully resolvable GLSM uniquely. The number of dimensions that the target space has in a given phase is a very crucial piece of information: It is determined by the types of coordinates that necessarily have to vanish in a given phase. When there are fixed points the target space is certainly singular. In the final three columns we indicate whether certain characteristic divisors, namely $D_{a1}$ and $E$, exist in a given phase. The final column indicates in which phases the curve $C_a = D_{a+1,1} D_{a+2,1}$ exists. 
\label{tb:MinimalPhases}}
\end{table}

\subsection{Phases of the minimal fully resolvable GLSM}
\label{sc:MinimalZ3Phases}

In most of the analysis in this Section we have only explored a small part of the moduli space of the corresponding GLSMs: We have concentrated on the orbifold and the blow--up phases, as defined in Subsection \ref{sc:Phases}. The analysis of the full moduli space is quite involved and becomes more and more complicated as the number of \Kh\ parameters which are explicitly realized in the GLSM increases. Therefore, we are not able to present a full analysis in this paper. Instead in this and the next Subsection we have selected two very interesting examples to illustrate the general phenomenon that one may encounter while exploring the whole moduli space of an orbifold resolution GLSM. 

In this Subsection we investigate the moduli space of the simplest $T^6/\Intr_3$ resolution GLSM: the minimal fully resolvable model of Aspinwall and Plesser, defined in Subsection \ref{sc:SingleGauging}. In this model there are four \Kh\ parameters realized within the GLSM: the three two--tori radii $a_a$ and the size of the blow--up cycle $b = b_{111}$. (Similarly, we denote $x_{111}=x$ for brevity.) Consequently, the dimension of the ambient space equals $D = 3\cdot 3 + 3 +1 - 4 = 9$. For presentational and computational convenience, we take the three torus radii equal, i.e.\ set $a = a_a$ for $a=1,2,3$, so that the moduli space becomes two dimensional. 

Following the general strategy described in Subsection \ref{sc:Phases} we determine the complete set of equations which can be generated from the D--term equations: 
\begin{subequations} 
\label{DtermsMinimal} 
\equ{
\sum_i |z_{ai}|^2 - 3\, |c_a|^2 = a\,, 
\qquad
\sum_b |z_{b1}|^2 - 3\, |x|^2 = b\,, 
\\[1ex] 
\sum_{j\neq 1} |z_{aj}|^2 + 3\, |x|^2 
- \sum_{b\neq a} |z_{b1}|^2 - 3\, |c_a|^2 = a-b\,, 
\\[1ex] 
|z_{a1}|^2+\sum_{b\neq a, j\neq 1} |z_{bj}|^2 + 3\, |x|^2 
- 3 \sum_{b\neq a} |c_b|^2 = 2\, a -b\,, 
\\[1ex] 
\sum_{j\neq 1,a} |z_{aj}|^2 + 3\, |x|^2 
- 3 \sum_{a} |c_a|^2 = 3\, a -b\,. 
}
\end{subequations}
This collection of equations generated by the D--terms divide the moduli space into seven regions. In each of these phases different combinations of coordinates cannot vanish all at the same time. 

All these different ambient space phases descent to the target space of the minimal fully resolvable GLSM. To show this explicitly we have to investigate the consequences of the complete set of F--terms 
\begin{subequations}
\label{FtermsMinimal}
\begin{align}
c_a \, z_{a1}\,  x = c_a \, z_{a2} = c_a \, z_{a3} &= 0\,, 
\label{FtermsMinimal1}
\\[1ex] 
z_{a1}^3 \, x + z_{a2}^3 + z_{a3}^3 &= 0\,, 
\label{FtermsMinimal2}
\\[1ex] 
c_1\, z_{11}^3 + c_2\, z_{21}^3 + c_3\, z_{31}^3 &= 0\,. 
\label{FtermsMinimal3}
\end{align}
\end{subequations}
The equations on the first line are reduced in the sense that we have dropped the squares on the coordinates $z_{ai}$. (Obviously this does not affect the space of solutions to these equations in any way.) 

As the analysis of the various phases is rather involved, we have summarized some of the essential information in Table \ref{tb:MinimalPhases}. From this table one can obtain a schematic picture of the moduli space as given in Figure~\ref{fg:MinimalPhases}. Below we describe the distinction between the various phases in some detail. After the non--geometrical regime, we first discuss some additional properties of the orbifold and blow--up phases which were ignored in Subsection \ref{sc:SingleGauging}. Next, we turn to the singular and smooth over--blow--up phases. Finally, we close the circle of phases by explaining how the blow--up and over--blow--up are connected within the critical blow--up phase. Quite surprisingly, we will see that in the over--blow--up phase the target space dimension is one. Moreover, the critical blow--up phase contains both one and three dimensional components.

\subsubsection*{$\mathbf{a,b < 0}$: \qquad\qquad~ Non--geometric regime}

The collection of D--terms \eqref{DtermsMinimal}  implies that in this regime the coordinates $c_a \neq 0$, $a=1,2,3$ and $x\neq 0$. Consequently, the F--terms \eqref{FtermsMinimal1} force all $z_{ai}=0$, $a,i = 1,2,3$, after which all F--term equations \eqref{FtermsMinimal} are fulfilled. The dimension of the target space in this phase becomes: $d= 9 - 3\cdot 3 = 0$. As there is no phase ambiguities, we concluded that the target space in this phase is just a single point.

\subsubsection*{$\mathbf{b < 0<a}$: \qquad\qquad  Orbifold regime}

The orbifold regime has already been discussed in Subsection \ref{sc:SingleGauging} so we can be brief here. In this regime the D--terms \eqref{DtermsMinimal} give the following sets of coordinates that cannot all vanish simultaneously: 
$\{x \neq 0\}$, $\{z_{ai}\}_{i=1,2,3}\neq 0$, $a =1,2,3$. The F--terms \eqref{FtermsMinimal1} imply $c_a = 0$, $a=1,2,3$ as was asserted in Subsection \ref{sc:SingleGauging}. The three remaining non--trivial F--terms are given in \eqref{eq:DFAspinwallPlesser1}, and hence the target space dimension equals: $d= 9 - 3 -3 = 3$. As was argued in Subsection \ref{sc:SingleGauging} the target space is the orbifold $T^6/\Intr_3$. From the equations \eqref{eq:DFAspinwallPlesser1} we see that in the limit $a \downarrow 0$ the target space collapses to a point. For negative $a$ we return to the non--geometric regime. 

In this phase the divisors $S_a$ and $E$ do not exist. Since $x \neq 0$ the equation defining $E$ can never be satisfied. On the other hand, since $c_a=0$, $S_a$ does not define a divisor, but is rather part of the definition of the target space geometry itself. The other divisors $D_{ai}$ are realized in this phase. As can be read off from \eqref{DtermsMinimal}, the sizes of the divisors $D_{a1}$ are encoded in the equations
\equ{
\sum_{b\neq a} \Big( |z_{b2}|^2 + |z_{b3}|^2 \Big) + 3\, |x|^2 = 2\,a -b\,.
\label{SizeDa1}
}
For the other divisors $D_{aj}$, $j \neq 1$ we find instead that their sizes are set by $3\,a -b$, e.g.\ for $D_{12}$ have 
\equ{
|z_{13}|^2 +|z_{22}|^2 +|z_{23}|^2 +|z_{32}|^2 +|z_{33}|^2 
+ 3\, |x|^2 = 3\, a -b\,. 
}

\subsubsection*{$\mathbf{0 < b < a}$: \qquad\qquad  Blow--up regime I -- $\mathbf{D_{11}\cap D_{21} \neq \emptyset}$}

Also the blow--up regime has already been briefly considered in Subsection \ref{sc:SingleGauging}. So here we focus only on those aspects that explain how this phase fits within the complete moduli space. 

The D--terms \eqref{DtermsMinimal} imply that the following sets of coordinates do not all vanish at the same time: 
$\{z_{ai}\}_{i=1,2,3}\neq 0$, $a=1,2,3$; $\{z_{a1}\}_{a=1,2,3}\neq 0$; $\{x, z_{a2}, z_{a3}\} \neq 0$, $a=1,2,3$. When we assume that $c_a \neq 0$, we find $x=0$ and $z_{a2}=z_{a3}=0$, but in this phase not all of them can be zero at the same time. Hence, in this phase $c_a=0$, $a=1,2,3$ as claimed in Subsection \ref{sc:SingleGauging}. 

Next we describe the divisors that characterize this phase. The divisor $E := \{x=0\}$ with 27 components has appeared. Its size is determined by 
\equ{
|z_{11}|^2 + |z_{21}|^2 + |z_{31}|^2 = b\,.
} 
The curve $C_a$, that arises by intersecting the divisors $ = D_{a+1,1}$ and $D_{a+2,1}$ (the sum of the index calculated mod 3), has a size given by 
\equ{
3\, |x|^2 + |z_{a2}|^2 + |z_{a3}|^2 = a - b\,. 
}
When $b \uparrow a$ these curves collapse to zero volume, hence this curve defines the boundary between the first and second blow--up regimes.

\subsubsection*{$\mathbf{0 < a < b < 2\, a}$: \quad\, Blow--up regime II -- $\mathbf{D_{11}\cap D_{21} = \emptyset}$}

The non--vanishing sets of coordinates are in this case: $\{z_{ai}\}_{i=1,2,3}\neq 0$, $a=1,2,3$; $\{z_{a1}\}_{a=1,2,3}\neq 0$; $\{z_{11},z_{21}, c_3\}\neq 0$; $\{z_{11},z_{31}, c_2\}\neq 0$; $\{z_{21},z_{31}, c_1\}\neq 0$; 
$\{x, z_{12},z_{13},z_{22},z_{23}\}\neq 0$;
$\{x, z_{12},z_{13},z_{32},z_{33}\}\neq 0$; 
$\{x, z_{22},z_{23},z_{32},z_{33}\}\neq 0$. 
This still forces $c_a = 0$ for all $a=1,2,3$. However, in this phase the curves $C_a$ are no longer present; between this and the previous phase a flop--like transition occurred. As \eqref{SizeDa1} shows the divisors $D_{a1}$ still exist; they shrink to zero size when $b \uparrow 2\,a$. This marks the end of the blow--up phases.

\subsubsection*{$\mathbf{a< 0 < b}$: \qquad\qquad  Singular over--blow--up regime}

Next, we proceed from the other end of the moduli space so to say, where $a < 0 < b$. The D--terms \eqref{DtermsMinimal} show that the non--vanishing coordinate sets are: $\{c_a\neq 0\}$, $a=1,2,3$ and $\{z_{11}, z_{21}, z_{31}\}\neq 0$. The F--terms \eqref{FtermsMinimal1} force $x=0$ and $z_{ai}=0$, $a=1,2,3$; $i=2,3$. The only non--trivial remaining F--term is 
\equ{
c_1\, z_{11}^3 + c_2\, z_{21}^3 + c_3\, z_{31}^3 = 0\,. 
\label{FsingOverBlow}
}
From this it follows that the dimension of the target space equals: $d= 9 -1-3\cdot 2-1 =1$. 

By a change of $U(1)$ basis, we infer that the target space geometry in this phase can be thought of as  
\equ{
\frac{ 
\left[ \arry{c|c}{\mathbbm{P}^2 & 1 \\ 
\mathbbm{P}^2 & 3 }
\right]
}{(\Cplx^*)^2}\,. 
}
The $\mathbbm{P}^2$'s are spanned by the coordinates $c_a$ and $z_{a1}$ with $\Cplx^*$--actions corresponding to the $U(1)$ generators $R = E - \sum_b R_b$ and $E$, respectively. The remaining two independent $U(1)$ directions give rise to the additional modding by $(\Cplx^*)^2$. 

The size of the geometry is determined by the D--terms which set the sizes of the two $\mathbbm{P}^2$'s:
\equ{
|z_{11}|^2 + |z_{21}|^2 + |z_{31}|^2 = b\,, 
\qquad  
|c_{1}|^2 + |c_{2}|^2 + |c_{3}|^2 = \sfrac 13\, b - a\,, 
} 
We see that in the limit $b \downarrow 0$ this space shrinks to a point; in that limit we approach the non--geometrical phase with a point--like target space. The second equation shows that this geometry can extend to other phases in principle until $b = 3\,a$. 

Since the target space is only one dimensional, its divisors are mere points. Because each $c_a\neq 0$, $a=1,2,3$, the only divisors are $D_{a1}$. Moreover, this implies that there are three independent $\Intr_3$ actions on the three coordinates $\gth_a :~ z_{a1}\mapsto \gz\, z_{a1}$, $a=1,2,3$. Hence their fixed points are located at
\equ{
\arry{c}{
z_{11} = 0~:~~ c_2\, z_{21}^3 + c_3\, z_{31}^3 = 0\,, 
\\[1ex] 
z_{21} = 0~:~~ c_1\, z_{11}^3 + c_3\, z_{31}^3 = 0\,, 
\\[1ex] 
z_{31} = 0~:~~ c_1\, z_{11}^3 + c_3\, z_{21}^3 = 0\,. 
}
\label{FPoverblowup}
}
Since these are cubic equations, one might expect three solutions. But since the other two $\Intr_3$ act on the other coordinates, the three fixed points get identified. We concluded that the target space is complex one dimensional with three singular points determined by the equations \eqref{FPoverblowup}.

\subsubsection*{$\mathbf{0 < 3\, a < b}$: \qquad~~\, Over--blow--up regime}

In the over--blow--up phase the following sets of coordinates are not allowed to vanish simultaneously: $\{z_{ai}\}_{i=1,2,3}\neq 0$, $a=1,2,3$; $\{z_{a1}\}_{a=1,2,3}\neq 0$; 
$\{c_a\}_{a=1,2,3} \neq 0$, $\{z_{21}, z_{31}, c_1\} \neq 0$; $\{z_{11}, z_{31}, c_2\} \neq 0$; $\{z_{11}, z_{21} c_3\} \neq 0$; $\{z_{11}, c_2, c_3\} \neq 0$; $\{z_{21}, c_1, c_3\} \neq 0$; $\{z_{31}, c_1, c_2\} \neq 0$. Because of this, if we assume that $x\neq 0$ we find that all $c_a=0$, but that is not possible in this phase. Hence we conclude that $x=0$ and thus the divisor $E$ does not exist in this geometry. 

This time the reduced set of F--term equations  
\equ{
z_{a2}^3 + z_{a3}^3 = 0\,, 
\quad 
c_1\, z_{11}^3 + c_2\, z_{21}^3 + c_3\, z_{31}^3  = 0\,, 
\quad 
c_a\, z_{a2} = c_a\, z_{a3} = 0\,, 
}
still contains monomial equations with multiple coordinates. The consequence of this is that we have to distinguish four different components in the geometry: either all $c_a\neq 0$ or just a single one. (The vanishing of two or more $c_a$ leads to contradictions.)

\subsubsection*{The component: all $\mathbf{c_a \neq 0}$:}

When all $c_a\neq 0$, the geometry is characterized as in the singular over--blow--up regime: $z_{ai}=0$, $a=1,2,3$ and $i=2,3$, and it is one dimensional. However, contrary to the singular over--blow--up geometry, the fixed points are removed: Some of the D--terms show that 
\equ{
|z_{a1}|^2 = a + |c_a|^2\,, 
}
which means that the $z_{a1}$ cannot vanish. The size of this component is confined by the following two equations: 
\equ{ 
|z_{11}|^2 + |z_{21}|^2 + |z_{31}|^2  = b\,, 
\qquad 
|c_1|^2 + |c_2|^2 + |c_3|^2 = \sfrac 13\, b - a\,. 
\label{DtermsOBU1}
}
These show that it exists in the singular and smooth over--blow--up regime. 

In this regime the divisor $D_{a1}$ is present because the equation 
\equ{
\sum_{b\neq a} |c_b|^2 = \sfrac 13\, b - \sfrac 23\, a
}
can be fulfilled. In fact, this equation tells us that even though for $b < 3\, a$ this component has disappeared, the divisor exists also in the critical blow--up region (which is discussed below).

\subsubsection*{The component: $\mathbf{c_a=0, c_{b\neq a}\neq 0}$:}

In the component, where say, $c_3 =0$ but $c_1, c_2 \neq 0$, the situation is different: In this case the F--terms imply that 
$z_{12}= z_{13}= z_{22}= z_{23}= 0$. There are two remaining F--terms 
\equ{
c_1\, z_{11}^3 + c_2\, z_{21}^3 = 0\,, 
\qquad 
z_{32}^3 + z_{33}^3 = 0\,, 
\label{FtermsOBU2}
}
hence the dimension of this component equals again: 
$d = 9 - 1-1 -4- 2 = 1$. The size of this component is determined by the equations 
\equ{
|z_{11}|^2 + |z_{21}|^2 + |z_{31}|^2  = b\,, 
\quad 
|c_1|^2 + |c_2|^2 + |c_3|^2 = \sfrac 13\, b - a\,, 
\quad 
|z_{31}|^2 + |z_{32}|^2 + |z_{33}|^2  = a\,. 
}
The first two equations are the same as the D--terms \eqref{DtermsOBU1} which encode the size of the first component of the target space geometry. The final equation here signifies that this component only exists in the smooth over--blow--up region. 

Like in the singular over--blow--up phase the first equation of \eqref{FtermsOBU2} has effectively only one solution because $c_1,c_2 \neq 0$ induce $\Intr_3$ identifications. The second equation has four solutions. The trivial one $z_{33}=z_{23}=0$ just defines the missing point on the first component of the geometry. The other three solutions $z_{33} = - \gz^n\, z_{32}$, $n=0,1,2$ correspond to the blow--up cycles associated with the $\Intr_3$ fixed points of singular over--blow--ups.

\subsubsection*{$\mathbf{0 < 2\,a < b < 3\, a}$: \quad Critical blow--up regime}

Finally, we come to the critical blow--up regime. The sets of coordinates that cannot all be zero at the same time are: 
$\{z_{ai}\}_{i=1,2,3}\neq 0$, $a=1,2,3$; $\{z_{a1}\}_{a=1,2,3}\neq 0$; 
$\{z_{21}, z_{31}, c_1\} \neq 0$; $\{z_{11}, z_{31}, c_2\} \neq 0$; $\{z_{11}, z_{21} c_3\} \neq 0$; $\{z_{11}, c_2, c_3\} \neq 0$; $\{z_{21}, c_1, c_3\} \neq 0$; $\{z_{31}, c_1, c_2\} \neq 0$; 
$\{x,z_{12},z_{13},z_{22},z_{23},z_{32},z_{33}\}\neq 0$. It follows that if all $c_a\neq 0$ or when two of them are zero but the third is not, we run into a conflict between these conditions and the F--terms. Hence, like in the over--blow--up regime the geometry consists of various components. But contrary to the case above, these components have different dimensions now:

\subsubsection*{The component: all $\mathbf{c_a=0}$:}

In this component the only remaining F--terms are: 
\equ{
z_{a1}^3\, x + z_{a2}^3 + z_{a3}^3 = 0\,. 
}
Hence its dimension is $d=9 - 3 -3 = 3$ as in the orbifold and blow--up phases. Its size is set by the equations 
\equ{
\sum_i |z_{ai}|^2 = a\,, 
\qquad 
\sum_{j\neq 1, a} |z_{ai}|^2 + 3\, |x|^2 = 3\, a -b\,. 
}
Hence in the limit $b \uparrow 3\, a$ this component disappears.

\subsubsection*{The component $\mathbf{c_a=0, c_{b\neq a}\neq 0}$:}

This component is described by the same F--term equations \eqref{FtermsOBU2} as the corresponding component in the over--blow--up regime: 
\equ{
c_1\, z_{11}^3 + c_2\, z_{21}^3 = 0\,, 
\qquad 
z_{32}^3 + z_{33}^3 = 0\,, 
}
and $z_{12}= z_{13}= z_{22}= z_{23}= 0$, hence it is also one dimensional. However, its size is governed by different combinations of the D--terms. In particular, 
\equ{
3\, |c_1|^2 + 3\, |c_2|^2 + |z_{31}|^2 = b - 2\, a\,, 
}
shows that it disappears in the limit $b \downarrow 2\,a$.

\subsection{A duality of the fully resolvable GLSM with coordinates $\mathbf{x_{111},x_{222},x_{333}}$}
\label{sc:Z3dualGLSM}

\begin{figure}[t]
\[
\includegraphics[width=7.0cm]{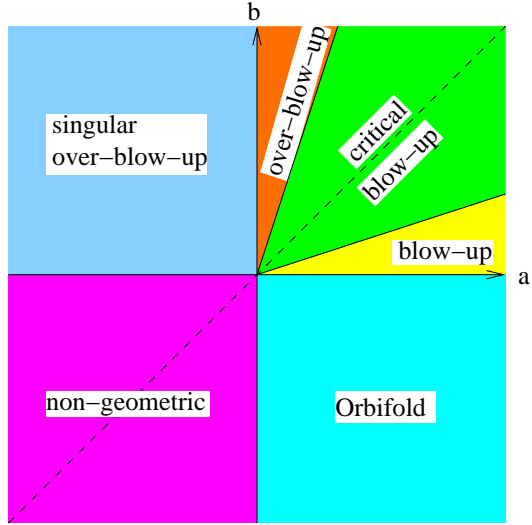} 
\]
\caption{This picture displays the different phases of the fully resolvable GLSM that possesses a duality symmetry between the torus radii $a_a$ and the blow--up cycle radii $b_i$.
\label{fg:DualPhases}}
\end{figure}

The fully resolvable GLSM model with the exceptional coordinates $x_i = x_{iii}$, i.e.\ the one on the non--factorized $E_6$ lattice, is very interesting because in this model there is a complete symmetry between the two--tori radii measured by $a_a$ and the sizes of the blow--up cycles described by $b_i = b_{iii}$. This symmetry becomes manifest when we write down the resulting D--terms 
\equ{
\sum_j |z_{aj}|^2 - 3\, |c_a|^2 = a_a\,, 
\qquad 
\sum_b |z_{bi}|^2 - 3\, |x_i|^2 = b_i\,, 
}
and the superpotential 
\equ{
W = \sum_{b,j} \cC_b \, \cZ_{bj}^3\, \cX_j\,. 
}
In this form it is obvious that the model is symmetric under the simultaneous interchange: 
\equ{ 
\cC_a ~\leftrightarrow~ \cX_i\,, 
\qquad 
\cZ_{ai} ~\leftrightarrow~ \cZ_{ia}\,, 
\qquad 
a_a ~\leftrightarrow~ b_i\,. 
}
In particular, this duality shows that in this model the orbifold and the singular over--blow--up are identical, and so are the blow--up and over--blow--up phases. 

This  has some interesting consequences for the investigation of the full moduli space: As we saw in the previous Subsection the determination of the target space geometry can be rather involved, and can have surprises, like a change in target space dimension. Because of this duality we immediately know the properties of the singular over--blow--up: It is the orbifold $T^6/\Intr_3$. This allows us to also give a full picture of the moduli space of this model under the simplifying assumption that all $a_a =a$ and $b_i = b$, see Figure~\ref{fg:DualPhases}. The duality between the torus radii and the blow--up cycle radii corresponds to the diagonal dashed symmetry axis in the picture of the moduli space.


\section{GLSM Resolutions of $\mathbf{T^6/\Intr_4}$ orbifolds}
\label{sc:ResT6Z4}

In this section we present GLSMs for the $T^6/\Intr_4$ orbifold and its resolution. Since this orbifold has more structure than the $\Intr_3$ case, we encounter a few novel features: In the minimal model we need to introduce exceptional coordinates and gaugings for the different types of fixed points. Secondly, even in its fully resolvable GLSM there are two fixed tori that are not distinguished by our procedure and therefore blown up/down simultaneously. Thirdly, using our GLSM techniques we are able to describe genuine non--factorizable $T^6/\Intr_4$ orbifolds. 

To illustrate these issues and our formalism in general, we first construct the minimal fully resolvable GLSM on the factorized lattice $D_2^2 \times A_1^2$. Next we turn to the maximal fully resolvable model and show that one pair of fixed points remains indistinguishable. We conclude this Section by explaining how we can obtain GLSMs associated with non--factorizable lattices $A_3 \times D_2 \times A_1$ and $A_3 \times A_3$.

\subsection[The minimal $T^6/\Intr_4$ model on $D_2^2 \times A_1^2$]{The minimal $\mathbf{T^6/\Intr_4}$ model on $\mathbf{D_2^2 \times A_1^2}$}

\begin{table}[tb]
\centering
\renewcommand{\arraystretch}{\arrystrch}
\begin{tabular}{|c||c|c|c|c|c||c|c|c|c||c|c|}
\hline
Superfield 
& $\mathcal{Z}_{1i}$ & $\mathcal{Z}_{13}$ & $\mathcal{Z}_{2j}$ & $\mathcal{Z}_{23}$ & $\mathcal{Z}_{3k}$ 	
& $\mathcal{C}_1$ & $\mathcal{C}_2$ & $\mathcal{C}_3$& $\mathcal{C}'_3$
& $\mathcal{X}_{1,111}$	& $\mathcal{X}_{2,ij}$	
\\
$U(1)$ charges 
& $z_{1i}$ & $z_{13}$ & $z_{2j}$ &$z_{23}$ & $z_{3k}$ 	
& $c_1$ & $c_2$ & $c_3$& $c'_3$
& $x_{1,111}$	& $x_{2,ij}$	
\\
\hline
\hline
${R_1}$ & $1$	& $2$ & $0$ & $0$	& $0$ 
& $-4$ & $0$ & $0$ & $0$
& $0$& $0$ 
\\
\hline
${R_2}$ & $0$ & $0$ & $1$ & $2$ & $0$ 
& $0$ & $-4$ & $0$ & $0$
& $0$ & $0$ 
\\
\hline
${R_3}$ & $0$	& $0$ & $0$ & $0$ & $1$ 
& $0$	& $0$ & $-2$	& $-2$
& $0$ & $0$ 
\\
\hline
\hline
${E_{1,111}}$ & $\delta_{i1}$	& $0$ & $\delta_{j1}$ & $0$ & $2\, \delta_{k1}$ 
& $0$ & $0$ & $0$ & $0$
& $-4$ & $0$ 
\\
\hline
${E_{2,\alpha\beta}}$ & $\delta_{\alpha i}$ & $0$ & $\delta_{\beta j}$ & $0$	& $0$ 
& $0$ & $0$ & $0$ & $0$
& $0$& $-2\,\delta_{\alpha i}\delta_{\beta j}$ 
\\
\hline
\end{tabular}
\renewcommand{\arraystretch}{1}
\caption{The superfield content and charge assignment for the minimal fully resolvable GLSM of $T^6/\Intr_4$. The indices $\alpha,\beta, i,j$ run from 1 to 2 and $k$ from 1 to 4.}
\label{tab:Z4Minimal}
\end{table}

The $\Intr_4$ orbifold action on the 3 complex six--torus coordinates $(u_1,u_2,u_3)$ is given by 
\begin{align}
 \theta:~ (u_1,u_2,u_3) \mapsto (i\, u_1, i\, u_2, -u_3 ) \,. 
\label{eq:Z4OrbifoldActionFlat}
\end{align}
The only factorized lattice compatible with this action is of the form $D_2^2 \times A_1^2$, i.e.\ the complex structure of the first two tori is fixed to $i$, whereas the complex structure of the third torus stays a modulus $\gt$. The $\theta$ sector has $16$ fixed points whose structure factorizes as $2 \times 2 \times 4$. The $\theta^2$ element acts like a $\Intr_2$ on the first two tori therefore naively we expect $16$ fixed two--tori. But the $\theta$ action folds four of them to $T^2 / \Intr_2$'s, whereas the remaining twelve get identified pairwise, i.e.\ they give rise to six fixed two--tori.

To understand how to obtain a global resolution model, we briefly review the local GLSM resolution of $\Cplx^3/\Intr_4$ with the local orbifold action $\gth(z_1,z_2,z_3) = (i\,z_1,i\,z_2,-z_3)$. Then, following Subsection \ref{sc:LocalRes}, to resolve the singularity we have to introduce two exceptional coordinates $x_1$ and $x_2$, together with two appropriate gaugings. The charge Table of the corresponding $(2,2)$ superfields reads:
\begin{center}
 \begin{tabular}{c|c|c|c|c|c}
$U(1)$ charges & $\mathcal{Z}_1$ &    $\mathcal{Z}_2$ &    $\mathcal{Z}_3$ &    $\mathcal{X}_1$ &    $\mathcal{X}_2$ \\
\hline
${E_1}$ & 1 & 1 & 2 & $-4$ & 0 \\  
\hline
${E_2}$ & 1 & 1 & 0 & 0 & $-2$ \\  
 \end{tabular}
\end{center}
When $x_1 \neq 0$, the $U(1)_{E_1}$ reduces to the $\mathbb{Z}_4$ orbifold action on the coordinates $z_a$. However, by only introducing $x_1$ a $\Intr_2$ sub--singularity remains; it gets resolved by $x_2$. The discrete action induced from $U(1)_{E_2}$ when $x_2 \neq 0$ is a $\Intr_2$ subgroup of $\Intr_4$.

To construct the global models we first have to choose three appropriate elliptic curves. Since the orbifold $\Intr_4$ acts to fourth order in the first and second two--torus, and to second order in the third torus, we write the factorized six torus as $T^6 = T^2(\Intr_4) \times T^2(\Intr_4) \times T^2(\Intr_2)$. Each of these two--tori is described as hypersurfaces in appropriately chosen weighted projective spaces as summarized in Table \ref{tab:SummaryTori} of Section \ref{sc:twotori}. Hence, we need the homogeneous coordinates $z_{1i}$, $z_{2j}$ with labels $i,j=1,2,3$ and $z_{3k}$ with  $k=1,\ldots,4$. To enforce the hyperplane constraints in the orbifold and blow--up regimes, we need four superfields $\mathcal{C}_1$, $\mathcal{C}_2$, $\mathcal{C}_3$ and $\mathcal{C}_3'$ which lead to the superpotential 
\equa{
W_\text{torus} = & \cC_1 ( \cZ_{11}^4 + \cZ_{12}^4 + \cZ_{13}^2) + 
 \cC_2 ( \cZ_{21}^4 + \cZ_{22}^4 + \cZ_{23}^2) 
 \non \\[1ex] 
 &+ \cC_3 ( \gk\, \cZ_{31}^2 + \cZ_{32}^2 + \cZ_{33}^2) + 
   \cC_3' ( \cZ_{31}^2 + \cZ_{32}^2 + \cZ_{34}^2)\,, 
   \label{SPtorusZ4}
}
for the corresponding superfields. Here $\kappa$ encodes the complex structure modulus $\gt$ in the elliptic curve description of the third torus. 

In this formulation the orbifold action \eqref{eq:Z4OrbifoldActionFlat} acts on the homogeneous coordinates as 
\begin{align}
 \theta:~ (z_{11},z_{21},z_{31}) \mapsto (i\, z_{11}, i\, z_{21}, -z_{31} ) \,. 
\label{eq:Z4OrbifoldAction}
\end{align}
This orbifold action can be transferred to the other coordinates by using the discrete $U(1)_{R_a}$ rotations, 
\begin{align}
 \theta_{R_a} : (z_{a,1},z_{a,2},z_{a,3}) \mapsto (i\,z_{a,1},i\,z_{a,2},-z_{a,3}) \,, \qquad a = 1,2 \,,
\end{align}
and the involutions $\ga_1, \ga_2$ and $\ga_3, \ga_3'$ of these two--tori defined in Subsections \ref{sc:Z4torus} and \ref{sc:Z2torus}, respectively. 

Next we investigate how many exceptional coordinates and gaugings we need to construct the minimal fully resolvable model, which by definition should be able to resolve all types of singularities. For the local resolution of a $\Cplx^3/\Intr_4$ singularity we need two exceptional coordinates $x_{1,111}$ and $x_{2,11}$ with gaugings $U(1)_{E_{1,111}}$ and $U(1)_{E_{2,11}}$, respectively, to resolve the $\Intr_4$ fixed points. Their positions are determined by the equations $z_{11}=z_{21}=z_{31}=0$. The $F_{c_a}$--terms obtained from the superpotential \eqref{SPtorusZ4}, 
\begin{align}
\begin{split}
 z_{1,1}^4 + z_{1,2}^4 + z_{1,3}^2 &= 0 \,, 
 \\[1ex] 
 z_{2,1}^4 + z_{2,2}^4 + z_{2,3}^2 &= 0 \,, 
 \\[1ex] 
 \gk\, z_{3,1}^2 + z_{3,2}^2 + z_{3,3}^2 &= 0 \,, 
 \\[1ex] 
 z_{3,1}^2 + z_{3,2}^2 + z_{3,4}^2 &= 0 \,, 
\end{split}
\end{align}
have $2\cdot2\cdot2\cdot2 = 16$ roots, hence this corresponds to the 16 $\Intr_4$ fixed points. The additional six fixed two--tori are located at $z_{11}=z_{22}=0$, $z_{21}=z_{21}=0$ and $z_{12}=z_{22}=0$: E.g.\ $\theta^2 \theta_{R_1}^2$ has fixed tori at $z_{1,2}=z_{2,1}=0$. For them the first two F--terms factorize in two times two roots. However $\theta$ identifies the solutions pairwise, hence we find two true fixed two--tori. Similarly, there are two fixed tori of $\theta \theta_{R_2}^2$ at $z_{1,1}=z_{2,2}=0$ and two $\theta^2\theta_{R_1}^2\theta_{R_2}^2$ fixed tori at $z_{1,2}=z_{2,2}=0$. Hence, to resolve  all six fixed two--tori we need three additional exceptional coordinates $x_{2,ij}$ and $U(1)_{E_{2,ij}}$ with $(i,j)=(2,1), (1,2), (2,2)$. All these gaugings together define the minimal fully resolvable model; the superfield content and the charge assignment are displayed in Table \ref{tab:Z4Minimal}.

To show that this is indeed the minimal fully resolvable GLSM, we have to confirm that indeed all the fixed points get resolved in the blow--up regime. In this regime we have $b_r>0$, yet sufficiently smaller than all the torus radii $a_a$ to avoid running into the critical or over--blow--up regimes. From the charges in Table \ref{tab:Z4Minimal} we find the following  D--terms
\begin{align}
\begin{split}
 |z_{11}|^2 + |z_{12}|^2 + 2 |z_{13}|^2 - 4\, |c_1|^2 &= a_1 \,, 
 \\[1ex] 
 |z_{21}|^2 + |z_{22}|^2 + 2 |z_{23}|^2 - 4\, |c_2|^2 &= a_2 \,, 
 \\[1ex] 
 |z_{31}|^2 + |z_{32}|^2 + |z_{3,3}|^2  + |z_{34}|^2 - 2\, |c_3|^2 - 2\, |c'_3|^2 &= a_3 \,, 
 \\[1ex] 
|z_{11}|^2 + |z_{21}|^2 + 2\, |z_{31}|^2 - 4\, |x_{1,111}|^2 &= b_{1,111}\,, 
\\[1ex] 
|z_{1i}|^2 + |z_{2j}|^2  - 2\, |x_{2,ij}|^2 &= b_{2,ij} \,,\qquad i,j\in\{1,2\}\,.
\end{split}
\label{eq:Z4MinimalDterms}
\end{align}
These equations clearly forbid setting $z_{11}=z_{22}=0$, $z_{21}=z_{21}=0$ or $z_{12}=z_{22}=0$, because $b_{1,111}, b_{2,ij}$ are all positive in the blow--up phase. Consequently, all the orbifold fixed points have been removed and replaced by the corresponding blow--up cycles $E_{1,111} := \{x_{1,111}=0\}$ and $E_{2,ij} := \{x_{2,ij}=0\}$. 

To show that these blow--up cycles are glued in at the appropriate positions, we look at the F--terms in blow--up. In this regime the $b_{1,111}, b_{2,ij}$ are sufficiently smaller than $a_a$, such that the F--terms associated with $z_{ai}$ force $c_a = 0$. Hence the only remaining F--terms are:
\begin{align}
\begin{split}
 z_{11}^4\, x_{1,111} x_{2,11} x_{2,12} + z_{12}^4\, x_{2,21} x_{2,22} + z_{13}^2 &= 0 \,, 
 \\[1ex] 
 z_{21}^4\, x_{1,111} x_{2,11} x_{2,21}+ z_{22}^4\,x_{2,12} x_{2,22} + z_{23}^2 &= 0 \,, 
 \\[1ex] 
\gk\, z_{31}^2\, x_{1,111} + z_{32}^2 + z_{33}^2 = 0 \,, \quad 
 z_{31}^2\, x_{1,111} +  z_{32}^2 + z_{34}^2 &= 0 \,. 
\end{split}
\label{Z4MinimalFterms}
\end{align}
Thus in blow--up one recovers the resolved $\Intr_4$ singularities by setting $x_{1,111} = 0$. The factorization of the F--term \eqref{Z4MinimalFterms} then shows that there are the same 16 components as in the orbifold case discussed above. Similarly, by setting $x_{2,ij} = 0$ with $(i,j) = (2,1), (2,1), (2,2)$, we find the roots associated with the $\Intr_2$ fixed two--tori. This confirms that this model indeed resolves all singularities of the $T^6/\Intr_4$ orbifold.

\subsection[The Maximal $T^6/\Intr_4$ model on $D_2^2\times A_1^2$]{The Maximal $\mathbf{T^6/\Intr_4}$ model on $\mathbf{D_2^2\times A_1^2}$}

\begin{table}
\centering
\renewcommand{\arraystretch}{1.1}
\begin{tabular}{|c||c|c|c|c|c||c|c|c|c||c|c|}
\hline
Superfield 
& $\mathcal{Z}_{1i}$ &$\mathcal{Z}_{13}$ & $\mathcal{Z}_{2j}$ & $\mathcal{Z}_{23}$ & $\mathcal{Z}_{3k}$ 	
& $\mathcal{C}_1$ & $\mathcal{C}_2$ & $\mathcal{C}_3$ & $\mathcal{C'}_3$
& $\mathcal{X}_{1,ijk}$ & $\mathcal{X}_{2,ij}$ 
\\
$U(1)$ charges 
& $z_{1i}$ & $z_{13}$ & $z_{2j}$ & $z_{23}$ & $z_{3k}$ 	
& $c_1$ & $c_2$ & $c_3$& $c'_3$
& $x_{1,ijk}$ & $x_{2,ij}$	
\\
\hline
\hline
${R_1}$ & $1$	& $2$ & $0$ & $0$ & $0$ 
& $-4$ & $0$ & $0$ & $0$
& $0$ & $0$ 
\\
\hline
${R_2}$ & $0$	& $0$ & $1$ & $2$ & $0$ 
& $0$ & $-4$ & $0$ & $0$
& $0$ & $0$ 
\\
\hline
${R_3}$ & $0$	& $0$ & $0$ & $0$ & $1$ 
& $0$ & $0$ & $-2$ & $-2$
& $0$ & $0$ 
\\
\hline
\hline
${E_{1,\alpha\beta\gamma}}$ & $\delta_{i\alpha}$	& $0$ & $\delta_{j\beta}$ & $0$ & $2\, \delta_{k\gamma}$	
& $0$ & $0$ & $0$ & $0$
& $-4\,\delta_{i\alpha}\delta_{j\beta}\delta_{k\gamma}$ & $0$ 
\\
\hline
${E_{2,\alpha\beta}}$ & $\delta_{i\alpha }$ & $\delta_{3\alpha}$ & $\delta_{j\beta }$ & $\delta_{3\beta}$ & $0$ 
& $0$ & $0$ & $0$ & $0$
& $0$ & $-2\,\delta_{\alpha i}\delta_{\beta j}$	 
\\
\hline
\end{tabular}
\renewcommand{\arraystretch}{1}
\caption{The superfield content and charge assignment for the maximal GLSM model of $T^6/\Intr_4$. The indices $k,\gamma$ run from 1 to 4. The indices $i,j,\alpha,\beta$ run from 1 to 2 for $x_{1,ijk}$ and $U(1)_{E_{1,ijk}}$, and from 1 to 3 for $x_{2,ij}$ and $U(1)_{E_{2,ij}}$. }
\label{tab:Z4Maximal}
\end{table}

In the maximal model we have 16 separate exceptional coordinates $x_{1,ijk}$ with gaugings $U(1)_{E_{1,ijk}}$, where $i,j=1,2$ and $k=1,\ldots,4$ in the $\Intr_4$ sector, see Table \ref{tab:Z4Maximal}. In addition, we have 9 coordinates $x_{2,ij}$ and gaugings $U(1)_{E_{2,jk}}$ with $i,j=1,2,3$ in the $\Intr_2$ sector. The resulting D--terms are given by 
\begin{align}
\begin{split}
 |z_{11}|^2 + |z_{12}|^2 + 2 |z_{13}|^2 - 4\, |c_1|^2 &= a_1 \,, \\
 |z_{21}|^2 + |z_{22}|^2 + 2 |z_{23}|^2 - 4\, |c_2|^2 &= a_2 \,, \\
 |z_{31}|^2 + |z_{32}|^2 + |z_{33}|^2  + |z_{34}|^2 - 2\, |c_3|^2 - 2\,|c'_3|^2 &= a_3 \,, \\
|z_{1i}|^2 + |z_{2j}|^2 + 2\, |z_{3k}|^2 - 4\, |x_{1,ijk}|^2 &= b_{1,ijk} \,, \qquad i,j=1,2\,, \quad k=1,\ldots,4\,, \\
|z_{1i}|^2 + |z_{2j}|^2 - 2\, |x_{2,ij}|^2 &= b_{2,ij} \,, \qquad i,j=1,2,3\,.
\end{split}
\label{eq:Z4MaximalDterms}
\end{align}
In the orbifold and blow--up phases the F--terms reduce to 
\begin{align}
\begin{split}
 z_{11}^4  \prod_{j} x^2_{2,1j} \prod_{j,k} x_{1,1jk}  + z_{12}^4  \prod_{j} x^2_{2,2j} \prod_{j,k} x_{1,2jk}  +  z_{1,3}^2 \prod_{j} x_{2,3j} &= 0 \,, \\
 z_{21}^4  \prod_{i} x^2_{2,i1} \prod_{i,k} x_{1,i1k}  + z_{22}^4  \prod_{i} x^2_{2,i2} \prod_{i,k} x_{1,i2k}  +  z_{23}^2 \prod_{i} x_{2,i3} &= 0 \,, \\
\gk\, z_{31}^2 \prod_{i,j} x_{1,ij1} + z_{32}^2\prod_{i,j} x_{1,ij2} + z_{33}^2\prod_{i,j} x_{1,ij3} &= 0 \,, \\
 z_{31}^2 \prod_{i,j} x_{1,ij1} +  z_{32}^2\prod_{i,j} x_{1,ij2} + z_{34}^2\prod_{i,j} x_{1,ij4} &= 0 \,.
\end{split}
\label{eq:Z4MaximalFterms}
\end{align}

As expected in the maximal model each $\Intr_4$ fixed point has its own exceptional coordinate and gauging. Indeed, in the blow down regime we find $\Intr_4$ fixed points at $z_{1,i}=z_{2,j}=z_{3,k}=0$, $i,j \neq 3$; and  in blow--up their exceptional divisor counterparts $x_{1,ijk}=0$.   Inserting either of those conditions into the F--terms \eqref{eq:Z4MaximalFterms}, we find that in order to factorize them we would have to introduce enough square roots to identify all distinct solutions via the sign ambiguities. Thus each index triplet $i,j,k$ describes just one single $\Intr_4$ fixed point. 

However, some of the $\Intr_2$ fixed tori at $z_{1i} = z_{2j} = 0$ cannot be uniquely identified. As long as $(i,j) \neq (3,3)$, for each $i,j$ we obtain a single fixed torus in a similar manner as above for the $\Intr_4$ fixed points. Contrary, in the case $z_{1,3}=z_{2,3}=0$ we find two distinct solutions to the relevant F--terms. To see this, we first fix all the values of all the non--zero $x$ in the orbifold regime and absorb them in the $z$. It is then easy to see that each F--term has four solutions, $z_{a1} = i^{n+1/2} z_{a2}$, $n=1,2,3,4$, i.e.\ 16 solutions altogether. The discrete actions in each of the two tori $(z_{a1},z_{a2}) \mapsto (i\,z_{a1},-i\,z_{a2})$ identify the solutions in four quadruplets. The orbifold $\Intr_4$ further identifies pairs of them, so that in the end we still have two independent solutions. This shows that two $\Intr_2$ fixed two--tori remain indistinguishable.

\subsection[Non--Factorizable $T^6 / \Intr_4$ models]{Non--Factorizable $\mathbf{T^6 / \Intr_4}$ models}
\label{subsec:Z4NonFactorizableCases}

In the previous Subsections we have discussed the minimal and maximal fully resolvable GLSMs of the $T^6/\Intr_4$ orbifold on the factorized lattice $D_2^2\times A_1^2$. As for the $T^6/\Intr_3$ it is possible to construct a variety of models which have more gaugings than the minimal, yet less than the maximal models. Since the classification of all $\Intr_4$ models is straightforward but a little more complicated than for the $\Intr_3$ case, we refrain from presenting it here. Instead we demonstrate that in two special cases a novel possibility arises which was not present for the $\Intr_3$ models: We can construct $\Intr_4$ models on genuine non--factorizable lattices. 

For both orbifolds the construction of non--factorized lattices is the same: We choose exceptional gaugings that act within more than one two--torus simultaneously. As we discussed in Subsection \ref{sc:NonFact} in the $\Intr_3$ case it is always possible to find an off--diagonal \Kh\ deformation to bring the non--factorized lattice back to a the factorized form. In the $\Intr_4$ case there are some non--factorized lattices that cannot be deformed to a factorized version, because the required off--diagonal \Kh\ deformations simply do not exist. Hence, these are genuinely non--factorizable. As a consequence, these models have a different amount of fixed tori than the factorizable case, and thus correspond to truly different topologies. 

Below we present two GLSMs which yield $\Intr_4$ orbifolds based on the lattices $A_3\times D_2 \times A_1$ and $A_3 \times A_3$. 
In particular, we show that as the lattice becomes more non--factorizable the number of fixed tori decreases, whereas the amount of fixed $T^2/\Intr_2$'s with $\Intr_4$ fixed points stays the same.  Together with the factorizable case on $D_2^2 \times A_1^2$ we thus can realize all $\Intr_4$ orbifolds classified in Theorem 1 of \cite{Erler:1992ki} (see also Table 2.1 of \cite{Lust:2006zh}).

\subsubsection*{Non--factorizable GLSM on $\mathbf{A_3 \times D_2 \times A_1}$}

In the first model we just add one extra coordinate to the minimal model:
We choose the coordinates $x_{1,111}$ and $x_{1,122}$ with their $U(1)$ gaugings. Moreover, in order to resolve the $\Intr_2$ singularities we also need all $x_{2,ij}$ with $i,j=1,2$ plus gaugings, even though they do not introduce further discrete symmetries. In this setup we focus on the action of $\theta_{1,111}\theta_{1,122}^3$:
\begin{align}
\tilde\theta_1 =  \theta_{1,111}\theta_{1,122}^3 :~ (z_{21},z_{22},z_{31},z_{32}) \longmapsto (i\, z_{21},-i\, z_{22},-z_{31},-z_{32}) \,.
\label{Z4NonFactDiscreteAction1}
\end{align}
This is a freely acting $\Intr_2$ element on the homogeneous torus coordinates, thus a simultaneous shift in the second and third torus. In appendix \ref{appendixZ4NonfactLattices} we show that modding it out of the torus results in the root lattice of the Lie algebra $A_3 \times D_2 \times A_1$. Contrary to the $\Intr_3$ orbifold, the off--diagonal K\"ahler modulus which would mix the second and third torus is not orbifold invariant and thus does not exist. Therefore, in this case our construction gives a truly non--factorizable lattice. 

The topology of the resulting non--factorizable orbifold is truly different from its factorized counterpart. To show this, we inspect the fixed point and fixed torus structure. There are eight $\theta_{1,111}$ fixed points at $z_{1,1}=z_{2,1}=z_{3,1}=0$ and eight $\theta_{1,122}$ fixed points at $z_{1,1}=z_{2,2}=z_{3,2}=0$. The fixed tori in the $\Intr_2$ sector are described by the same zero loci as in the minimal model. The factorization of the F--terms shows that there are two fixed tori at $z_{1,1}=z_{2,1}=0$ and two at $z_{1,1}=z_{2,2}=0$, but just one at $z_{1,2}=z_{2,1}=0$ and $z_{1,2}=z_{2,2}=0$ each. One can easily convince oneself that there are no more fixed points when $\theta^2$ actions are combined with $U(1)_R$ rotations. This is exactly the right amount of fixed tori expected on this lattice, see e.g.\ \cite{Lust:2006zh}. Note that due to the presence of the coordinates $x_{2,ij}$, all fixed tori can be blown up, thus this model is fully resolvable.

\subsubsection*{Non--factorizable GLSM on  $\mathbf{A_3 \times A_3}$}

In the case discussed above we went from the minimal model on the factorized lattice $D_2^2\times A_1^2$ to the non--factorizable case $A_3 \times D_2 \times A_1$ by modding out one discrete $\Intr_2$ element. We can repeat this operation again to obtain a $\Intr_4$ orbifold on the $A_3 \times A_3$ lattice. The GLSM realization of this, as a fully resolvable model, requires the fields $x_{1,111}$, $x_{1,122}$, $x_{1,213}$ and $x_{1,224}$. Again, resolving the $\Intr_2$ singularities requires the coordinates $x_{2,ij}$ with $i,j=1,2$, which are the same ones as for the minimal model.

Let us look at the discrete symmetries that get induced on the $z_{ak}$ in this setup. They form a $\Intr_4 \times \Intr_2 \times \Intr_2$ group, generated by $\theta_{1,111}$, $\tilde\theta_1$ as defined in \eqref{Z4NonFactDiscreteAction1} and 
\begin{align}
\tilde\theta_2 :=  \theta_{1,111}\theta_{1,213}^3 : (z_{11},z_{12},z_{31},z_{33}) \longmapsto (i\,z_{11},-i\, z_{12},-z_{31},-z_{33}) \,.
\label{Z4NonFactDiscreteAction2}
\end{align}
The $\theta_{1,224}$ induced by a VEV of $x_{1,224}$ is generated as $\theta_{1,111}\tilde\theta_1\tilde\theta_2$ together with some $U(1)_R$ rotations. Therefore, including $x_{1,224}$ does not enhance the discrete symmetry group. Rather the $U(1)_{E_{1,224}}$ gauging is necessary to be able to resolve the fixed points which would be present at $z_{1,2}=z_{2,2}=z_{3,4}=0$ otherwise. 

The topology of this non--factorizable orbifold is again different from the previous case. Inspecting the F--terms we find a multiplicity of four for each of these fixed points, i.e.\ again there are 16 fixed points altogether. Furthermore, we find fixed tori of the various $\theta^2_{1,ijk}$ actions, with $(i,j,k)=(1,1,1), (1,2,2), (2,1,3)$ and $(2,2,4)$, at $z_{1,i}=z_{2,j}=0$, respectively. Each of them has multiplicity one, i.e.\ there are four fixed tori, each of which is orbifolded to $T^2/\Intr_2$ by the residual $\Intr_2$. One can check that there are indeed no further fixed tori than the ones identified here. This again agrees with the results on this lattice by \cite{Lust:2006zh}.


\begin{table}[tb]
\centering
\footnotesize
\renewcommand{\arraystretch}{\arrystrch}
\begin{tabular}{|c||c|c|c||c|c|c|c||c|c|c|c|}
\hline
Superfield & $\mathcal{Z}_{1i}$ & $\mathcal{Z}_{2j}$ & $\mathcal{Z}_{3k}$ 
& $\mathcal{C}_1$  & $\mathcal{C}_2$ & $\mathcal{C}_3$ & $\mathcal{C}'_3$
&  $\mathcal{X}_{1,1jk}$ &  $\mathcal{X}_{2,i'j}$ &  $\mathcal{X}_{3,i''k}$ &  $\mathcal{X}_{4,i'j}$ 
\\
$U(1)$ charges & $z_{1i}$ & $z_{2j}$ & $z_{3k}$ 
& $c_1$  & $c_2$ & $c_3$ & $c'_3$
&  $x_{1,1jk}$ &  $x_{2,i'j}$ &  $x_{3,i''k}$ &  $x_{4,i'j}$ 
\\
\hline
\hline
$R_1$ & $i$ & 0 & 0 
& $-6$ & 0 & 0 & 0
& 0 & 0 & 0 & 0 
\\
\hline
$R_2$ & 0 & $1$ & 0 
& 0 & $-3$ & 0 & 0 
& 0 & 0 & 0 & 0 
\\
\hline
$R_3$ & 0 & 0 & $1$ 
& 0 & 0 & $-2$ & $-2$
& 0 & 0 & 0 & 0 
\\
\hline\hline
$E_{1,1\beta\gamma}$ & $\delta_{1 i}$ & $2\,\delta_{\beta j}$ & $3\,\delta_{\gamma k}$ 
& 0 & 0 & 0 & 0
& $-6\,\delta_{\beta j}\delta_{\gamma k}$ & 0 & 0 & 0 
\\
\hline
$E_{2,\alpha'\beta}$ & $\delta_{\alpha' i}$ & $2\,\delta_{\beta,j}$ & 0 
& 0 & 0 & 0 & 0  
& 0 & $-3\,\delta_{\alpha' i'}\delta_{\beta j}$ & 0 & 0 
\\
\hline
$E_{3,\alpha''\gamma}$ & $\delta_{\alpha'' i}$ & 0 & $\delta_{\gamma k}$ 
& 0 & 0 & 0 & 0  
& 0 & 0 & $-2\delta_{\alpha'' i''}\delta_{\gamma k}$ & 0 
\\
\hline
$E_{4,\alpha'\beta}$ & $2\,\delta_{\alpha' i}$ & $\delta_{\beta j}$ & 0 
& 0 & 0 & 0 & 0 
& 0 & 0 & 0 & $-3\,\delta_{\alpha' i'}\delta_{\beta j}$ 
\\
\hline
\end{tabular}
\renewcommand{\arraystretch}{1}
\normalsize
\caption{$U(1)$ charge assignment for the maximal GLSM resolution of the $T^6/\Intr_\text{6--II}$ orbifold. The indices have the following ranges: $i$, $j$ run from 1 to 3 and $k$ runs from 1 to 4. $\ga',i'$ takes values 1 or 2 and $\ga'', i''$ takes values 1 or 3. }
\label{tab:Z6IIMaximalCharges}
\end{table}

\section{GLSM Resolutions of $\mathbf{T^6/\Intr_\text{6--II}}$ orbifolds}
\label{sc:ResT6Z6II}

The orbifold $T^6/\Intr_\text{6--II}$ has received a lot of attention recently, since from it a large collection of MSSM--like models, the so--called mini--landscape, can be constructed, see e.g.\ \cite{Buchmuller:2006ik,Lebedev:2006tr}. In fact, there are also $T^6/\Intr_\text{6--II}$ orbifolds based on non--factorizable lattices, see e.g.\ \cite{Katsuki:1990bf,Erler:1992ki}. Unlike the $\Intr_4$ orbifolds discussed in Section \ref{sc:ResT6Z4} we cannot obtain the non--factorized orbifolds from GLSMs which describe the factorized one:  The truly non--factorizable descriptions cannot occur as $\Intr_6=\Intr_2\times\Intr_3$: With 2 and 3 being relative prime, the $\Intr_6$ will always factorize back into the $\Intr_2$ and the $\Intr_3$. This means that the description of the $\Intr_{6-\text{II}}$ orbifold on non--factorizable lattices is beyond the GLSM techniques introduced in this paper. Fortunately, the mini--landscape models are all based on the factorized lattice, $G_2 \times A_2\times A_1^2$, and can therefore be studied using our methods.  

The study of these mini--landscape models might benefit from the partially resolvable GLSMs discussed in Subsection \ref{sc:PartialGLSMs}:   In these models generically at least some of the fixed points have to be blown up in order to cancel the anomalous $U(1)$ on the orbifold and to decouple unwanted exotics. However, all of these models necessarily break the hypercharge (or the $SU(2)$ subgroup) in full blow--up \cite{Nibbelink:2009sp}, because some of the orbifold fixed points only support states charged under the SM group. Hence these models can neither be considered purely from the orbifold perspective nor purely from the CY perspective. For this reason, it is desirable to have a framework where some fixed points are blown up whereas the SM breaking fixed points stay singular. This can be achieved in at least two ways from the GLSM perspective. One way is to take the maximal model, which will be discussed in section \ref{subsec:Z6MaximalModel}, but leaves the K\"ahler parameters $b_{r,ijk}$ of the relevant fixed points negative. An easier approach is probably to study partially resolvable GLSMs in which the problematic fixed points simply cannot be resolved due to the lack of K\"ahler parameters in this description. This could be considered the more natural approach, as in this case the presence of the singularities is not due to manifestly artificial fine--tuning of resolution cycle volumes. 

This Section is structured as follows: We first construct the maximal GLSM resolution of the $T^6/\Intr_\text{6--II}$ orbifold. This description illustrates an important advantage of the GLSM description over the conventional gluing procedure, namely that  the identification of the fixed points in the first torus is built into the GLSM via $U(1)_{R}$ actions. The minimal GLSM resolution, constructed next, can easily be extended to partially resolvable GLSMs, relevant for some benchmark mini--landscape models, in which the SM breaking singularities are not blown up.

\subsection{The maximal fully resolvable model}
\label{subsec:Z6MaximalModel}

To obtain $T^6/\Intr_\text{6--II}$ resolution GLSMs we choose the appropriate torus descriptions for the two--tori out of which the $\Intr_\text{6--II}$ orbifold is built. On the  torus coordinates $(u_1,u_2,u_3)$ the orbifold action reads
$\gth :~ (u_1,u_2,u_3)  \mapsto (- \gz^2\, u_1, \gz\, u_2, - u_3)$, where as throughout this paper $\gz = e^{2\pi i/3}$. Hence, since the orbifold acts with $\Intr_6=\Intr_2\times\Intr_3$ in the first torus, with $\Intr_3$ in the second torus, and with $\Intr_2$ in the third torus, we choose the elliptic curve description $T^6=T^2(\Intr_6)\times T^2(\Intr_3)\times T^2(\Intr_2)$. From the discussion in Section \ref{sc:twotori} we know that the orbifold action is translated to the action
\equ{
\gth :~ (z_{11}, z_{21}, z_{31}) \mapsto (- \gz^2\, z_{11}, \gz\, z_{21}, - z_{31})\,,
\label{eq:Z6IIOrbifoldActionInduced}
}
on the elliptic curve coordinates $z_{ai}$. This action can be transferred to the other coordinates of the elliptic curves using the $U(1)_{R_a}$ transformations. 

From this we can infer the fixed point structure. As can be read off from Table \ref{tab:SummaryTori} the elliptic curve $T^2(\Intr_6)$ is described by three homogeneous coordinates $z_{11}, z_{12}, z_{13}$ which satisfy the holomorphic condition 
\equ{
z_{11}^6 + z_{12}^3 + z_{13}^2 = 0\,.
}
The root $z_{11} = z_{12} = z_{13} = 0$ corresponds to the $\Intr_6$ fixed point. The two roots of $z_{12} =0$:  $z_{13} + i \, z_{11}^3=0$ and $z_{13} - i\, z_{11}^3=0$ with $z_{11} \neq 0$, which define the $\Intr_3$ tori, get identified under the orbifold action $\Intr_6$, hence this gives a single $\Intr_3$ fixed point. Finally, the three roots of $z_{13}=0$: $z_{12} + \gz^n\, z_{11}^2=0$, $z_{11}\neq 0$, $n=0,1,2$, get identified, giving a single $\Intr_2$ fixed point. Hence, the first two--torus possesses three types of fixed points: one $\Intr_6$ fixed point, one $\Intr_3$ fixed point and one $\Intr_2$ fixed point. In our convention the $\Intr_6$, $\Intr_3$ and $\Intr_2$ fixed points are labeled by $i=1,2$ and $3$, respectively. In addition to this, the second and third tori have three $\Intr_3$ fixed points labeled by $j=1,2,3$, and four $\Intr_2$ fixed points labeled by $k=1,\ldots, 4$, respectively.  

Let us compare the identification of fixed points in our GLSM description with the blow--up procedure outlined in \cite{Lust:2006zh}. There the fixed point structure is analyzed on the covering $T^6$ torus. Even though the fixed point structures of the second and third torus are identical to ours, on the $T^2(\Intr_6)$ torus there seem to be differences: In addition to the single $\Intr_6$ fixed point, their description gives three $\Intr_2$ fixed points and two $\Intr_3$ fixed points in the $\gth^3$ and $\gth^2$ sectors, respectively. We only encounter one pure $\Intr_2$ fixed point and one pure $\Intr_3$ fixed point: Their three $\Intr_2$ fixed points correspond to the three roots of $z_{12}=0$ that got identified by the residual $\Intr_3$ orbifold action induced by $\theta$. Similarly, their two $\Intr_3$ fixed points correspond to two roots of $z_{13}=0$ which again got identified. Hence in the GLSM approach the fixed point identifications are automatically built in. 

In the maximal fully resolvable GLSM all these singularities are resolved separately, leading to the charge assignment given in Table~\ref{tab:Z6IIMaximalCharges}. This in turn gives rise to a total of 35 D--term equations 
\begin{subequations}
\begin{align}
&\makebox[250pt][l]{$\sum\limits_{i}i |z_{1i}|^2-6\,|c_1|^2=a_1\,,\quad \sum\limits_{j}|z_{2j}|^2-3\,|c_2|^2=a_2\,,\quad \sum\limits_{k}|z_{3k}|^2-2\,|c_1|^2-2\,|c_2|^2=a_3\,,$}\label{eq:Z6DTerms1} 
\\[1ex] 
&|z_{11}|^2+2\,|z_{2j}|^2+3\,|z_{3k}|^2-6\,|x_{1,1jk}|^2=b_{1,1jk}\,, &{}~ j\,&=1,2,3\,,&{}~k&=1,2,3,4\,,\label{eq:Z6DTerms2} \\[1ex]
&|z_{1i'}|^2+2\,|z_{2j}|^2-3\,|x_{2,i'j}|^2=b_{2,i'j}\,,&{} i'&=1,2\,,&{}~ j&=1,2,3\,,\label{eq:Z6DTerms3} \\[1ex]
&|z_{1i''}|^2+|z_{3k}|^2-2\,|x_{3,i''k}|^2=b_{3,i''k} \,,&{} i''\!&=1,3\,,&{}~ k&=1,2,3,4\,,\label{eq:Z6DTerms4} \\[1ex]
&2\,|z_{1i'}|^2+|z_{2j}|^2-3\,|x_{4,i'j}|^2=b_{4,i'j}\,, &{}i'&=1,2\,,&{}~ j&=1,2,3\,. \label{eq:Z6DTerms5} 
\end{align}
\label{eq:Z6DTerms}
\end{subequations}
Moreover, given the charge assignment of Table~\ref{tab:Z6IIMaximalCharges}, we can write down the following superpotential 
\begin{align}
 W=&\!~~~~\mathcal{C}_1\left[\mathcal{Z}_{11}^6\prod\limits_{\beta,\gamma}\mathcal{X}_{1,1\beta \gamma} \prod\limits_{\beta}(\mathcal{X}_{2,1\beta}^2 \mathcal{X}_{4,1\beta}^4) \prod\limits_{\gamma}\mathcal{X}_{3,1\gamma}^3 + \mathcal{Z}_{12}^3\prod\limits_{\beta}(\mathcal{X}_{2,2\beta} \mathcal{X}_{4,2\beta}^2) + \mathcal{Z}_{13}^2\prod\limits_{\gamma}\mathcal{X}_{3,3\gamma} \right] 
 \non \\
 & +
\mathcal{C}_2\left[\sum\limits_{j=1}^3\left(\mathcal{Z}_{2j}^3\prod\limits_\gamma \mathcal{X}_{1,1j\gamma}\prod\limits_{\alpha'}(\mathcal{X}_{2,\alpha'j}^2\mathcal{X}_{4,\alpha'j})\right)\right]
\\
 & +
\mathcal{C}_3\left[\kappa \mathcal{Z}_{31}^2\prod\limits_{\beta} 
\mathcal{X}_{1,1\beta1}\prod\limits_{\alpha''}\mathcal{X}_{3,\alpha''1}+\mathcal{Z}_{32}^2\prod\limits_{\beta} \mathcal{X}_{1,1\beta2}\prod\limits_{\alpha''}\mathcal{X}_{3,\alpha''2}+\mathcal{Z}_{33}^2\prod\limits_{\beta} \mathcal{X}_{1,1\beta3}\prod\limits_{\alpha''}\mathcal{X}_{3,\alpha''3}\right]
\non \\
 & +
\mathcal{C}'_3\left[\phantom{a} \mathcal{Z}_{31}^2\prod\limits_{\beta} \mathcal{X}_{1,1\beta1}\prod\limits_{\alpha''}\mathcal{X}_{3,\alpha''1}+\mathcal{Z}_{32}^2\prod\limits_{\beta} \mathcal{X}_{1,1\beta2}\prod\limits_{\alpha''}\mathcal{X}_{3,\alpha''2}+\mathcal{Z}_{34}^2\prod\limits_{\beta} \mathcal{X}_{1,1\beta4}\prod\limits_{\alpha''}\mathcal{X}_{3,\alpha''4}\right]\,.
\non 
\end{align}
From this it is straightforward but lengthy to compute the complete set of F--term equations. Together with the D--terms \eqref{eq:Z6DTerms} they determine the phase structure of this GLSM. Below, we restrict our attention to the orbifold and blow--up phases only. In the orbifold regime, $b_r < 0 < a_a$, the singularities are recovered.

\subsection{The minimal model}
\label{sc:MinimalZ4} 

\begin{table}[tb]
\centering
\renewcommand{\arraystretch}{\arrystrch} 
\begin{tabular}{|c||c|c|c||c|c|c|c||c|c|c|c|}
\hline
Superfield & $\mathcal{Z}_{1i}$ & $\mathcal{Z}_{2j}$ & $\mathcal{Z}_{3k}$ 
& $\mathcal{C}_1$  & $\mathcal{C}_2$ & $\mathcal{C}_3$ & $\mathcal{C}'_3$
&  $\mathcal{X}_{1,111}$ &  $\mathcal{X}_{2,i'1}$ &  $\mathcal{X}_{3,i''1}$ &  $\mathcal{X}_{4,i'1}$ 
\\
$U(1)$ charges & $z_{1i}$ & $z_{2j}$ & $z_{3k}$ 
& $c_1$  & $c_2$ & $c_3$ & $c'_3$
&  $x_{1,111}$ &  $x_{2,i'1}$ &  $x_{3,i''1}$ &  $x_{4,i'1}$ 
\\
\hline
\hline
$R_1$ & $i$ & 0 & 0 
& $-6$ & 0 & 0 & 0
& 0 & 0 & 0 & 0 
\\
\hline
$R_2$ & 0 & $1$ & 0 
& 0 & $-3$ & 0 & 0 
& 0 & 0 & 0 & 0 
\\
\hline
$R_3$ & 0 & 0 & $1$ 
& 0 & 0 & $-2$ & $-2$
& 0 & 0 & 0 & 0 
\\
\hline\hline
$E_{1,111}$ & $\delta_{1i}$ & $2\,\delta_{1j}$ & $3\,\delta_{1k}$ 
& 0 & 0 & 0 & 0
& $-6$ & 0 & 0 & 0 
\\
\hline
$E_{2,\alpha'1}$ & $\delta_{\alpha' i}$ & $2\,\delta_{1 j}$ & 0 
& 0 & 0 & 0 & 0  
& 0 & $-3\,\delta_{\alpha' i'}$ & 0 & 0 
\\
\hline
$E_{3,\alpha''1}$ & $\delta_{\alpha'' i}$ & 0 & $\delta_{1 k}$ 
& 0 & 0 & 0 & 0  
& 0 & 0 & $-2\,\delta_{\alpha'' i''}$ & 0 
\\
\hline
$E_{4,\alpha'1}$ & $2\,\delta_{\alpha' i}$ & $\delta_{1 j}$ & 0 
& 0 & 0 & 0 & 0
& 0 & 0 & 0 & $-3\,\delta_{\alpha' i'}$ 
\\
\hline
\end{tabular}
\renewcommand{\arraystretch}{1} 
\normalsize
\caption{$U(1)$ charge assignment for the minimal GLSM resolution of the $T^6/\Intr_\text{6--II}$ orbifold. The indices $i',\alpha'$ take values $1$ or $2$ and the indices $i'',\alpha''$ take values $1$ or $3$.}
\label{tab:Z6IIMinimalCharges}
\end{table}

After the maximal fully resolvable GLSM for the $\Intr_{6-\text{II}}$ orbifold, we focus on the minimal fully resolvable model. In contrast to the $\Intr_3$ orbifold discussed in Section \ref{sc:ResT6Z3}, the geometry cannot be fully resolved using one exceptional divisor only. This can be understood from the number of gaugings needed to resolve the local singularities as discussed in Subsection \ref{sc:LocalRes}: The GLSMs describing the resolutions of $\Cplx^3/\Intr_\text{6--II}$, $\Cplx^2/\Intr_3$ and $\Cplx^2/\Intr_2$ require 4, 2 and 1 gaugings, respectively. Thus, given that we have $\Intr_\text{6--II}$ fixed points and $\Intr_3$ and $\Intr_2$ fixed tori, at least seven gaugings and exceptional fields $\cX$ have to be introduced. All fields necessary for the minimal fully resolved model are summarized together with their charges in Table \ref{tab:Z6IIMinimalCharges}.

In order to check that this model reproduces the expected fixed points and fixed tori in the orbifold regime, we investigate the factorization of the F--terms. In the orbifold regime we have $c_a=0$ and all $x$--fields have non--vanishing VEVs. Consequently, the relevant F--terms are
\begin{subequations}
\begin{align}
z_{11}^6\, x_{1,111}x_{2,11}^2x_{3,11}^3x_{4,11}^4+z_{12}^3\, x_{2,21}x_{4,21}^2+z_{13}^2\, x_{3,31}  &= 0\,,\label{eq:Z6FTerms1}
\\[1ex] 
z_{21}^3 \, x_{1,111}x_{2,11}^2x_{2,21}^2x_{4,11}x_{4,21}+z_{22}^3+z_{23}^3 &= 0\,,\label{eq:Z6FTerms2}
\\[1ex] 
 \gk\, z_{31}^2\, x_{111}x_{3,11}x_{3,31}+z_{32}^2+z_{33}^2 &= 0\,,\label{eq:Z6FTerms3}
 \\[1ex] 
 z_{31}^2\, x_{111}x_{3,11}x_{3,31}+z_{32}^2+z_{34}^2 &= 0\,.\label{eq:Z6FTerms4}
\end{align}
\end{subequations}
The $\Intr_6$ fixed points are determined by $z_{11}=z_{21}=z_{31}=0$. Here the F--term \eqref{eq:Z6FTerms1} of the first torus does not factorize. The F--term \eqref{eq:Z6FTerms2} of the second torus factorizes into three parts as in the minimal $\Intr_3$ torus case. Likewise, the F--terms for the third torus \eqref{eq:Z6FTerms3}--\eqref{eq:Z6FTerms4} factorize such that they have four solutions $z_{32}=\pm z_{33}=\pm z_{34}$. Hence, in total we find $1\cdot3\cdot4=12$ $\Intr_6$ fixed points as expected.

Under the $\theta^2$ (or $\theta^4$) action the $\Intr_2$ torus is fixed. The fixed tori are at $z_{1i'}=z_{21}=0$. For $i'=1$ the discussion is parallel to the one of the $\Intr_6$ given above: The F--term solution in the first torus is unique, while in the second torus the F--terms factorize into three parts, yielding three solutions. For $i'=2$ the result is the same: the F--term of the first torus does not factorize and the F--term in the second torus yields three solutions. Hence we find $3+3=6$ fixed tori in the second and fourth twisted sectors.

The last independent sector is the $\theta^3$ sector, which leaves the second torus fixed.  The fixed tori of this action are at $z_{1i''}=z_{31}=0$. Again, the F--term of the first torus does not factorize for $i''=1,3$ and the F--terms in the third torus have $4$ solutions each, resulting in a total of $4+4=8$ $\Intr_2$ fixed tori. Hence, by combining these results we have recovered all fixed points/tori of the $\Intr_{6-\text{II}}$ orbifold.

Finally, we confirm that all singularities are indeed resolved in our minimal model in the blow--up regime. We analyze the D--terms \eqref{eq:Z6DTerms} in the regime where all $a_a,b_{r,ijk}>0$, but the $b_{r,ijk}$ are taken to be parametrically smaller than the $a_a$.  From the second D--term \eqref{eq:Z6DTerms2} we immediately conclude that the 12 $\Intr_6$ fixed points $z_{11}=z_{21}=z_{31}=0$ are removed. Similarly, \eqref{eq:Z6DTerms3} and \eqref{eq:Z6DTerms5} forbid the six $\theta^2$ and $\theta^4$ fixed tori, and \eqref{eq:Z6DTerms4} forbids the eight $\theta^3$ fixed points. This way, all fixed points/tori are removed in the blow--up regime.

\subsection[Partially resolved $\Intr_{6-\text{II}}$ model]{Partially resolved $\boldsymbol{\Intr_{6-\text{II}}}$ model}

As explained above the aim of studying partially resolvable GLSMs is to have models where the SM breaking singularities simply cannot be blown up. Therefore, we need to determine which singularities are the problematic ones. For the sake of concreteness we focus on the two mini--landscape benchmark models \cite{Lebedev:2007hv}. In these models the gauge symmetry breaking pattern is as follows:  The gauge shift breaks the original $E_8$ down to $SO(10)$, the $\Intr_3$ Wilson line in the second torus to $SU(5)$ and finally a $\Intr_2$ Wilson line on the third torus to the SM gauge group. (The other $\Intr_2$ Wilson line is not switched on.) Consequently, two of the four fixed points in the third torus cannot be blown up without breaking the SM group. 

In order to obtain a corresponding partially resolvable GLSM, we introduce the fields $x_{1,111}$, $x_{2,\alpha'j}$, $x_{3,\alpha''1}$, $x_{3,\alpha''2}$, $x_{4,\alpha'j}$. The fields $x_{2,\alpha'j}$ and $x_{4,\alpha'j}$ only serve to get a fully resolvable model in the $\Intr_3$ sectors. Hence we concentrate on the effect of the four fields $x_{3,\alpha''1}$ and $x_{3,\alpha''2}$. The VEVs of these fields induce the discrete actions
\begin{align}
\arry{c}{
 \theta_{1,1}: (z_{11},z_{31})\mapsto(-z_{11},-z_{31})\,,
 \\[1ex]  
 \theta_{1,2}: (z_{11},z_{32})\mapsto(-z_{11},-z_{32})\,,
 \\[1ex] 
 \theta_{3,1}: (z_{13},z_{31})\mapsto(-z_{13},-z_{31})\,,
 \\[1ex] 
 \theta_{3,2}: (z_{13},z_{32})\mapsto(-z_{13},-z_{32})\,.
 }
\end{align}
Now, the story is similar to the $\Intr_3$ case, studied in Subsection \ref{sc:FactResZ3}. The above actions have four fixed points at $z_{1\alpha''}=z_{3\rho}=0$, $\alpha''=1,3$, $\rho=1,2$. We can define two new effective actions $U(1)_{\text{eff},1}$ and $U(1)_{\text{eff},2}$, generated by the GLSM charges 
$T_{\text{eff},1} = R_{3}-E_{3,11}-E_{3,12}$ 
and 
$T_{\text{eff},2}=R_{3}-E_{3,31}-E_{3,32}$, respectively. 
Their D--terms are
\begin{subequations}
\begin{align}
-2|z_{11}|^2+|z_{33}|^2+|z_{34}|^2+2\,|x_{3,11}|^2+2\,|x_{3,12}|^2=a_3-b_{3,11}-b_{3,12}\,,\\
-2|z_{13}|^2+|z_{33}|^2+|z_{34}|^2+2\,|x_{3,31}|^2+2\,|x_{3,32}|^2=a_3-b_{3,31}-b_{3,32}\,.
\end{align}
\end{subequations}
This show that the 4 fixed tori at $z_{1\alpha''}=z_{3\sigma}=0$, $\alpha''=1,3$, $\rho=3,4$ are present in the orbifold or the blow--up regime for any value of the $a$'s and $b$'s and are thus not resolved\footnote{There may exist solutions in other phases beyond the blow--up regime not studied here.}. Hence such a description might be useful to study partially resolved orbifolds associated to the mini--landscape benchmark MSSMs. 

A similar pattern can be identified for all mini--landscape models: In all of them the Wilson line in the second torus is always switched on, and in the third torus, either one or both Wilson lines are used. Hence it could be that it is possible to construct partially resolvable GLSMs for all the mini--landscape models in a similar fashion. For a full--fledged analysis of such models, however, one has to leave the $(2,2)$ GLSM description for the standard embedding and investigate the gauge bundle in $(2,0)$ models, which is beyond the scope of the paper.


\section{Concluding remarks}
\label{sc:concl}

\subsection*{Summary}

The main purpose of this work was to give a solid description of toroidal orbifolds and their resolutions from a worldsheet perspective. To this end we described them using two dimensional GLSMs with (2,2) worldsheet supersymmetry. The resulting models are interesting both for formal studies of CY compactifications as well as for phenomenological considerations.  

The procedure can be summarized as follows: Our starting point are factorizable toroidal orbifolds. We reformulate their three two--tori in an algebraic way as elliptic curves, i.e.\ as hypersurfaces in weighted projective spaces. Their descriptions are unique provided that one requires that they manifestly reflect the orbifold symmetries. In this formulation the hypersurfaces are sums of monomials of single homogeneous coordinates. The zeros of the homogeneous coordinates determine the positions of the orbifold singularities. The individual orbifold singularities are resolved using a GLSM guise of non--compact toric resolutions, i.e.\ replacing the discrete orbifold actions by Abelian gaugings and introducing additional exceptional coordinates. In the phase where these coordinates are necessarily non--zero, the orbifold symmetries are recovered. Since we perform this construction on the level of  homogeneous coordinates, this procedure results in a global description of toroidal orbifold resolutions. 

Given that our GLSM formalism takes place within a single framework, it puts the gluing procedure, discussed in \cite{Lust:2006zh}, on solid footing. Both the construction of the elliptic curve and the blow--up procedure are described as Abelian gaugings on the worldsheet, and can therefore be discussed in the language of toric geometry. Topological properties of the geometry can be directly read off from the GLSM. For example, some linear equivalence relations correspond to monomials in the superpotential. The intersection numbers are determined by the number of solutions the F-- and D--terms admit. We confirmed this explicitly for the GLSM resolution of the $T^6/\Intr_3$ orbifold: In the blow--up phase the intersection numbers agreed with those given in \cite{Lust:2006zh}.

A given resolution GLSM possesses a large variety of phases: Our starting point is the orbifold phase. A fully resolvable GLSM possesses one or a multitude of fully resolved regimes. In these phases the model admits an interpretation as a smooth geometry where a supergravity analysis is applicable. However, the target space geometries which these phases describe can be very different. We have provided a tentative list of possible phases of toroidal orbifold resolution GLSMs. Besides the orbifold and blow--up phases, we identified the following ones: In the critical blow--up regimes the torus radii and blow--up parameters have comparable sizes. In the over--blow--up phases the blow--up cycles have become much larger than the torus radii. In certain cases different phases are related by flop--like transitions. More violent transitions are also possible: Divisors might emerge or disappear, and even the target space dimension might jump. Given that these different regimes are distinguished by the relative values of the corresponding \Kh\ parameters, these transitions are in principle smooth processes in the GLSM framework. 

For any factorized orbifold we define its maximal fully resolvable GLSM as the GLSM that realizes the maximal number of geometrical \Kh\ variables as FI--parameters. By reducing the number of gaugings one can construct a whole tree of GLSMs which are related to this maximal model. In the so--called partially resolvable GLSMs not all singularities can be blown up for any values of the FI--parameters. In addition it might happen that the originating geometry is no longer the factorizable $T^6$, but one of its non--factorized cousins. For the $T^6/\Intr_3$ all non--factorized lattices can be turned into factorized ones by \Kh\ deformations. However, the GLSMs describing the resolutions of the  $T^6/\Intr_4$ orbifold can lead to genuine non--factorizable lattices resulting in distinct geometries with different Hodge numbers. 

The maximal fully resolvable models are also of interest for another reason. According to a result of Beasley and Witten \cite{Beasley:2003fx}, the \Kh\ and complex structure parameters that appear as FI-- or superpotential parameters in the GLSM, are protected against worldsheet instanton effects even when (2,0) deformations are considered. Hence, within the setting of orbifold resolutions our maximal fully resolvable GLSMs provide us with the maximum of protected \Kh\ parameters. 

One of the striking features of the resolution GLSMs we have presented in this paper is that the \Kh\ parameters associated with the two--tori and the blow--up parameters associated with the exceptional cycles are in principle on equal footing. We have constructed one particular $T^6/\Intr_3$ GLSM which takes this to the extreme: It is completely symmetric under the interchange of two--tori radii and the blow--up cycle radii.

\subsection*{Outlook}

As mentioned above, partially resolvable GLSM do not correspond to a completely smooth geometry: A certain number of the orbifold singularities remain. We think that precisely such GLSMs might be very important for certain phenomenological studies. The mini--landscape MSSMs constructed in \cite{Lebedev:2006kn} on the $T^6/\Intr_\text{6-II}$ orbifold need to have a certain number of singlet VEVs switched on in order to cancel the universal FI--term and to decouple all the unwanted exotics. In \cite{Nibbelink:2009sp} it was shown that such mini--landscape models cannot be completely blown up without breaking the hypercharge. This means that precisely the phenomenologically interesting regime of these models is neither an orbifold nor a completely smooth geometry. We think that the partially resolvable GLSMs, described in Section \ref{sc:ResT6Z6II}, might provide a valid description for this situation, and it would therefore be interesting to develop their properties further. 

In the current paper we have worked only with (2,2) worldsheet theories. For the description of the geometry of toroidal orbifolds and their resolutions, this formulation is very convenient. However, we need to be able to consider other bundles on these geometries than just the standard embedding. From the worldsheet point of view this means that we have to consider (2,0) theories. Unfortunately, (2,0) GLSMs are far less understood than their (2,2) relatives. A reflection of this is that the construction of gauge bundles on Calabi--Yaus is a complicated endeavor. Even locally, i.e.\ for non--compact orbifold singularities and their blow--ups the techniques for this are far less developed: Only recently a systematic mapping between non--compact heterotic orbifold models with twisted VEVs switched on and corresponding GLSMs has been proposed \cite{Nibbelink:2010wm}. However, in that case the gauge anomaly conditions on the worldsheet are not automatically solved, and five--branes might have to be included in the worldsheet description \cite{Adams:2006kb,Carlevaro:2009jx,Blaszczyk:2011ib,Quigley:2011pv,Mertens:2011ha}. It would therefore be very interesting to try to extend our GLSM construction of toroidal orbifold resolutions in this direction.

\subsection*{Acknowledgments}

We would like to thank Sebastian Halter, Ilarion Melnikov, Thorsten Rahn, Emanuel Scheidegger, and Patrick Vaudrevange for very useful discussions. SGN would like to thank the Bethe Center for Theoretical Physics for its kind hospitality. 
This work was partially supported by the SFB--Tansregio TR33 ``The Dark Universe'' (Deutsche Forschungsgemeinschaft) and the European Union 7th network program ``Unification in the LHC era'' (PITN-GA-2009-237920). Furthermore, this work was supported by the LMUExcellent Programme.


\appendix 
\def\theequation{\thesection.\arabic{equation}} 
\setcounter{equation}{0}


\section{Equivalence of discrete group actions on elliptic curves and translations on tori}
\label{app:DiscreteActions}

In this appendix we show the equivalence of some discrete group actions on elliptic curves and translations on  tori they correspond to. We focus here on the $\Intr_3$, $\Intr_4$ and $\Intr_2$ tori discussed in Subsection \ref{sc:Z3torus}--\ref{sc:Z2torus}. To establish this equivalence the following addition formulae for the Weierstrass function and its derivative are of crucial importance:
\begin{align}
\begin{aligned}
\wp_\tau( u_1+ u_2) = \frac 14 \Big( \frac{\wp_\tau'( u_1)-\wp_\tau'( u_2)}{\wp_\tau( u_1)-\wp_\tau( u_2)} \Big)^2 
- \wp_\tau( u_1) - \wp_\tau( u_2)\,, 
\\[2ex] 
\wp_\tau'( u_1+ u_2) =  -\frac 14 \Big( \frac{\wp_\tau'( u_1)-\wp_\tau'( u_2)}{\wp_\tau( u_1)-\wp_\tau( u_2)} \Big)^3
+ 3\, \frac{\wp_\tau'( u_1)-\wp_\tau'( u_2)}{\wp_\tau( u_1)-\wp_\tau( u_2)} \, \wp_\tau( u_2) +  \wp_\tau'( u_1) - 2\, \wp_\tau'( u_2)\,.
\label{WeierstrassADD}
\end{aligned}
\end{align}
Note that the first relation can be used to show that the sum of the three roots $\gve_1$, $\gve_2$ and $\gve_3$ vanishes.

\subsection{Generic two--torus}
\label{app:Z2torus}

To establishes that the actions of \eqref{WL_Z2} and \eqref{Z2translation} are identical we first consider the effect of the translations \eqref{Z2translation} on the Weierstrass function and its derivative using \eqref{WeierstrassADD} and the fact that the $\gve_i$ define the zeros of $\wp_\tau'$ as follows from \eqref{eq:WeierstrassGeneral2}: 
\begin{subequations} 
\begin{align}
\wp_\gt\Big(u+ \sfrac 12 e_i \Big) & = 
\frac {\gve_i \wp_\tau(u) + \gve_{i+1}\gve_{i+2} + \gve_i^2}{\wp_\tau(u) - \gve_i}\,, 
\\[2ex] 
\wp_\gt^\prime \Big(u+ \sfrac 12 e_i \Big) & = 
- (\gve_{i+1}\gve_{i+2} + 2\,\gve_i^2) 
\frac{\wp_\gt^\prime(u) }{(\wp_\tau(u) - \gve_i)^2}
\end{align}
\end{subequations} 
where $e_1 =1, e_2 = \tau$ and $e_3 = 1+\tau$. Using the inverse of the mapping \eqref{eq:Z2Mapping} 
\equ{
\wp_i(u) = \frac xv\,, 
\qquad 
\wp_i^\prime(u) = 2 \, \frac y{v^2}\,, 
\label{Z2invmapping}
}%
and $\Cplx^*$--scalings with $\gl = \cP(u) - \gve_i$, one can show that these actions  on $(x,y,v)$ are given by \eqref{WL_Z2}. 

The fixed point mappings \eqref{Z2fixedpointmapping} follows from applying \eqref{WL_Z2} to the factors in \eqref{eq:Z2TorusWeierstrass}. In detail we have 
\begin{subequations} 
\begin{align}
\ga_i &:& v \mapsto & \frac{ x - \gve_i v~}{\sqrt{\gve_{i+1}\gve_{i+2} + 2\,\gve_i^2}}\,, 
&
x - \gve_i \mapsto &  \sqrt{\gve_{i+1}\gve_{i+2} + 2\,\gve_i^2}\,\cdot v\,, 
\\[2ex] 
\ga_i &:& x - \gve_{i+1} v \mapsto& \frac{gve_i - \gve_{i+1}}{\sqrt{\gve_{i+1}\gve_{i+2} + 2\,\gve_i^2}}(x - \gve_{i+2} v)\,, 
& 
x - \gve_{i+2} v \mapsto & \frac{\gve_i - \gve_{i+2}}{\sqrt{\gve_{i+1}\gve_{i+2} + 2\,\gve_i^2}}(x - \gve_{i+1} v)\,. 
\label{Z22ShiftP121}
\end{align}
\end{subequations} 
This show how the factors and therefore the zeros are pairwise permuted.

\subsection{$\mathbf{\mathbbm{Z}_3}$ torus} 
\label{app:Z3torus}

To establishes that the actions of \eqref{WL_Z3} and \eqref{Z3translation} represent the same operation, we show that they are identical using the Weierstrass function and its derivative. To this end we first give the mapping of the $\gth_{\rm WL}$ action on the coordinates $(x,y,v)$ 
\begin{align}
 \alpha:~  
\begin{pmatrix}
 y \\ v \\
 \end{pmatrix}
\mapsto -\frac12
 \begin{pmatrix}
 1 & 3 i \\   i & 1 \\
 \end{pmatrix}
 \begin{pmatrix}
  y \\  v \\
 \end{pmatrix} \,, 
 \qquad 
 x \mapsto x\,. 
\end{align}
Together with the inverse of the mapping \eqref{WeierstrassMappingZ3} 
\equ{
\wp_\gz(u) = \gve_1\, \frac xv\,, 
\qquad 
\wp_\gz^\prime(u) = 2 \gve_1^{3/2}\, \frac {y}v\,. 
\label{WeierIden}
}
this gives after some algebra:
\begin{align}
 \wp_\gz(u) \mapsto
 \frac{- 4 \gve_1^{3/2} \wp_\gz(u)}{  i \wp_\gz'(u) + 2\gve_1^{3/2} } \,,
\qquad 
 \wp_\gz'(u) \mapsto
 \frac{2 \gve_1^{3/2} \wp_\gz'(u) + 12 i \gve_1^3}{  i \wp_\gz'(u) + 2\gve_1^{3/2} } \,,
\end{align}
On the other hand, since $\gz\, \frac {\gz-1}3 = \frac{\gz-1}3 -\gz$ the periodicities of the Weierstrass function \eqref{periodicities} and $\wp_\gz(\gz u) = \gz \wp_\gz(u)$ imply that 
\equ{
\wp_\gz\big( \sfrac{\gz-1}3 \big) = 0\,, 
\qquad 
\wp_\gz^\prime\big(\sfrac{\gz-1}3\big) = -2 i\, \gve_1^{3/2}\,. 
}
Using these identities and making extensive use of the Weierstrass addition formulae \eqref{WeierstrassADD} and differential equation \eqref{eq:WeierstrassGeneral} we obtain identical results
\begin{align}
  \wp_\gz\big(u+\sfrac{\gz-1}3\big) =  
  \frac{ -4 \gve_1^{3/2}\wp_\gz(u)}{ i \wp_\gz'(u) +2 \gve_1^{3/2} }\,,  
\qquad 
 \wp_\gz'\big(u + \sfrac{\gz-1}3\big) = 
 \frac{2 \gve_1^{3/2} \wp_\gz'(u) + 12 i \gve_1^3}{  i \wp_\gz'(u) + 2\gve_1^{3/2} } \,, 
\end{align}
showing that \eqref{WL_Z3} and \eqref{Z3translation} represent the same operation.

\subsection{$\mathbf{\mathbbm{Z}_4}$ torus} 
\label{app:Z4torus}

For the $\Intr_4$ torus we have to clarify two issues: 1) how does the $\Intr_4$ orbifold action act on both the torus and the elliptic curve. In particular we explain the necessary shift in the arguments of the Weierstrass function and its derivative in \eqref{eq:Z4Mapping}. 2) how does the $\Intr_2$ involution act.

\subsubsection*{$\mathbf{\mathbbm{Z}_4}$ orbifold group action}

Let us start with explaining issue 1), i.e.\ how the $\Intr_4$ orbifold element acts on both the torus and the elliptic curve. Extrapolating from the $\Intr_3$ torus case one may naively expect that these actions read $u \mapsto i\, u$ on the  torus and $(z_1,z_2,z_3) \mapsto (i\, z_1, z_2,z_3)$ on the elliptic curve taking into account the mapping 
\begin{align}
\arry{rcl}{
T^2 = \Cplx/\Lambda &\ra& \mathbbm{CP}_{1,2,1}^2 
 \\ 
 u &\mapsto& (x,y,v) =  
 \begin{cases} 
  \Big(\gve_1^{-1/4}\wp_i(u),\wp_i^\prime(u)/2,\gve_1^{3/4}\Big)\,, & \text{away from lattice points~,}  
    \\
   \big(1,0,0\big) & u \in \Lambda\,, 
  \\ 
   \Big(1,\gve_1^{1/2} \frac{\wp_i^\prime(u)}{2(\wp_i(u))^2},\frac{\gve_1}{\wp_i(u)}\Big)\,, & \text{near lattice points~.}
\end{cases} 
 }
 \label{eq:Z4Mapping_Naive}
\end{align}
and \eqref{eq:Z4TorusFieldRedefinition}. However, if one takes this action on the torus and goes through these mappings, one finds that the action on homogeneous coordinates $(z_1,z_2,z_3)$ is not even diagonal. Hence we see that the correspondence between the orbifold action on the  torus and the elliptic curve is more complicated: Using the mapping \eqref{eq:Z4Mapping_Naive} between the  torus and the elliptic curve, the $\Intr_4$ orbifold action $\gth :~ (z_1,z_2,z_3) \mapsto (i\, z_1, z_2,z_3)$ on the elliptic curve corresponds to the action 
\equ{
\theta: u \mapsto  i u + \sfrac 12
\label{Z4flattorus}
}
on the  torus. 

To confirm this we start with the $\Intr_4$ orbifold action on the elliptic curve and write it in terms of the coordinates $(x,y,v)$ using \eqref{eq:Z4TorusFieldRedefinition}:
 \begin{align}
  \theta :  
\begin{pmatrix}
 x \\ v \\
 \end{pmatrix}
\mapsto \frac{1+ i}2
 \begin{pmatrix}
 1 & -1 \\  1 & 1 \\
 \end{pmatrix}
 \begin{pmatrix}
  x \\ v \\
 \end{pmatrix} \,, \qquad y \mapsto y\,.
\end{align}
Using the inverse of the mapping \eqref{eq:Z4Mapping_Naive}, i.e.\ 
\equ{
\wp_i(u) = \gve_1\, \frac xv\,, 
\qquad 
\wp_i^\prime(u) = 2 \gve_1^{3/2}\, \frac y{v^2}\,, 
\label{Z4invmapping}
}
this leads to the following mappings of the Weierstrass function and its derivative 
\begin{align}
 \wp_i(u)  \mapsto 
 \gve_1\,  \frac{\wp_i(u)-\gve_1}{\wp_i(u)+\gve_1} \,,
 \qquad 
 \wp_i^\prime(u) \mapsto 
 -2i\,\gve_1^2\, \frac{\wp_i^\prime(u)}{(\wp_i(u) + \gve_1)^2}\,. 
\end{align}

Next we show that we obtain the same transformations of the Weierstrass function and its derivative by inserting the transformation \eqref{Z4flattorus} into them. To this end we employ the following identities 
\begin{align}
 \wp_i(\sfrac12) = \gve_1\,, 
 \qquad  
 \wp_i'(\sfrac12) = 0\,, 
 \qquad  
 \wp_i( i u) = -\wp_i(u)\,, 
 \qquad  
 \wp_i'( i u) =  i \wp_i(u) \,,
\end{align}
together with the addition formulae \eqref{WeierstrassADD} and the defining differential equation \eqref{eq:WeierstrassGeneral} to obtain 
\begin{align}
\wp_i(u)  &\mapsto  \wp_i( i u + \sfrac12) 
= \gve_1\, \frac{\wp_i(u)-\gve_1}{\wp_i(u)+\gve_1}\,, 
\\
\wp_i^\prime(u) &\mapsto \wp_i^\prime(i u + \sfrac12) 
= -2 i\, \gve_1^2\, \frac{\wp_i^\prime(u)}{(\wp_i(u) + \gve_1)^2}\,. 
\end{align}
Hence we have shown that the transformation on the elliptic curve and the  torus are identical. 

Let us finally give an interpretation of this result: The fact that on the  torus the $\Intr_4$ orbifold action is \eqref{Z4flattorus} rather than $u \mapsto i\, u$ simply means that on the  torus the orbifold action rotates by 90 degrees not around the origin but rather around the point $u_0 = \frac 14(1+i)$. Since it is conventional in the orbifold literature that the orbifold action rotates around the origin, we have included a shift of $u$ over $u_0$ in the arguments of the Weierstrass function and its derivative in \eqref{eq:Z4Mapping}.

\subsubsection*{$\mathbf{\mathbbm{Z}_2}$ Involution}

Next we establish 2), i.e.\ the correspondence between the actions of \eqref{WL_Z4} and \eqref{Z4translation} by showing that their actions on  the Weierstrass function and its derivative are identical. To translate the action on the  torus \eqref{Z4translation} to the action on the Weierstrass function and its derivative, we need the identities
\begin{align}
 \wp_i(\frac{1+i}2) = \gve_3 = 0\,, 
 \qquad  
 \wp_i'(\frac{1+i}2) = 0\,. 
\end{align}
Using these and the usual addition formulae \eqref{WeierstrassADD} and the defining differential equation \eqref{eq:WeierstrassGeneral}, we find 
\begin{align}
  \wp_i(u) \mapsto  \wp_i( u + \frac{e_1+e_2}{2}) =  -\frac{\gve_1^2}{\wp_i(u)} \,, 
\qquad 
 \wp_i'(u) \mapsto  \wp_i'( u + \frac{e_1+e_2}{2}) = \gve_1^2 \, \frac{\wp_i'(u)}{(\wp_i(u))^2}\,. 
\end{align}
On the elliptic curve the transformation \eqref{WL_Z4} translates to 
\begin{align}
  \ga :~  
\begin{pmatrix}
 x \\ v \\
 \end{pmatrix}
\mapsto 
 \begin{pmatrix}
 0 & -1 \\  1 & 0 \\
 \end{pmatrix}
 \begin{pmatrix}
  x \\ v \\
 \end{pmatrix} \,, 
 \qquad 
 y \mapsto y
\end{align}
for the coordinates $(x,y,v)$. Using \eqref{Z4invmapping} this gives us the following mappings of the Weierstrass function and its derivative: 
\begin{align}
 \wp_i(u) \mapsto -\frac{\gve_1^2}{\wp_i(u)}\,,
 \qquad 
 \wp_i'(u) \mapsto 2 \gve_1^2\, \frac{\wp_i'(u)}{(\wp_i(u))^2}\,, 
\end{align}
which agrees with the action on the  torus.

\subsection{$\mathbf{\mathbbm{Z}_2}$ torus} 
\label{app:Z2torusDiscreteActions}

We want to translate the maps \eqref{Z22ShiftP121} to the $\mathbbm{P}^3[2,2]/\Intr_2 \times \Intr_2$ torus, i.e.\ the $T^2(\Intr_2)$. For this we define the map
\begin{align}
 \mathbbm{P}^2_{1,2,1}[4] \rightarrow \mathbbm{P}^3[2,2]/\Intr_2 \times \Intr_2 \,: 
\qquad 
(x,y,v) \mapsto (z_1,z_2,z_3,z_4) \,,
\end{align}
as follows: On $\mathbbm{P}^3[2,2]/\Intr_2 \times \Intr_2$ we have three sign ambiguities, one from $U(1)_R$ and two from $\Intr_2 \times \Intr_2$. Thus we can define
\begin{align}
 z_1 = \sqrt{(\epsilon_3 - \epsilon_1)v} \,, \qquad z_2 = \sqrt{-x+\epsilon_1 v} \,, \qquad z_3 = \sqrt{x-\epsilon_2} \,, \qquad z_4 = \frac{-iy}{2z_1 z_2 z_3}.
\end{align}
One could also define $z_4 = \pm \sqrt{x - \epsilon_3 v}$ where the sign has to be chosen such that $y=2iz_1z_2z_3z_4$, as there are no more sign ambiguities. This way we find that the discrete $\Intr_2 \times \Intr_2$ shifts, given in \eqref{Z22ShiftP121} translate to the actions given in \eqref{involutionsZ2Torus}.


\section{Compactification lattice analyses for the $\boldsymbol{T^6/\Intr_3}$ orbifold}
\label{app:LatticeAnayses}

The aim of this Appendix is three--fold: 1) We briefly describe the factorized Lie--algebra lattice $A_2^3$ and the non--factorized lattices underlying the $T^6/\Intr_3$ orbifolds encountered in Section \ref{sc:ResT6Z3}. To this end we work out in detail how dividing out the free discrete actions on the  torus induce refined and often non--factorized lattices. 2) Where possible we classify the resulting lattice as some Lie algebra lattice. 3) Finally, we show that each of the non--factorized lattices can be obtained from the factorized $A_2^3$ lattice by a change of \Kh\ structure. 

In our conventions for $\Intr_3$ compatible tori defined in Section \ref{sc:twotori}, the lattice underlying the two--tori is spanned by $1$ and the complex structure $\gt = \gz$. Therefore, our reference basis vectors $e_1, \ldots, e_6$ for the lattice $A_2^3$ in $\Real^6$ are given by 
\begin{align*}
 e_1 = 
\begin{pmatrix}
 1 \\ 0 \\ 0 \\ 0 \\ 0 \\ 0 \\
\end{pmatrix}\,, \quad 
 e_2 = 
\begin{pmatrix}
 -\frac12 \\ \frac{\sqrt{3}}{2} \\ 0 \\ 0 \\ 0 \\ 0 \\
\end{pmatrix}\,, \quad 
 e_3 = 
\begin{pmatrix}
 0 \\ 0 \\ 1 \\ 0 \\ 0 \\ 0 \\
\end{pmatrix}\,, \quad 
 e_4 = 
\begin{pmatrix}
 0 \\ 0 \\ -\frac12 \\ \frac{\sqrt{3}}{2} \\  0 \\ 0 \\
\end{pmatrix}\,, \quad 
 e_5 = 
\begin{pmatrix}
 0 \\ 0 \\ 0 \\ 0 \\ 1 \\ 0 \\ 
\end{pmatrix}\,, \quad 
 e_6 = 
\begin{pmatrix}
 0 \\ 0 \\ 0 \\ 0 \\-\frac12 \\ \frac{\sqrt{3}}{2} \\ 
\end{pmatrix}\,.
\end{align*}
These basis vectors have norm squared $1$ rather than $2$, which would be the conventional choice for simple roots.

The basis vectors, which are obtained by dividing out the free discrete actions, are referred to as $\hat e_1, \ldots, \hat e_6$. Where necessary we include normalization factors to ensure that we can identify the corresponding Lie algebra lattice by computing the Cartan matrix
\equ{ 
A_{mn} = \frac {2\, \hat e_m \cdot \hat e_n}{~~ \hat e_m \cdot \hat e_m}\,,
\label{Cartan} 
}
where $x\cdot y $ denotes the standard inner product on $\Real^6$. 

To establish that the refined (non--factorized) lattice can be turned into a $A_2^3$ lattice via \Kh\ deformations, we have to find a basis of this lattice such that the $\mathbb{Z}_3$ action on it is the standard $\Intr_3$ action 
\begin{align}
e_{2a-1} \mapsto e_{2a}\,,
\qquad 
e_{2a} \mapsto - e_{2a} - e_{2a-1}\,,
\label{eq:Z3ActionOnA2}
\end{align}
with $a=1,2,3$ and $A_2^3$ basis vectors $e_1, \ldots, e_6$. We refer to the non--factorized lattice basis as $\tilde e_1,\ldots, \tilde e_6$. To discuss \Kh\ deformations we define the inner product 
\equ{
\langle x, y \rangle_G = x^T \, G \, y\,, 
\qquad 
x, y \in \Real^6\,, 
}
w.r.t.\ a metric $G$; the standard inner product $x\cdot y$ is obtained when the metric $G$ is taken to be the identity. To establish that the basis vectors  $\tilde e_1, \ldots, \tilde e_6$ of the non--factorized lattice are equivalent to our standard $A_2^3$ basis up to a \Kh\ deformation, we need to show that there is a \Kh\ metric $G$ such that 
\equ{ 
\langle e_m, e_n \rangle_G = \tilde e_m \cdot \tilde e_n\,, 
\label{KhEquiv}
}
for all $m,n = 1, \ldots, 6$. 

To make \eqref{KhEquiv} explicit, we write out our reference basis and give a general  parameterization of a \Kh\ metric $G$. To construct the general metric $G$ that includes all K\"ahler moduli, we start from a generic K\"ahler form
\begin{align}
 J = \frac{i}{2}  J_{a b}\, \d \bar u_{a} \wedge \d u_b \,. 
\end{align}
Its reality condition reads $J_{a b}=\bar J_{ba}$, so we rewrite it in terms of real parameters defined as 
\begin{align}
 J_{aa} = b_a \,, \qquad J_{ab} = c_{ab} + i d_{ab} \quad \text{for } a<b \,.
\end{align}
Then, decomposing into real coordinates, $u_a = x_{2a-1} + i x_{2a}$ the metric becomes
\begin{align}
 G = 
\begin{pmatrix}
 b_1 & 0 & c_{12} & d_{12} & c_{13} & d_{13} \\
 0 & b_1 & -d_{12} & c_{12} & -d_{13} & c_{13} \\
 c_{12} & -d_{12} & b_2 & 0 & c_{23} & d_{23} \\
 d_{12} & c_{12} & 0 & b_2 & -d_{23} & c_{23} \\
 c_{13} & -d_{13} & c_{23} & -d_{23} & b_3 & 0 \\
 d_{13} & c_{13} & d_{23} & c_{23} & 0 & b_3\\
\end{pmatrix} \,.
\label{eq:generalZ3Metric}
\end{align}
The inner product of the $A_2^3$ lattice vectors $e_i$ using this metric reads 
\begin{align}
\langle e_m, e_n \rangle_G =
\left(\!\!\!
\begin{array}{cccccc}
b_1 & -\frac{b_1}{2} & c_{12} & -\frac{c_{12}-\sqrt{3}d_{12}}{2} & c_{13} & -\frac{c_{13}-\sqrt{3} d_{13}}{2} 
\\
-\frac{b_1}{2} & b_1 & -\frac{c_{12}+\sqrt{3}d_{12}}{2} & c_{12} & -\frac{c_{13}+\sqrt{3} d_{13}}{2} & c_{13} 
\\
c_{12} & -\frac{c_{12}+\sqrt{3}d_{12}}{2} & b_2 & -\frac{b_2}{2} & c_{23} & -\frac{c_{23}-\sqrt{3} d_{23}}{2} 
\\
-\frac{c_{12}-\sqrt{3}d_{12}}{2} & c_{12} & -\frac{b_2}{2} & b_2 & -\frac{c_{23}+\sqrt{3} d_{23}}{2} &c_{23} 
\\
c_{13} & -\frac{c_{13}+\sqrt{3}d_{13}}{2} & c_{23} & -\frac{c_{23}+\sqrt{3}d_{23}}{2} & b_3 & -\frac{b_3}{2} \\
-\frac{c_{13}-\sqrt{3}d_{13}}{2} & c_{13} & -\frac{c_{23}-\sqrt{3}d_{23}}{2} & c_{23} & -\frac{b_3}{2} & b_3 \\
\end{array}
\!\!\!\right)\,. 
\label{GeneralMetric}
\end{align}
We obtain the standard $A_2^3$ metric for $b_i=2$, $c_{ij}=d_{ij}=0$.

The free discrete $\Intr_3$ actions $\ga_a$ given in \eqref{Z3translation} act on the torus coordinates as 
\begin{align}
u_b \mapsto u_b+\sfrac{\zeta-1}{3}\, \gd_{ab}\,. 
\label{eq:Z3WLActionOnA2}
\end{align}
Therefore, modding out such free actions leads to a refinement of the $A_2^3$ lattice. The refined lattice is spanned by the old $A_2^3$ basis vectors $e_1,\ldots, e_6$ and combinations of the vectors 
\equ{
\hat \ga_1 = \sfrac 13 ( e_2 - e_1)\,, 
\qquad 
\hat \ga_2 = \sfrac 13 ( e_4 - e_3)\,, 
\qquad 
\hat \ga_3 = \sfrac 13 ( e_6 - e_5)\,, 
}
arising from modding out the corresponding combinations of $\ga_1, \ga_2, \ga_3\,$. Using this technique we will construct each non--factorized $\Intr_3$ lattice and compute the standard inner products of all its basis vectors and thereby simply read off the metric $G$ using \eqref{GeneralMetric}.

\subsection[The ${F_4\times A_2}$ lattice]{The $\boldsymbol{F_4\times A_2}$ lattice}
\label{app:Z3LatticeF4}

When the discrete action \eqref{eq:Z3WLActionOnA2} acts on two of the three tori (e.g.\ the first and the second), this results in a new non--factorized lattice with basis vectors 
\begin{align}
\hat{e}_1:=-\frac{e_2}{\sqrt 2}\,,
\quad  
\hat{e}_2:=\frac13\Big(\frac{e_2-e_1}{\sqrt 2}+ e_4-e_3 \Big)\,,
 \quad 
 \hat{e}_3:=-e_4\,,
 \quad 
 \hat{e}_4:=e_3+e_4\,,
 \label{eq:F4LatticeBasis1}
\end{align}
and $\hat e_5 = e_5$, $\hat e_6 = e_6$. In order to match the resulting lattice to a Lie lattice, we have changed the normalization of the basis vectors in the first two--torus to obtain the Lie lattice of $F_4\times A_2$.

Next we demonstrate that it is possible to factorize this non--factorized $F_4\times A_2$ lattice into a factorized $A_2^3$ lattice. To do so, we have to find a basis of the  $F_4\times A_2$ lattice such that the $\mathbb{Z}_3$ action acts as in \eqref{eq:Z3ActionOnA2}. Taking the vectors \eqref{eq:F4LatticeBasis1} such a basis is given by
\begin{align}
\tilde{e}_1 := \sqrt{2}\,\hat{e}_2\,, \quad \tilde{e}_2 := \sqrt{2}(\hat{e}_1+\hat{e}_2+\hat{e}_3)\,,\quad \tilde{e}_3 := \hat{e}_3+\hat{e}_4\,,\quad \tilde{e}_4 :=-\hat{e}_3\,.
\end{align}
In this case we find the product matrix
\begin{align}
\tilde e_i \cdot \tilde e_j = 
\left(
\begin{array}{cccccc}
 1 & -\frac{1}{2} & -\frac{1}{\sqrt{2}} & \frac{1}{\sqrt{2}} & 0 & 0 \\
 -\frac{1}{2} & 1 & 0 & -\frac{1}{\sqrt{2}} & 0 & 0 \\
 -\frac{1}{\sqrt{2}} & 0 & 1 & -\frac{1}{2} & 0 & 0 \\
 \frac{1}{\sqrt{2}} & -\frac{1}{\sqrt{2}} & -\frac{1}{2} & 1 & 0 & 0 \\
 0 & 0 & 0 & 0 & 1 & -\frac{1}{2} \\
 0 & 0 & 0 & 0 & -\frac{1}{2} & 1
\end{array}
\right)\,.
\end{align}
This can be obtained from \eqref{GeneralMetric} by setting $b_i=1$, $c_{12}=-1/\sqrt{2}$, $d_{12}= 1/\sqrt{6}$, and the rest to zero. Thus, we obtain the $F_4\times A_2$ lattice from the $A_2^3$ lattice upon switching on some of the off--diagonal K\"ahler moduli.

\subsection[The ${E_6}$ lattice]{The $\boldsymbol{E_6}$ lattice}
\label{app:Z3LatticeE6}

When the discrete action \eqref{eq:Z3WLActionOnA2} acts in all three tori simultaneously, we obtain a new lattice with basis vectors 
\begin{align}
\begin{aligned}
 \hat{e}_1&:=e_3+e_4\,,
&\hat{e}_2&:=-e_4\,,
&\hat{e}_3&:=\sfrac13(-e_1+e_2-e_3+e_4-e_5+e_6)\,,
\\[2ex] 
 \hat{e}_4&:=-e_6\,,
&\hat{e}_5&:=e_5+e_6\,,
&\hat{e}_6&:=-e_2\,.
\label{eq:E6LatticeBasis1}
\end{aligned}
\end{align}
We find their metric to be (one half times) the Cartan matrix of $E_6$.

We again demonstrate that it is possible to factorize the $E_6$ lattice into $A_2^3$. For this we have to find a basis of the $E_6$ lattice such that the $\mathbb{Z}_3$ action acts as in \eqref{eq:Z3ActionOnA2}. Taking the vectors \eqref{eq:E6LatticeBasis1} such a basis is given by
\begin{align}
\begin{aligned}
 \tilde{e}_1 &:= \hat{e}_1+\hat{e}_2 \,, 
&\tilde{e}_2 &:= -\hat{e}_2\,,
&\tilde{e}_3 &:= \hat{e}_3\,,
\\[2ex]  
 \tilde{e}_4 &:= \hat{e}_3+\hat{e}_2+\hat{e}_4+\hat{e}_6\,,
&\tilde{e}_5 &:= \hat{e}_4+\hat{e}_5\,,
&\tilde{e}_6 &:=-\hat{e}_4\,.
\end{aligned}
\end{align}
In this case we find the product matrix
\begin{align}
\tilde e_i \cdot \tilde e_j = 
\left(
\begin{array}{cccccc}
 1 & -\frac{1}{2} & -\frac{1}{2} & 0 & 0 & 0 \\
 -\frac{1}{2} & 1 & \frac{1}{2} & -\frac{1}{2} & 0 & 0 \\
 -\frac{1}{2} & \frac{1}{2} & 1 & -\frac{1}{2} & -\frac{1}{2} & \frac{1}{2} \\
 0 & -\frac{1}{2} & -\frac{1}{2} & 1 & 0 & -\frac{1}{2} \\
 0 & 0 & -\frac{1}{2} & 0 & 1 & -\frac{1}{2} \\
 0 & 0 & \frac{1}{2} & -\frac{1}{2} & -\frac{1}{2} & 1
\end{array}
\right)\,.
\end{align}
Comparing this with \eqref{GeneralMetric} we find $b_i=1$, $c_{12}=c_{23}=-1/2$, $d_{12}=-d_{23}= -1/(2\sqrt{3})$, and the rest to zero. Thus, also the $E_6$ lattice can be deformed into the $A_2^3$ lattice using off--diagonal K\"ahler moduli.

\subsection{A non--Lie lattice}
\label{app:Z3LatticeNonLie}

It is also possible to have two discrete actions \eqref{eq:Z3WLActionOnA2}.  The first one mixes the first and second two--torus and the second one mixes the second and third two--torus. Each action by itself is similar to the $F_4$ case, but they get ``intermingled'' in such a way that this lattice does not correspond to any Lie algebra lattice. 

From the two free discrete actions we obtain two new lattice vectors
\begin{align}
\begin{aligned}
 \hat{e}_1&:=\frac13\Big(\frac{e_2-e_1}{\sqrt 2}+e_4-e_3)\,,
&\hat{e}_2&:= \frac{e_2}{\sqrt 2}\,,
&\hat{e}_4&:=e_4\,,
\\[2ex] 
 \hat{e}_3&:=\frac13\Big(\frac{e_6-e_5}{\sqrt 2}+e_4-e_3\Big)\,, 
&\hat{e}_5&:= \frac{e_5}{\sqrt 2}\,,
&\hat{e}_6&:=\frac{e_6}{\sqrt 2}\,.
\end{aligned}
\label{eq:NonLieLatticeBasis1}
\end{align}
As in the $F_4$ case, we have scaled the first and the third torus via $(e_1,e_2,e_5,e_6)\mapsto 1/\sqrt{2}(e_1,e_2,e_5,e_6)$. 

It is also possible to factorize the non--Lie lattice into $A_2^3$. To do so, we have to find a basis of the non--Lie lattice such that the $\mathbb{Z}_3$ action acts as in \eqref{eq:Z3ActionOnA2}. Taking the vectors \eqref{eq:NonLieLatticeBasis1} such a basis is given by
\begin{align}
\tilde{e}_1 := \hat{e}_1\,, \quad \tilde{e}_2 := \hat{e}_1-\hat{e}_2-\hat{e}_4\,,\quad \tilde{e}_3 := \hat{e}_3\,,\quad \tilde{e}_4 := \hat{e}_3-\hat{e}_4-\hat{e}_6\,,\quad \tilde{e}_5:= \hat{e}_5\,,\quad \tilde{e}_6:=\hat{e}_6\,.
\end{align}
Here, the product matrix 
\begin{align}
\tilde e_i \cdot \tilde e_j = 
\left(
\begin{array}{cccccc}
 1 & -\frac{1}{2} & \frac{2}{3} & -\frac{1}{3} & 0 & 0 \\
 -\frac{1}{2} & 1 & -\frac{1}{3} & \frac{2}{3} & 0 & 0 \\
 \frac{2}{3} & -\frac{1}{3} & 1 & -\frac{1}{2} & -\frac{1}{2} & \frac{1}{2} \\
 -\frac{1}{3} & \frac{2}{3} & -\frac{1}{2} & 1 & 0 & -\frac{1}{2} \\
 0 & 0 & -\frac{1}{2} & 0 & 1 & -\frac{1}{2} \\
 0 & 0 & \frac{1}{2} & -\frac{1}{2} & -\frac{1}{2} & 1
\end{array}
\right)\,,
\end{align}
is obtained from \eqref{GeneralMetric} by setting $b_i=1$, $c_{12}=2/3$, $c_{23}=-1/2$, $d_{23}=1/(2\sqrt{3})$, and the rest to zero. Thus, we obtain the non--Lie lattice from the $A_2^3$ lattice via off--diagonal K\"ahler moduli.


\section{Non--factorizable lattices for the $\boldsymbol{T^6/\Intr_4}$ orbifold}
\label{appendixZ4NonfactLattices}

In this appendix we describe the non--factorizable lattices we find for $\Intr_4$ orbifolds in section \ref{subsec:Z4NonFactorizableCases}. We start with the factorizable lattice $D_2^2 \times A_1^2$. The torus $T^2(\Intr_4) \times T^2(\Intr_4) \times T^2(\Intr_2)$ obtained from it is mapped to the elliptic curves as described in Appendix \ref{app:Z4torus}. When we add exceptional coordinates and $U(1)$ gaugings, we induce discrete actions on this torus which we can identify on the  torus as we did for the $\Intr_3$ orbifold. Let us describe the factorizable lattice by the lattice vectors $e_i$ , $i=1,\ldots,6$. The inner product of this basis is\footnote{We neglect possible complex structure deformations in the third torus since they are not relevant here.}
\begin{align}
 G = 
\begin{pmatrix}
 b_1 & 0 & c_{12} & d_{12} & 0& 0\\
 0 & b_1 & -d_{12} & c_{12} & 0 & 0 \\
 c_{12} & -d_{12} & b_2 & 0 & 0 & 0 \\
 d_{12} & c_{12} & 0 & b_2 & 0 &0 \\
 0 & 0 &0 & 0 & b_3 & 0 \\
 0 & 0 & 0 & 0 & 0 & b_3\\
\end{pmatrix} \,.
\label{eq:generalZ4Metric}
\end{align}
The $\Intr_4$ acts as
\begin{align}
 e_1 \mapsto e_2 \,, \qquad  e_2 \mapsto -e_1 \,, \qquad  e_3 \mapsto e_4 \,, \qquad  e_4 \mapsto -e_3 \,, \qquad  e_5 \mapsto -e_5 \,, \qquad  e_6 \mapsto -e_6 \,. 
\end{align}

\subsection[The $A_3 \times D_2 \times A_1$ lattice]{The $\mathbf{A_3 \times D_2 \times A_1}$ lattice}

In the first truly non--factorizable model we find the discrete action
\begin{align}
 \tilde\theta_1 : (z_{21},z_{22},z_{31},z_{32}) \longmapsto (i z_{21},-i z_{22},-z_{31},-z_{32}) \,.
\end{align}
Via the Weierstrass map to the  torus it translates into a shift by $\frac{e_3 + e_4 + e_5}{2}$, i.e.\ a shift in two two--tori simultaneously. For the non--factorizable torus obtained in this way we choose a basis 
\begin{subequations}
\begin{align}
 \hat e_1 &:= \frac{e_3 + e_4 + e_5}{2} \,, & \hat e_2 &:= \frac{-e_3 + e_4 - e_5}{2} \,, & \hat e_3 &:= \frac{-e_3 - e_4 + e_5}{2} \,, \label{eqn:Z4NonFactBasis11} \\
 \hat e_4 &:= e_1 \,, & \hat e_5 &:= e_2 \,, & \hat e_6 &:= e_6 \,. \label{eqn:Z4NonFactBasis12} 
\end{align} 
\end{subequations}
On the first three basis vectors their inner product reads
\begin{align}
 \hat e_i \cdot \hat e_j = 
\begin{pmatrix}
b_2/2 + b_3 / 4 & - b_3 / 4 & -b_2/2 + b_3/4 \\
 - b_3 / 4 & b_2/2 + b_3 / 4 & - b_3/4 \\
 -b_2/2 + b_3/4 &  - b_3 / 4 & b_2/2 + b_3 / 4 \\
\end{pmatrix} \,.
\end{align}
This product cannot be given a factorized structure. Choosing $b_3 = 4$ and $b_2 = 2$ it becomes the Cartan matrix for the Lie algebra $A_3$. The $\Intr_4$ acts as 
\begin{align}
\hat e_1 \mapsto \hat e_2\,,\quad \hat e_2 \mapsto \hat e_3\,,\quad\hat e_3 \mapsto - \hat e_1 - \hat e_2 - \hat e_3\,, 
\end{align}
i.e.\ as a rotation of the extended Dynkin diagram.

\subsection[The $A_3 \times A_3$ lattice]{The $\mathbf{A_3 \times A_3}$ lattice}

Analogously to the case above, we find here in addition the discrete action
\begin{align}
 \tilde\theta_2 : (z_{11},z_{12},z_{31},z_{33}) \longmapsto (i z_{11},-i z_{12},-z_{31},-z_{33}) \,.
\end{align}
which translates into a shift by $\frac{e_1 + e_2 + e_6}{2}$. Now the basis consists of $\hat e_i$, $i=1,2,3$ as defined in \eqref{eqn:Z4NonFactBasis11} but we replace \eqref{eqn:Z4NonFactBasis12} by
\begin{align}
 \hat e_4 &:= \frac{e_1 + e_2 + e_6}{2} \,, & \hat e_5 &:= \frac{-e_1 + e_2 - e_6}{2} \,, & \hat e_6 &:= \frac{-e_1 - e_2 + e_6}{2} \,.
\end{align}
Altogether this gives the inner product,
\begin{align}
 \hat e_i \cdot \hat e_j = \frac12
\begin{pmatrix}
b_2 + b_3 / 2 & - b_3 / 2 & -b_2 + b_3/2 & c & d & -c\\
 - b_3 / 2 & b_2 + b_3 / 2 & - b_3/2 & -d & c & d\\
 -b_2 + b_3/2 &  - b_3 / 2 & b_2 + b_3 / 2 & -c & -d & c\\
c & -d & -c & b_1 + b_3 / 2 & - b_3 / 2 & -b_1 + b_3/2 \\
d & c & -d &  - b_3 / 2 & b_1 + b_3 / 2 & - b_3/2 \\
-c & d & c &  -b_1 + b_3/2 &  - b_3 / 2 & b_1 + b_3 / 2 \\
\end{pmatrix} \,,
\end{align}
which for  $b_3 = 4$, $b_1=b_2 = 2$ and $c=d=0$ becomes the Cartan matrix of $A_3 \times A_3$.


\clearpage
{
\small
\providecommand{\href}[2]{#2}\begingroup\raggedright\endgroup

}

\end{document}